\documentclass[%
 reprint,
superscriptaddress,
 amsmath,amssymb,
 aps,
floatfix,
]{revtex4-1}

\usepackage{graphicx}
\usepackage{dcolumn}
\usepackage{bm}
\usepackage{epsfig}
\usepackage{amsmath}
\usepackage{pgf}
\usepackage{float}
\usepackage{amssymb}
\usepackage{subcaption}
\usepackage{tikz}
\usepackage{mathptmx}

\newcommand{\E}{\mathbf{E}}
\newcommand{\R}{\mathbf{R}}

\newcommand{\G}{\mathcal{G}}

\newcommand{\sL}{\mathcal{L}}
\newcommand{\dt}{\textrm{dt}}

\newcommand{\dx}{\textrm{d}x}

\newcommand{\sX}{\mathcal{X}}

\renewcommand{\d}{\textrm{d}}
\renewcommand{\epsilon}{\varepsilon}

\begin{document}
\title{Two network Kuramoto-Sakaguchi model under tempered stable L\'evy noise}

\author{Alexander C. Kalloniatis}
\email{alexander.kalloniatis@dst.defence.gov.au}
\affiliation{
Defence Science and Technology Group, Canberra,
ACT 2600, Australia
}

\author{Timothy A. McLennan-Smith}
\email{timothy.mclennan-smith@anu.edu.au}
\affiliation{%
Australian National University, Canberra,
ACT 2601, Australia
}%

\author{Dale O. Roberts}
\email{dale.roberts@anu.edu.au}
\affiliation{%
Australian National University, Canberra,
ACT 2601, Australia
}%

\author{Mathew L. Zuparic}
\email{mathew.zuparic@dst.defence.gov.au}
\affiliation{
Defence Science and Technology Group, Canberra,
ACT 2600, Australia
}

\begin{abstract}
We examine a model of two interacting populations of phase oscillators labelled `Blue' and `Red'. To this we apply tempered stable L{\'e}vy noise, a generalisation of Gaussian noise where the heaviness of the tails parametrised by a power law exponent $\alpha$ can be controlled by a tempering parameter $\lambda$. This system models competitive dynamics, where each population seeks both internal phase synchronisation and a phase advantage with respect to the other population, subject to exogenous stochastic shocks. We study the system from an analytic and numerical point of view to understand how the phase lag values and the shape of the noise distribution can lead to steady or noisy behaviour. Comparing the analytic and numerical studies shows that the bulk behaviour of the system can be effectively described by dynamics in the presence of tilted ratchet potentials. Generally, changes in $\alpha$ away from the Gaussian noise limit, $1< \alpha < 2$, disrupts the locking between Blue and Red, while increasing $\lambda$ acts to restore it.
However we observe that with further decreases of
$\alpha$ to small values, $\alpha\ll 1$, with $\lambda\neq 0$,
locking between Blue and Red may be restored. This is
seen analytically in a restoration of metastability through the
ratchet mechanism, and numerically in transitions between periodic and noisy regions in a fitness landscape using a measure of
noise. This non-monotonic transition back to an ordered regime
is surprising for a linear variation of a parameter such as the
power law exponent and provides a novel mechanism for guiding the
collective behaviour of such a complex competitive dynamical system.

\end{abstract}



\maketitle

\section{Introduction}

Many biological, social and economic examples of complex systems contain both cooperative and competitive phenomena in tension that lead to both ordered and disordered behaviours. Various stylised features of these systems have been studied through the Kuramoto model~\cite{Kur84} of synchronising oscillators and its many extensions; see the review articles~\cite{Strog2000,Aceb2005,DGM2008,ADKMZ2008,DorBull2014}.

In this paper, we study synchronisation in what we call the `Blue vs Red' model~\cite{KallZup2015, HolZupKall2017} under the effect of heavy-tailed stochastic perturbations. This model extends the Kuramoto-Sakaguchi oscillator model (e.g., see~\cite{DGM2008,ADKMZ2008,DorBull2014}) to the case of two interacting populations of oscillators attempting to synchronise with each other under various frustrations. More precisely, we consider $N$ blue agents in a undirected network given by adjacency matrix ${\cal B}_{ij},\ ( i,j \in \{1,\dots,N\}= {\cal B} )$, and $M$ red agents in a undirected network given by adjacency matrix ${\cal R}_{ij},\ (i,j \in \{1,\dots, M \} = {\cal R})$. Let $\beta_i \in \R$ denote the phase of the blue agent $i \in {\cal B}$ and $\rho_i \in \R$ the phase of the red agent $i\in{\cal R}$. The nonlinear dynamical system we consider is
\begin{equation} \label{BR-eq}
\begin{split}
\dot{\beta}_i = \omega_i - \zeta_B \sum_{j\in {\cal B}}{\cal B}_{ij} \sin(\beta_i-\beta_j) 
+  f_{BR}(\beta_i, \rho) + \xi_i,\;\;  i \in {\cal B} \\
\dot{\rho}_i = \nu_i - \zeta_R \sum_{j \in {\cal R}} {\cal R}_{ij} \sin(\rho_i-\rho_j) 
+ f_{RB}(\rho_i, \beta) + \tilde \xi_i, \;\;  i\in {\cal R},
\end{split}
\end{equation}
where $\xi_i, \tilde \xi_i$ are time-varying independent heavy-tailed white noises and $f_{BR}$ and $f_{RB}$ are interaction terms between the red and blue networks. These nonlinear interaction terms are given by
\begin{equation}\label{model}
\begin{split}
  f_{BR}(\beta_i, \rho) &= -\zeta_{BR} \sum_{j \in {\cal R}} {\cal A}^{(BR)}_{ij} \sin(\beta_i-\rho_j-\phi)\\
  f_{RB}(\rho_i, \beta) &= -\zeta_{RB} \sum_{j \in {\cal B}} {\cal A}^{(RB)}_{ij} \sin(\rho_i-\beta_j-\psi),
\end{split}
\end{equation}
where the matrices ${\cal A}^{(BR)}$ and ${\cal A}^{(RB)}$ of size $N\times M$ and $M\times N$ respectively are adjacency matrices  representing the undirected and unweighted edges from `Blue to Red' (BR) and `Red to Blue' (RB). We assume that ${\cal A}^{(BR)} = {{\cal A}^{(RB)}}^T$ so that the edges between blue and red populations are symmetric. These interaction terms contain temporal lags $\phi$ and $\psi$, often called frustrations, between the populations similar to the Sakaguchi variation~\cite{SakKur1986,Cool2003,NVCDGL2013,KirkSev2015} of the Kuramoto model. This typically drives a tendency to disorder in tension with the ordering tendency of synchronisation. 

Extending the Kuramoto model to multi-networks offers a viable approach to modelling competitive dynamics between
ecological, social and organisational populations~\cite{Kall10,Bocc2014,MKB2004,BHOS2008,KNAKK2010,SkarRes2012,KallZup2015,HolZupKall2017}; hence our use of the labels `red' and `blue'. It has been shown that richer behaviour in the dynamics of such models are observed under various forms of Gaussian white or coloured noise perturbations~\cite{Aceb2005,Bag2007,Khosh2008,Toen2010,Zup2013,Esfah2012,Deville2012,Traxl2014,KNAKK2010b}. Recent studies have also shown relationships between the heaviness of the noise distribution's tails and synchronisation behaviour of the Kuramoto model~\cite{KallRob2017,RobKall2017}. The case of tempered stable noise, which we consider here, is particularly interesting due to its relation with tempered fractional diffusion~\cite{MetzKlaft2000,MetzKlaft2004, Mant1994,Kop95,Kullberg12,del-Castillo-Negrete12} and are part of broader class of processes called L\'evy processes. 
At the core is so-called stable L\'evy noise
where the stochastic process arises from a distribution with a heavy tail
and scale law parametrised by an exponent $\alpha$. As
$\alpha$ is decreased from $\alpha=2$ the noise deviates more
strongly from Gaussian noise, and the tails become heavier with
significant changes to the distribution below $\alpha=1$. From a modelling perspective, {\it tempered} stable noise offers a number of useful attributes such as a small number of parameters, finite moments of all order, and a parameter $\lambda$ that controls the tempering of the distribution's tails. Moreover, our approximations
hold in a regime at the interface between order and noisy behaviour, which is often where competing populations seek to operate.

A key question will be how the tempered stable noise
changes dynamics observed for the deterministic \cite{KallZup2015}
and Gaussian stochastic \cite{HolZupKall2017} versions of the Blue-Red system. As in these works, we explore the dynamics of our model from both an analytic and numerical point of view. Indeed, in \cite{KallZup2015}
we showed that much of the full dynamics of the system
could be understood through linearisation around
a fixed point describing respectively internally
synchronised Blue and Red clusters and examining
the system in the vicinity of the regime where
Blue and Red lose relative phase locking, and
the role of a projection of the system into
zero and normal modes of Laplacians of the networks.
In the locked configuration, whichever of Blue
or Red is ahead
in phase by an amount less than $\pi$ may be deemed
to have a competitive advantage over the other. 
The system is thus poised at this interface between order and noisy behaviour,
with a transition through periodic behaviour where
Blue and Red remain internally synchronised but drifting in relation to
each other.
It turns out that this regime corresponds to a 
particle in a 
ratchet potential, which is a superposition of
an oscillatory potential on a linear slope.
In such systems there is an interplay of the
size of the wells and the slope determining if
the particle may become trapped or rolls
down the slope. In the Blue-Red system this
maps to the property that Blue and Red clusters
lock, respectively drift, with respect to each other;
drift would mean that Blue and Red rotate and thus
the overall dynamics is periodic in time.
In \cite{HolZupKall2017} we went on to show
how this structure figured in the
stochastic system, such that noise could
trigger metastability, namely situations
where even though deterministically Blue and Red may
be locked with respect to each other, noise would
enable drift and thus periodic behaviour. Nevertheless
this analysis showed that the deterministic 
fixed point system leaves its fingerprint on the stochastic
dynamics. Nevertheless as the type of noise becomes more exotic
the potential for disruption of global synchronisation in
the pure Kuramoto model, or locking for
multiple populations, increases.

In this paper we uncover a genuinely novel type of behaviour that is not merely incremental to previous insights, namely transitions between order,
periodicity and noisy behaviour within the
aforementioned nonlinear mechanism of
metastability. 
Specifically, we discover a novel mechanism for restoring locking between Blue and Red in the presence of noise, or even diminishing the drift
between the two when the system is deterministically periodic
in behaviour.

The key results of this paper are several: that we may analytically derive transitions to and away from disruption of locking between Blue and Red
for particular parameter ranges of the tempered stable L{\'e}vy noise in a regime
where the deterministic system shows locked behaviour (in Fig.\ref{zeromoderatchet}); that in numerical computation
for the full nonlinear dynamics
both individual paths (Figs.\ref{fig:StableTS3},\ref{fig:TStableTS2}) 
of the fitness landscape 
for frustration parameters $\phi$ and $\psi$ the
imprint of the deterministic dynamics (Fig.\ref{ObjFuncPlots}) survives for weak
noise strength across a range of values
for the power and tempering indices of the 
tempered stable L{\'e}vy noise (Fig.\ref{ObjFuncStochPlots}); and that this
shows a transition from noisy behaviour back to locking 
(Fig.\ref{fig:01testTstable}) in the regime predicted in
the analytic approach. What is genuinely surprising is
that this restoration of locking or suppression of drift in
the presence of tempered stable L{\'e}vy noise occurs for $\alpha\ll 1$ where
naively the heavy tails of the stable noise case are extreme;
in the absence of tempering decreasing $\alpha$ only
destroys locking with monotonically increasing severity
(Fig.\ref{fig:01teststable}).
Both the generalisation of such a two network frustrated
Kuramoto model to
 tempered stable L{\'e}vy noise, and the detection of transitions 
 to/from noisy behaviour
of a nonlinear model of multiple competing populations of
synchronising and frustrated oscillators is, to our understanding, novel.

To this end, this paper is organised in the following manner. Section 2 provides the preliminary details of our setup, in particular this part gives details about the networks interactions and how we model the heavy-tailed noise injected into the system. Section 3 provides an analytical approximation of the dynamics of the system near certain locked states for the deterministic version of the model and details the choice of networks used later in the numerical analysis. Section 4 discusses analytical results for the stochastic model in some special cases. Here we establish the result that the zero modes of the system are described by dynamics in the presence of tilted ratchet potentials manifesting metastability in regimes that would otherwise have been expected to be noisy. Section 5 we focuses on the full nonlinear dynamics of the system using numerical simulations with aim to identify optimal internal synchronisation and phase lag of the Blue network against the Red network.We achieve these results using Bayesian optimisation to maximise an appropriately chosen objective function. We also study the presence of noisy paths given the choice of parameters in our noise through the implementation of the 0-1 test. Section 6 provides a discussion on the extension of our numerical optimisation to a wider parameter set.

\section{Preliminaries}\label{sec:det-BvsR}

In this section we give details on the various elements needed to specify and analyse the model in Eq.(\ref{BR-eq}), detailing the network interactions between the Blue and Red networks, the implementation of the tempered stable L\'evy noise and the measure of sychronisation within the two networks.

\subsection{Network interactions and coupling}

The $(N+M) \times (N+M)$ adjacency matrix for the corresponding external Blue-Red connections, labelled ${\cal M}$ has the following block off-diagonal form,
\begin{eqnarray*}
{\cal M} = \left( \begin{array}{cc}
0 & {\cal A}^{(BR)}\\
{\cal A}^{(RB)}& 0
\end{array} \right).
\end{eqnarray*}
The quantities $\omega_i, \nu_i$ in Eq.(\ref{BR-eq}) give the natural frequencies of the associated Blue and Red agents respectively; they may be drawn from some probability distribution. Finally, $\zeta_B$, $\zeta_R$, $\zeta_{BR}$, $\zeta_{RB}$ in Eqs.(\ref{BR-eq}) and (\ref{model}) are coupling constants, respectively, for intra-Blue, intra-Red, Blue to Red and Red to Blue interactions. Asymmetry between Blue and Red potentially exists in the coupling constants and frustrations. That is, $\zeta_{BR}$ and $\zeta_{RB}$, and $\phi$ and $\psi$ need not be equal. 

\subsection{Tempered stable L{\'e}vy noise}\label{TS}

The stochastic terms in our system can be rewritten by considering a stochastic differential equation formulation of our system by formally multiply both sides of Eq.(\ref{BR-eq}) by $\dt$ and assume $\xi_i\,\dt = \d L_i(t)$ where $L_i = (L_i(t))_{t \ge 0}$ is a L\'evy process for $i=1, \ldots, N$; see~\cite{Sato99} for a general introduction to L\'evy processes. We do the same for the $\tilde \xi_i$ for $i = 1, \ldots, M$ terms which are simply independent copies. This gives us $N + M$ noise terms that we relabel $L_i$ for $i=1, \ldots, N+M$. We now assume that these processes are tempered stable L\'evy processes which is a subclass that has 3 parameters governing the distribution of their increments: the stability parameter $\alpha$, the tempering parameter $\lambda$, and the asymmetry parameter $\theta$. From the L\'evy-Khintchine formula, the characteristic exponent $\Lambda$ satisfying $\E[e^{k L_i(t)]}] = e^{t k \Lambda(k)}$ of the process is given by $\Lambda(k) := C(k)$ for $0 < \alpha < 1$ and $\Lambda(k) := C(k) - 2 i k \alpha \theta \lambda^{\alpha-1}$ for $1 < \alpha < 2$ where
\begin{multline}
    C(k) := - \frac{1}{2 \cos(\alpha \pi / 2)} \left((1+\theta)(\lambda + ik)^\alpha \right. \\ \left. + (1-\theta)(\lambda - ik)^\alpha - 2\lambda^\alpha \right).
\end{multline}
The probability density $p(x,t)$ of each process $L_i$ for $i = 1, \ldots, N+M$ satisfies a tempered fractional diffusion equation $x=L_i(t)$,
\begin{equation}
    \partial_t p(x,t) = c \partial^{\alpha,\theta,\lambda}_x p(x,t),
\end{equation}
where the constant $c > 0$ for $1<\alpha<2$ and $c<0$ for $0<\alpha<1$ and the tempered-fractional-diffusion operator $\partial^{\alpha,\theta,\lambda}_x$ is given explicitly as \cite{Kullberg12,del-Castillo-Negrete12}
\begin{equation}
\partial^{\alpha,\theta,\lambda}_x = {\cal D}^{\alpha,\theta,\lambda}_x + v^{\alpha,\theta,\lambda} \frac{\partial}{\partial x} + \nu^{\alpha,\lambda}
\end{equation}
where $v^{\alpha,\theta,\lambda}$ and $\nu^{\alpha,\lambda}$ are additional drift and source/sink terms given by,
\begin{equation}
v^{\alpha,\theta,\lambda}= \left\{  \begin{array}{cl}
0, & \alpha \in (0,1)\\
\frac{\alpha \theta \lambda^{\alpha-1}}{|\cos(\pi \alpha/2)|} & \alpha \in (1,2)
\end{array}
\right. , \;\; \nu^{\alpha,\lambda} =  \frac{\lambda^{\alpha}}{\cos(\pi \alpha /2)}.
\end{equation}
Here $\alpha \in (0,1) \cup (1,2]$ is the fractional power or stability parameter, governing the heavy-tail law of the distribution generating the noise, 
$\theta \in [-1,1]$ is an asymmetry 
generating skew in the distribution, and $\lambda \in (0,\infty)$ is a tempering parameter that exponentially suppresses large jumps in the process $\eta$.
Importantly, we take the noise as {\it centred} in the Gaussian limit (zero mean).
For stability parameter $\alpha = 2$, the L\'{e}vy noise becomes Gaussian and the tempering $\lambda$ no longer has any effect. Importantly, the value $\alpha=1$
is excluded because it corresponds to the pathological Cauchy process.
The operator ${\cal D}^{\alpha,\theta,\lambda}_x$ is called the $\lambda$-truncated fractional derivative of order $\alpha$, given by,
\begin{equation}
{\cal D}^{\alpha,\theta,\lambda}_x = l(\theta) e^{-\lambda x} \,_{-\infty} D^{\alpha}_x e^{\lambda x} - r(\theta) e^{\lambda x} \,_x D^{\alpha}_{\infty} e^{-\lambda x}
\end{equation}
where the operators $ \,_{-\infty} D^{\alpha}_x$ and $\,_x D^{\alpha}_{\infty}$ are the Riemann-Liouville derivatives defined as 
\cite{del-Castillo-Negrete12}
\begin{align}
e^{-\lambda x} \,_{-\infty} D^{\alpha}_x e^{\lambda x} f(x) &= \frac{e^{-\lambda x}}{\Gamma(m-\alpha)} \frac{\partial^m}{\partial x^m}\int^{x}_{-\infty}\frac{d\zeta e^{\lambda \zeta}}{(x-\zeta)^{\alpha+1-m}} f(\zeta) \\
e^{\lambda x}\,_x D^{\alpha}_{\infty}e^{-\lambda x}f(x) &= \frac{(-1)^me^{\lambda x}}{\Gamma(m-\alpha)} \frac{\partial^m}{\partial x^m}\int^{\infty}_{x}\frac{d\zeta e^{-\lambda \zeta}}{(\zeta-x)^{\alpha+1-m}} f(\zeta)
\label{R-Lderiv}
\end{align}
for $\alpha-1< m < \alpha$. These can be equivalently defined in Fourier space \cite{Podlubny99,Samko93}
\begin{eqnarray}
\begin{split}
{\cal F} \left[e^{-\lambda x} \,_{-\infty} D^{\alpha}_x e^{\lambda x} f(x) \right] = (\lambda-i k)^{\alpha} \hat{f}(k)\\
{\cal F} \left[e^{\lambda x} \,_x D^{\alpha}_{\infty} e^{-\lambda x} f(x) \right] = (\lambda+i k)^{\alpha} \hat{f}(k)
\end{split}
\end{eqnarray}
where
\begin{eqnarray}
\begin{split}
{\cal F} \left[ f(x) \right] = \int^{\infty}_{-\infty}dx e^{i k x}f(x) =  \hat{f}(k)\\
{\cal F}^{-1} \left[ \hat{f}(k) \right] = \int^{\infty}_{-\infty} \frac{dk}{2\pi} e^{-i k x}\hat{f}(k) =  f(x).
\end{split}
\label{fourierdef}
\end{eqnarray}
The weighting factors
\begin{equation}
l(\theta) = \frac{\theta-1}{2 \cos(\pi \alpha/2)}, \;\; r(\theta) = \frac{\theta+1}{2 \cos(\pi \alpha/2)}
\end{equation}
determine the asymmetry imposed on each of the Riemann-Liouville derivatives. 
It is also useful to take the Fourier transform of the fractional derivative $ \partial^{\alpha,\theta,\lambda}_x$ giving
\begin{multline}
{\cal F} \left[  \partial^{\alpha,\theta,\lambda}_x  f(x) \right] = \\ \underbrace{\left\{l(\theta) (\lambda-i k)^{\alpha} - r(\theta)(\lambda+i k)^{\alpha} - i k v^{\alpha,\theta,\lambda} +  \nu^{\alpha,\lambda} \right\}}_{= \Lambda(k)/\sigma^2}\widehat{f}(k)
\label{lambda}
\end{multline}
where $\Lambda(k)$ is the characteristic exponent (given earlier) and and $\sigma^2$ becomes the variance of the noise in the Gaussian limit. The centred noise is specified by four parameters $\alpha$,$\sigma$,$\theta$, $\lambda$, generalising from the one parameter of mean-zero Gaussian noise, $\sigma$, that we obtain in the limit as $\alpha \to 2$.

\subsection{Measures of synchronisation}

To measure the degree of synchronisation within a given population we use local forms of Kuramoto's order parameter~\cite{Kur84} given by
\begin{eqnarray*}
O_B = \frac{1}{N}\left|  \sum_{j\in {\cal B}} e^{i\beta_j} \right|, \;\; O_R = \frac{1}{M} \left| \sum_{j\in {\cal R}} e^{i\rho_j} \right|.
\end{eqnarray*}
We emphasise that in many of our examples in the following the total system (`global')  order parameter 
$1/(N+M) | \sum_{i\in{\cal B}} e^{i\beta_i} +\sum_{j\in{\cal R}} e^{i\rho_j}|$
will be far from the value one.

\section{Bulk Dynamics Near Locked States}

In this section we make various approximations to understand the bulk dynamics analytically when the system is near certain locked states. In Section~\ref{sec:full} we return to the studying the full system (without approximations), albeit numerically.

There are various ways that the system can be deemed locked. We refer to the phase locking within Blue, that is $\beta_i \approx \beta_j\; \forall\; i,j \in {\cal B}$, 
and within Red $\rho_i \approx \rho_j\; \forall \; i,j \in {\cal R}$, as \emph{internal} or \emph{local locking}, and the
phase locking of Blue externally with Red, $\beta_i \approx \rho_j \; \forall \;\{i,j\} \in \{{\cal B},{\cal R}\}$, as \emph{external} or \emph{global phase locking}. If we denote by $B$ and $P$ the centroids of the Blue and Red phases given by
\begin{eqnarray}
B= {1\over{N}} \sum_{i\in {\cal B}} \beta_i , \qquad P= {1\over{M}} \sum_{i \in {\cal R}} \rho_i. \label{BPdef}
\end{eqnarray}
and consider the difference
\begin{eqnarray} \label{centroid}
\Delta(t) \equiv B(t)-P(t)  \label{fixpoint-alph},
\end{eqnarray}
then if $\Delta\neq 0$ is time-independent then we may speak of external \emph{frequency locking}.

\subsection{The two cluster ansatz}
\label{lin-technique}

In general, the system in Eq.(\ref{BR-eq}) can only be solved numerically. To gain analytic insight in a region of greatest relevance, given that Blue and Red may deem internal phase synchronisation ideal, we explore a fixed point given by the ansatz
\begin{equation}
\beta_i(t)  =  B(t) + b_i(t), \;\; \rho_j  =  P(t) + p_j(t), \;\; \forall \; \{i,j\} \in \{ {\cal B},{\cal R}\}, \label{fixpoint1}
\end{equation}
with $b_i,p_j$ small so that $b_i^2 \approx 0$,  $p^2_j \approx 0$. This leads to $\sum_{i\in{\cal B}} b_i = \sum_{i\in{\cal R}} p_i = 0$. To correctly take into account the noise terms, we rewrite our system in the stochastic differential equation formulation by formally multiply both sides of Eq.(\ref{BR-eq}) by $\dt$ and assume $\xi_i\,\dt = \d L_i(t)$ where $L_i(t)$ is a tempered stable L\'evy process as introduced in Section~\ref{TS}. We then linearise the system given by Eq. (\ref{BR-eq}), keeping only terms first order in a Taylor expansion, then using the ansatz given by Eqs. (\ref{fixpoint1}) and (\ref{fixpoint-alph}) we obtain the linearised dynamics
\begin{equation}\label{linsys}
\begin{aligned}
\d b_i(t)&= [-B(t) + \Omega_i(t) - \zeta_B \sum_{j \in {\cal B}}\sL^{(B)}_{ij}b_j(t)  \\
&\qquad - \zeta_{BR} \cos(\Delta(t)-\phi) \sum_{j \in {\cal B} \cup {\cal R}} \sL^{(BR)}_{ij} v_j]\dt+\d L_i(t),\\
\d p_i(t)&= [-P(t) + \Omega_i(t) - \zeta_R \sum_{j \in {\cal R}}\sL^{(R)}_{ij}p_j \\
&\qquad  - \zeta_{RB} \cos(\Delta(t)+\psi) \sum_{j \in {\cal B} \cup {\cal R}} \sL^{(RB)}_{ij} v_j]\dt + \d L_i(t),
\end{aligned}
\end{equation}
where
\begin{eqnarray*}
v_i = \left\{ \begin{array}{cc} b_i & i \in {\cal B} \\ p_i & i \in {\cal R} \end{array} \right. , \;\; \Omega_i(t) = \left\{ \begin{array}{cc}  \omega_i -  \zeta_{BR}\,d^{(BR)}_i \sin(\Delta(t) - \phi)\\ 
\nu_i +  \zeta_{RB}\,d^{(RB)}_i \sin(\Delta(t) + \psi).
\end{array}
\right. 
\end{eqnarray*}
The quantities $d$ and $\sL$, with superscripts $B, \ R, \ BR$ and $RB$,
represent respectively the degree and corresponding Laplacian for the Blue, Red, Blue-Red and Red-Blue networks using the matrix ${\cal M}$; see Appendix A. Note that the Laplacian is defined as $\sL := D-A$ where $D$ is the degree matrix and $A$ is the adjacency matrix of the respective networks; see \cite{Boll98,Pec-Car98,Kall10} for the significance of the Laplacian spectrum.


\subsection{Decoupling and projecting the dynamics}
\label{sec:decouple}

In order to proceed analytically we assume that 
\begin{eqnarray}
\sum_{j \in {\cal B} \cup {\cal R}} \sL^{(BR)}_{ij} v_j \approx \sum_{j \in {\cal B} \cup {\cal R}} \sL^{(RB)}_{ij} v_j \approx 0,
\label{interactionsaregone}
\end{eqnarray}
so that the equations for the fluctuations $b_i$ and $p_j$ in Eq.(\ref{linsys}) may be decoupled through the use of the two intra-population Laplacians $\sL^{(B)}$ and $\sL^{(R)}$. As mentioned in \cite{KallZup2015, HolZupKall2017}, this approximation is not guaranteed to completely hold, either deterministically or under the influence of noise, in model regimes which enable Eq.(\ref{fixpoint1}) to be satisfied; we apply it to generate qualitative analytic expressions to enable understanding of macroscopic model behaviours. Nevertheless, for regimes of weak noise
it is seen to hold sufficiently that such
insights are valid \cite{KallZup2015, HolZupKall2017}.

We factorise the Laplacian $\sL^{(B)}$ using an eigendecomposition and order the eigenvalues as $0 = \lambda_0^{(B)} \le \lambda_1^{(B)} \le \cdots \le \lambda_N^{(B)}$. Each eigenvalue has an associated eigenvector forming a complete spanning set of orthonormal eigenvectors for the $N$ dimensional subspaces that we label as $e^{(B,r)}$ for $r=0,1,\dots, N-1 \in {\cal B}_E$. We perform the same procedure for $\sL^{(R)}$, giving eigenvectors $e^{(R,r)}$ for $r=0,1,\dots, M-1 \in {\cal R}_E$ with associated eigenvalues $\lambda^{(R)}_r$. We distinguish between indices in node space $\{ {\cal B}, {\cal R} \}$, 
and those in eigenmode space $\{ {\cal B}^E, {\cal R}^E \}$ (which has the same dimensionality) by reserving labels $\{i,j\}$ for expressions involving graph nodes, and labels $\{r,s\}$ for expressions involving Laplacian eigenmodes. We assume that Blue and Red networks each consist of one component. Up to normalisation, the corresponding zero eigenvectors ${\vec e}^{(B,0)}$ and ${\vec e}^{(R,0)}$ consists of all unit valued entries~\cite{Boll98}. 

We now successively project the dynamics in Eq.(\ref{linsys}) onto the increasing eigenmodes $r=0,1,\dots, N-1 \in {\cal B}_E$ and $r=0,1,\dots, M-1 \in {\cal R}_E$. Combined with Eq.(\ref{BPdef}), we see that $B$ and $P$ are the zero-mode projections of the phases $\beta_i$ and $\rho_j$. Analogously, we denote by $x_r$ and $y_s$ the projections of $b_i$ and $p_i$ on the Blue and Red non-zero eigenvectors. We give explicit expressions in Appendix A.

First we assume that the noise is zero to understand the core dynamics of the system. Using Eq.(\ref{interactionsaregone}) and the eigenvector projections in Eq.(\ref{linsys}) combined with the orthonormality of the eigenvectors gives
\begin{align}\label{integrablesystem}
\dot{x}_r &= \omega^{(r)}-\zeta_B \lambda^{(B)}_r x_r - \zeta_{BR}d^{(BR)}_r\sin(\Delta-\phi), \\
\dot{y}_s &= \nu^{(s)}-\zeta_R \lambda^{(R)}_s y_s + \zeta_{RB}d^{(RB)}_s \sin(\Delta+\psi), \\
\dot{B} &= \bar{\omega} -\frac{\zeta_{BR}d^{(BR)}_T}{N}\sin(\Delta-\phi), \\
\dot{P} &= \bar{\nu} +\frac{\zeta_{RB}d^{(BR)}_T}{M}\sin(\Delta+\psi),
\end{align}
for $r \in {\cal B}^E / \{0\}$ and $s \in {\cal R}^E / \{0\}$. The terms $\omega^{(r)}$ and $d^{(BR)}_r$ are the projections onto the $r-$th eigenvector and $\bar{\omega}$ is the average over ${\cal B}$ with $d^{(RB)}_s$ the projections onto the $s-$th eigenvector and $\bar\nu$ the average frequency for Red. The difference of $\dot{B}$ and $\dot{P}$ gives then
\begin{eqnarray}
\dot{\Delta} = - V'(\Delta),
\label{Deltaeq}
\end{eqnarray}
where 
\begin{eqnarray}
V(\Delta) &=& -\mu \Delta -\sqrt{S^2+C^2}\cos\left( \Delta -  \varrho \right) \label{potential} \\
\mu &=& (\bar{\omega}-\bar{\nu})\\
\varrho &\equiv& \tan^{-1} \left( S/C\right), \label{rhodef} \\
C  &\equiv&  d^{(BR)}_T \left(\frac{ \zeta_{BR} \cos\phi}{N} +\frac{ \zeta_{RB} \cos\psi}{M}\right), \\
S &\equiv& d^{(BR)}_T \left(  \frac{ \zeta_{BR}\sin\phi}{N}  -  \frac{\zeta_{RB} \sin\psi   }{M}\right).
\end{eqnarray}
Thus $\Delta$ may be solved first from Eq.(\ref{Deltaeq}), and then
used in the forcing terms of the normal mode equations in Eqs.(\ref{integrablesystem}) which are otherwise solvable in their own right.

The substitution, $\Delta = \varrho + 2 \tan^{-1} \left( \vartheta + \frac{\sqrt{S^2+C^2}}{\mu} \right)$, leads to the solution
\begin{multline} 
\Delta(t) = \varrho+ 2 \tan^{-1} \left\{ \sqrt{\frac{S^2+C^2}{\mu^2}} \right. \\ 
\left. + \sqrt{\frac{ {\cal K}}{\mu^2} } \tanh \left( \frac{\sqrt{{\cal K}}}{2} (\textrm{const} - t)  \right) \right\}
\label{Deltasol}
\end{multline}
where $\textrm{const} = \frac{2}{\sqrt{{\cal K}}} \textrm{tanh}^{-1}\left\{ \frac{\mu}{\sqrt{{\cal K}}}\vartheta_0 \right\}$, and
\begin{eqnarray}\label{eq:special-K}
{\cal K} = C^2 + S^2 -\mu^2. 
\end{eqnarray}
The solutions to the normal modes are given in Appendix A. Ultimately their dynamics depends on the behaviour of $\Delta(t)$ which in turn is governed by
the potential $V(\Delta)$ in Eq.(\ref{Deltaeq}). This is commonly referred to as a tilted periodic~\cite{Lin-Kos-Sch2001,Khangjune,Challis} or tilted Smoluchowski-Feynman ratchet~\cite{Reimann2002} potential. Importantly, we have periodicity, $V'(\Delta) = V'(\Delta+ 2\pi)$, and the tilt refers to the constant forcing term in $V'(\Delta)$ given by the difference of frequency averages $\mu$. The sign of ${\cal K}$ is critical. If ${\cal K} > 0$ the solution asymptotes to the value $\varrho+ \sin^{-1} \left( \frac{\mu}{\sqrt{S^2 + C^2}} \right)$ mod $2 \pi$. If ${\cal K} < 0$ then Eq.(\ref{Deltasol}) gives oscillatory 
behaviour with period $\frac{2 \pi}{\sqrt{ |{\cal K}| }}$. 
Alternately, for ${\cal K} > 0$, $V(\Delta)$ is a series of local maxima  with unstable fixed points at $\Delta = \pi + \varrho - \sin^{-1} \left( \frac{\mu}{\sqrt{S^2 + C^2}} \right)+2 \pi n, \; n \in \mathbb{Z} $) and local minima with stable fixed points at $\Delta = \varrho+ \sin^{-1} \left( \frac{\mu}{\sqrt{S^2 + C^2}} \right)+2 \pi n, \; n \in \mathbb{Z}$ on a landscape which has an overall slope according to the sign of $\mu$.
For ${\cal K} = 0$ the hills and valleys of the potential become points of inflection, and hence unstable fixed points. For ${\cal K} <0$, the potential loses all of its fixed points, even the unstable ones. We may also
speak of {\it drift} when there is no stable fixed point.
This implies that Blue and Red clusters will be in motion
with respect to each other.

\subsection{Projecting the noise}
\label{LaplProjModel}

We now consider how the noise gets projected onto the eigenmodes of the Laplacians. At every time point $t$, the increments of the processes $L^{(B)}_i(t)$ and $L^{(R)}_i(t)$ may be decomposed in terms of eigenvectors of the Blue/Red Laplacians as
\begin{equation}\label{eq:noise-terms-sums}
\begin{split}
\d L_i^{(B)}(t)=\sum_{r\in{\cal B}_E }{\gamma_r^{(B)}e_i^{(B,r)}\d\eta_r^{(B)}(t)},\;\;i\in\mathcal{B}, \\
\d L_i^{(R)}(t)=\sum_{r\in{\cal R}_E }{\gamma_r^{(R)}e_i^{(R,r)}\d\eta_r^{(R)}(t)},\;\;i\in\mathcal{R}
\end{split}
\end{equation}
where $\eta_r^{(B)}$ and $\eta_r^{(R)}$ are also tempered stable L\'evy processes but with different parameters values. This follows from the fact that projections of L\'evy processes are also L\'evy processes; see Proposition 11.10 in \cite{Sato99}. 


In \cite{RobKall2017} we derived the general expression for projection onto an $N$-dimensional vector $u_j$ of the tempered stable L{\'e}vy process $\eta_i$
\begin{eqnarray}
\d L_{\rm{proj}}=\sum_{j=1}^N u_j \d \eta_j.
\end{eqnarray}
For the noise $\eta$ this is defined as a function of  the parameters $\alpha,\sigma,\theta,\lambda$, as in Eq.(\ref{lambda}). Then for the projected process $L_{\rm{proj}}$ the characteristic exponent is
\begin{multline}
\Lambda_{\rm{proj}}(k) = \sigma^2 \sum_{j=1}^N \left[ (1+\theta) (\lambda + i k u_j)^{\alpha} \right.\\
\left.+ (1-\theta) (\lambda-i k u_j)^{\alpha} - i k u_j v^{\alpha,\theta,\lambda} + \nu^{\alpha,\lambda} \right].
\label{projcharfunc}
\end{multline}
Essentially, the `momentum' $k$ is multiplied by the component of the vector $u_j$.

The system given in Eq.(\ref{integrablesystem}) with the noise re-introduced is a Langevin (aka. Ornstein-Uhlenbeck) system, written in SDE form as
\begin{align}
\label{eqn:integrablesystem-w-noise}
\d x_r(t) &= q_r^{(B)}(x_r(t),\Delta) \dt +\gamma_r^{(B)}\d\eta_r^{(B)}(t), \\
\d {y}_s(t) &= q_s^{(R)}(y_s(t),\Delta) \dt +\gamma_s^{(R)}\d\eta_s^{(R)}(t),\\
\d {\Delta}(t) &= -V'(\Delta)\dt+\gamma_0^{(B)}\d\eta_0^{(B)}(t) -\gamma_0^{(R)}\d\eta_0^{(R)}(t), \label{zeromodeLang}
\end{align}
for $r \in {\cal B}^E / \{0\}$, $s \in {\cal R}^E / \{0\}$, and where the variables and parameters are as used in Eq.\eqref{integrablesystem}.

\subsection{Measures of synchronisation and the fixed point approximation}

Something absent in our previous work \cite{KallZup2015,HolZupKall2017} is the centroid representation for order parameters
\begin{align}
O_B&=& \frac{1}{N} \sqrt{\sum_{i,j\in{\cal B}} \cos^2(\beta_i-\beta_j)} 
= \frac{1}{N} \sqrt{\sum_{i,j\in{\cal B}} \cos^2(b_i-b_j)}, \label{O_Btrunc} \\
O_R&=& \frac{1}{N} \sqrt{\sum_{i,j\in{\cal R}} \cos^2(\rho_i-\rho_j)}
= \frac{1}{N} \sqrt{\sum_{i,j\in{\cal R}} \cos^2(p_i-p_j)}. \label{O_Rtrunc}
\end{align}

In the fixed point approximation
when ${\cal K}>0$ (giving $\Delta$ constant), $b_i,p_i$ may be determined from the stationary solutions for $x_r, y_r$ in Eqs.(\ref{integrablesystem})
which may be substituted into Eqs.(\ref{O_Btrunc},\ref{O_Rtrunc}). Alternatively, and reintroducing the $\sL^{BR}, \sL^{RB}$ matrices,
Eq. (\ref{linsys}) in the static limit may also be compactly written in terms of a `super-Laplacian' matrix ${\cal L}$,
\begin{eqnarray}
{\vec\Omega} - {\cal L} {\vec v} = 0 \label{superLaplV},
\end{eqnarray}
where the off-diagonal parts of the matrix ${\cal L}$ incorporate the terms ignored thus far, Eq.(\ref{interactionsaregone}). The super-Laplacian, given explicitly
in \cite{KallZup2015} is not symmetric but has zero column sum and therefore a single zero right eigenvalue. Thus Eq.(\ref{superLaplV}) is invertible through the pseudo-inverse of the super-Laplacian (even though it may only be computed numerically, in general) and used in Eqs. (\ref{O_Btrunc}),(\ref{O_Rtrunc}).

Importantly, this brings together information about the network structure through the
Laplacian, but also on the nature of the threshold in ${\cal K}$ through the implicit dependence on $\Delta$. Thus any deterministic optimisation of $O_B$ (for example) in relation to parameters $\phi$ and $\psi$ will encode the transition to periodic behaviour at  ${\cal K}=0$. We will show a numerical example of this below when we consider the fitness landscape of the deterministic system.

\subsection{Example networks, frequencies and couplings}
To illustrate our otherwise general solutions and compare to numerical
simulation we consider an example of Blue agents in a hierarchy and Red agents on a random network.
As in \cite{KallZup2015,HolZupKall2017}, for ${\cal B}$ we consider a complete 4-ary tree, thus $N=21$.
For ${\cal R}$ we use a random Erd\"os-R\'enyi network also of $N=21$ with link of probability $0.4$.
The two networks, and further details of their interconnection, are shown in Appendix C in Fig.\ref{fig:BvsR-networks}.

The specific frequencies of each agent are drawn from a uniform distribution $\omega_i,\nu_j\in [0,1]$ and are plotted in Appendix C. In this instance
the average frequencies are $\bar{\omega}=0.503, \bar{\nu}=0.551$, 
giving $\mu=-0.048$. If cross-couplings were set to zero the Red population would lap Blue over time. The
sign of $\mu$ here implies a negative slope for the tilted ratchet potential $V(\Delta)$.

We choose coupling values
\begin{eqnarray}
\zeta_B=8,\quad \zeta_R=0.5,\quad\zeta_{BR}=\zeta_{RB} = 0.4. \label{coupls}
\end{eqnarray}
These values give high internal phase synchronisation \cite{KallZup2015,HolZupKall2017}. Note that $\zeta_B>\zeta_R$, reflecting the difficulty
of hierarchies to synchronise. In this respect we
have selected values that put quite different network 
structures on an approximately equal footing.
Nevertheless, we derive results for the general network case.
These couplings do 
allow for some changes in dynamics as the frustrations are varied; specifically we set $\psi=0$ and manipulate $\phi$.
Then ${\cal K}$ changes sign from positive to negative at $\phi_c=0.9498\pi$. In the vicinity
of this point, in the absence of
noise, for $\phi<\phi_c$ Blue locks to a fixed phase ahead of Red, and for $\phi\geq \phi_c $ Blue and Red remain internally
phase synchronised but lap each other with a period that decreases as $\phi$ gets larger. Thus we shall be interested in behaviours
slightly below and above this point, for example $\phi=0.94\pi$ and $\phi=0.96\pi$.

\section{Analytical Results}

In this section, we derive some analytic results when the system is perturbed by exogenous stochastic shocks. We seek to understand the bulk behaviour through a projection onto an expansion in terms of the Red/Blue Laplacians.

\subsection{Noisy normal modes subject to tempered stable noise}

We test the consequences of noise applied  to zero or normal modes via the selections:
\begin{itemize}
\item{$\gamma_0^{(B)}=\gamma_0^{(R)}=0$, $\gamma_r^{(B)}=\gamma_s^{(R)}=1$, $\{r,s \} \in \{ {\cal B}^E / \{0\}, {\cal R}^E / \{0\}  \}$ has noise applied to normal modes;} 
\item{$\gamma_0^{(B)}=\gamma_0^{(R)}=1$, $\gamma_r^{(B)}=\gamma_s^{(R)}=0$, $\{r,s \} \in \{ {\cal B}^E / \{0\}, {\cal R}^E / \{0\}  \}$ has noise applied to zero modes.}
\end{itemize}
These choices then specify the random variables in the stochastic system. In the full problem of noise applied generally, there are three sets of random variables, two vector and one scalar, corresponding to the deterministic variables $(\vec{x},\vec{y},\Delta)$.
These correspond to Blue and Red Laplacian projections and the phase difference between centroids.



We consider first the case $\gamma_0^B=\gamma_0^R=0$ where L{\'e}vy noise is introduced on normal mode fluctuations
on the background of the deterministic $\Delta$. The latter is governed by the solution Eq.\eqref{Deltasol} which tends to a fixed value as $t\to\infty$ or 
is periodic, according to the sign of ${\cal K}$ in Eq.\eqref{eq:special-K}. We show in the following how with tempered L{\'e}vy noise this devolves to a
system we have solved elsewhere \cite{ZupKall2017}; we will not give the solution here but explain the behaviour it implies and why it is less interesting than
the zero mode case.

As $\Delta$ is no longer a random variable, 
in this case, all the normal mode Langevin equations in Eq.(\ref{eqn:integrablesystem-w-noise}) are independent. 
We may therefore consider the fractional version of a Langevin equation for the Blue modes $x_s$ - the corresponding equations
for $y_s$ are easily translated from Eq.(\ref{eqn:integrablesystem-w-noise}) - 
\begin{equation}
\dot{x}_s = q_s^{(B)}(x_s,\Delta) + \dot L_s^{\alpha,\theta,\lambda}(t)
\label{defining-Lang}
\end{equation}
where $L^{\alpha,\theta,\lambda}(t)$ is a tempered stable L{\'e}vy process in time described by parameters $\alpha,\theta,\lambda$. 
For ease of notation we will first take the noise as if applied within the normal mode projected system, and translate back to
the original basis subsequently. Because, in the approximation that off-diagonal entries of the super-Laplacian are taken to be zero,
the $u_j$ in Eq.(\ref{projcharfunc}) are, for the Blue system, the eigenvector components $e^{(B,r)}_j, j=1,\dots,N$.

Thus the corresponding Fokker-Planck equation can be decomposed into a product of $N+M-2$ densities,
$${\cal P}(\vec{x},\vec{y},t)=\prod_{r\in{\cal B}^E / \{0\}}{{\cal P}^{(B,r)}(x_r,t)}\prod_{s\in{\cal R}^E / \{0\}}{{\cal P}^{(R,s)}(y_s,t)}$$
leading to a decoupling into separate fractional Fokker-Planck equations for each mode.

The probability density ${\cal P}(x,t)$ associated with the tempered-stable L\'{e}vy process given by Eq.(\ref{defining-Lang}) 
satisfies the fractional Fokker-Planck equation \cite{Cartea07},
\begin{align}
\frac{\partial}{\partial t} {\cal P}^{(B,r)}(x,t) =& \left\{ \sigma^2 \partial^{\alpha,\theta,\lambda}_{x_r} 
- \frac{\partial}{\partial {x_r}} q_r^{(B)}(x_r,t)  \right\} {\cal P}^{(B,r)}(x_r,t), \\
 {\cal P}^{(B,r)} (x_r,0) =& \delta(x_r-x'_r),
\label{defining-FP}
\end{align}
where $\sigma^2 \in (0,\infty)$ is the diffusion constant - which becomes the variance of the process in the Gaussian limit. 
Unpacking $q_r^{(B)}$ gives the explicit form of the tempered-fractional-Fokker-Planck equation (TFFP)
for the normal mode projection of the Blue-vs-Red system
\begin{eqnarray}
\frac{\partial}{\partial t} {\cal P}^{(B,r)}(x_r,t) &=& \big\{ \sigma^2 {\cal D}^{\alpha,\theta,\lambda}_{x_r} + 
 \left[ \zeta_B \lambda_r^{(B)} x_r + \sigma^2  v^{\alpha,\theta,\lambda} \right. \nonumber \\
&&\left. - (\omega^{(r)} - \zeta_{BR} d_r^{(BR)} \sin(\Delta(t) - \phi))
\right] \frac{\partial}{\partial {x_r}}  \nonumber \\
&& + \left[ \zeta_B \lambda_r^{(B)} + \sigma^2 \nu^{\alpha,\lambda}  \right] \big\} {\cal P}^{(B,r)}(x_r,t). 
\label{TFFP}
\end{eqnarray}
For the Gaussian limit, i.e. $\partial^{\alpha,\theta,\lambda}_x = \frac{\partial^2}{\partial x^2}$, section 1.8.3.6 of \cite{Polyanin2002} gives 
a set of nonlinear transformations which result in the Gaussian equivalent of Eq.(\ref{TFFP}) becoming the standard heat equation. Indeed, this technique enabled the analytical investigation of clustering effects in the frustrated Kuramoto model under the influence of Gaussian white noise \cite{HolZupKall2017}.  

As alluded several times, we have solved a version of this problem in \cite{ZupKall2017} where the sign of ${\cal K}$ and the shape of the
underlying noise distribution interplay in determining the solution. For ${\cal K}>0$, where $\Delta$ asymptotes to a constant value, the
probability density broadens from a delta function to a tempered stable density of corresponding characteristics given by
$\alpha,\theta,\lambda$. For ${\cal K}<0$, where $\Delta$ becomes oscillatory, the Fokker-Planck density evolves from the delta function to
 a tempered stable density of {\it fixed shape} (again, according to $\alpha,\theta,\lambda$) but whose centre oscillates according to the
 dynamics of $\Delta$. This is a natural generalisation of the behaviour seen for the Gaussian case in \cite{HolZupKall2017}.
 
 These behaviours will result in corresponding dynamics for the order parameters $O_B(t)$ and $O_R(t)$ for weak values of
 $\sigma$. For ${\cal K}>0$,
 these will exhibit noisy behaviour (showing tempered-heavy-tails) around a steady-state value, while for ${\cal K}<0$ there will be similar noisy
 behaviour around an oscillatory background. Importantly, if noise is applied across all nodes independently, we expect
 a residual imprint of the sign of ${\cal K}$ in the dynamics. In this respect, there is little that is surprising in the normal modes and so we turn to the
 potentially more interesting case of the zero modes.

\subsection{Noisy zero modes}

Taking $\gamma_i=0$ and $\gamma_0=1$, we consider the tempered fractional Fokker-Plank equation with the tilted ratchet potential from Langevin form Eq.(\ref{zeromodeLang}) and wrap their densities ${\cal P}(x,t)$ onto $\mathbb{S}^1$.

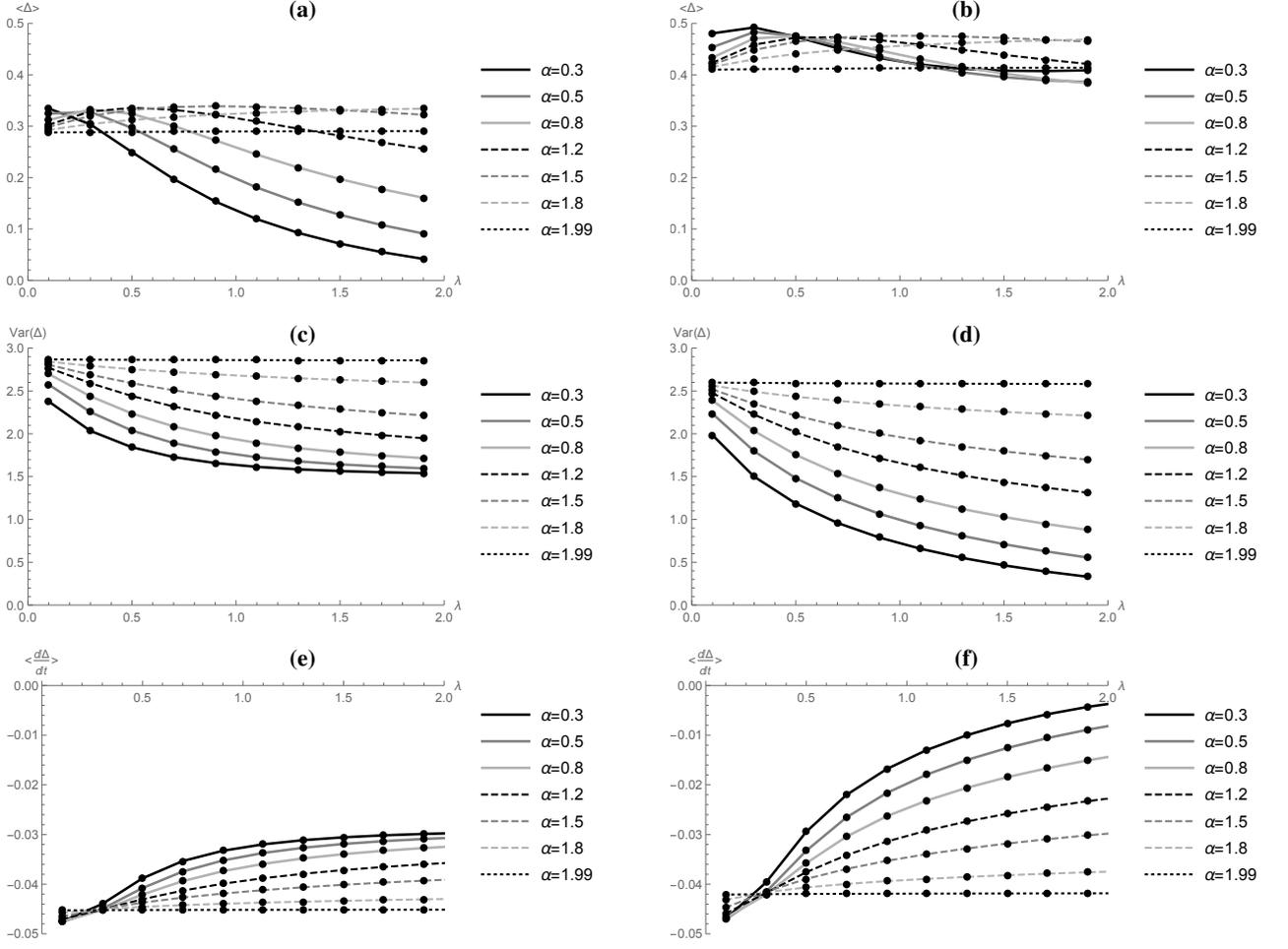
\begin{figure*}

\begin{subfigure}[b]{0.5\linewidth}
\centering
\begin{tikzpicture}
    \draw node[name=figure] {\includegraphics[width=8cm]{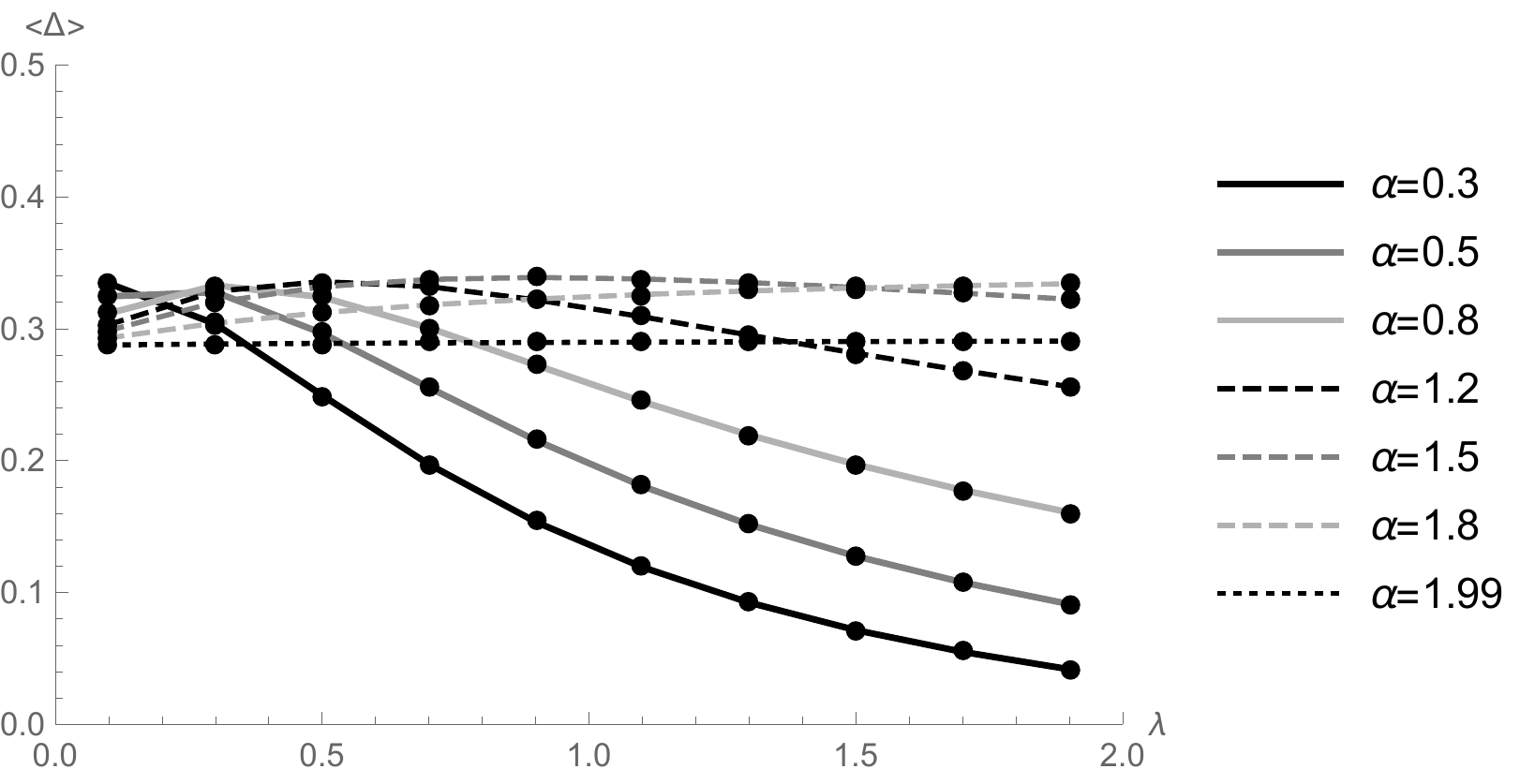}};
    \draw  (figure.north)  node[anchor=north ,yshift=-1,black]{\textbf{\small{(a)}}};
\end{tikzpicture}
\end{subfigure}\hfill
\begin{subfigure}[b]{0.5\linewidth}
\centering
\begin{tikzpicture}
    \draw node[name=figure] {\includegraphics[width=8cm]{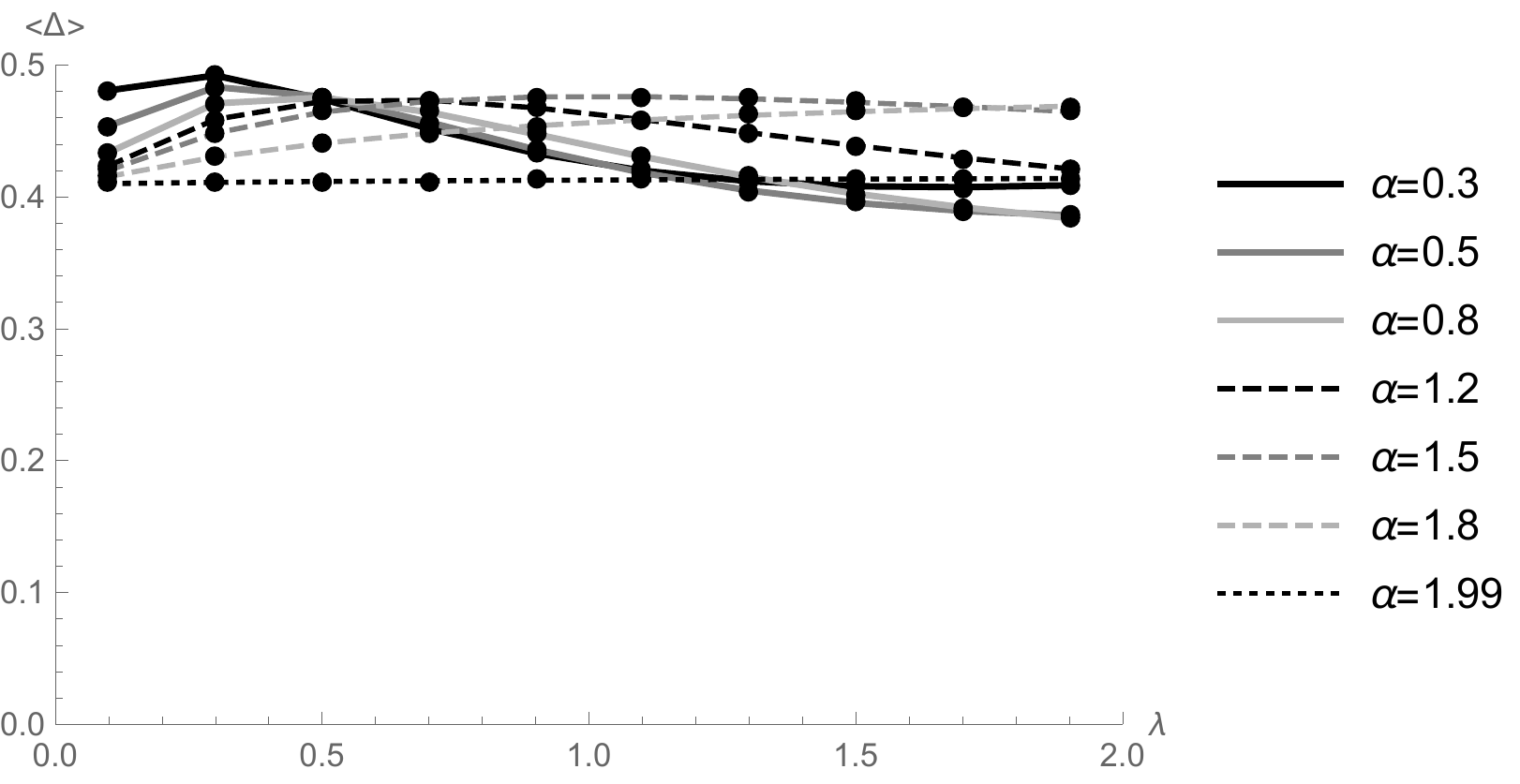}};
    \draw  (figure.north)  node[anchor=north,yshift=-1,black]{\textbf{\small{(b)}}};
\end{tikzpicture}
\end{subfigure}\\
\begin{subfigure}[b]{0.5\linewidth}
\centering
\begin{tikzpicture}
    \draw node[name=figure] {\includegraphics[width=8cm]{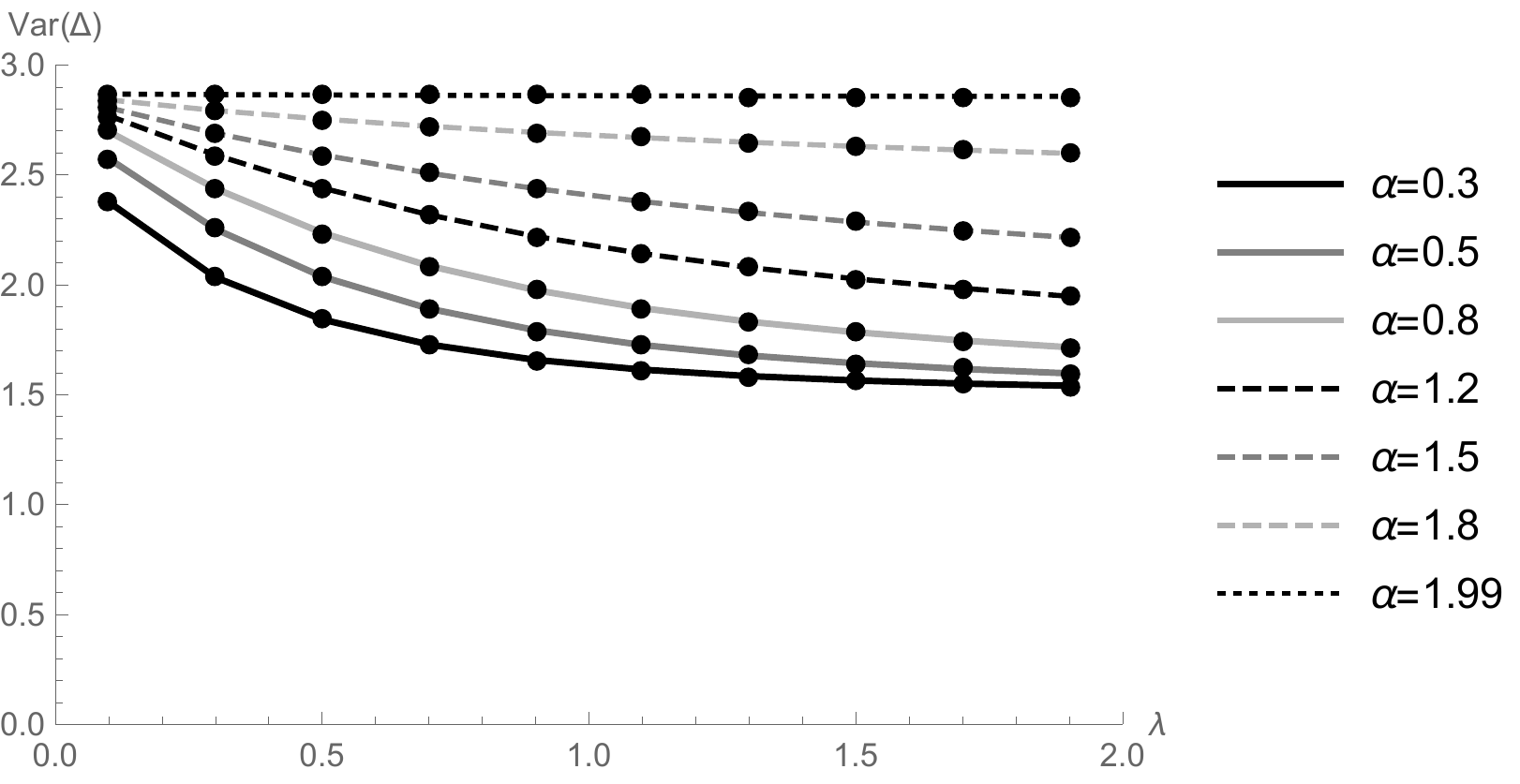}};
    \draw  (figure.north)  node[anchor=north,yshift=-1,black]{\textbf{\small{(c)}}};
\end{tikzpicture}
\end{subfigure}\hfill
\begin{subfigure}[b]{0.5\linewidth}
\centering
\begin{tikzpicture}
    \draw node[name=figure] {\includegraphics[width=8cm]{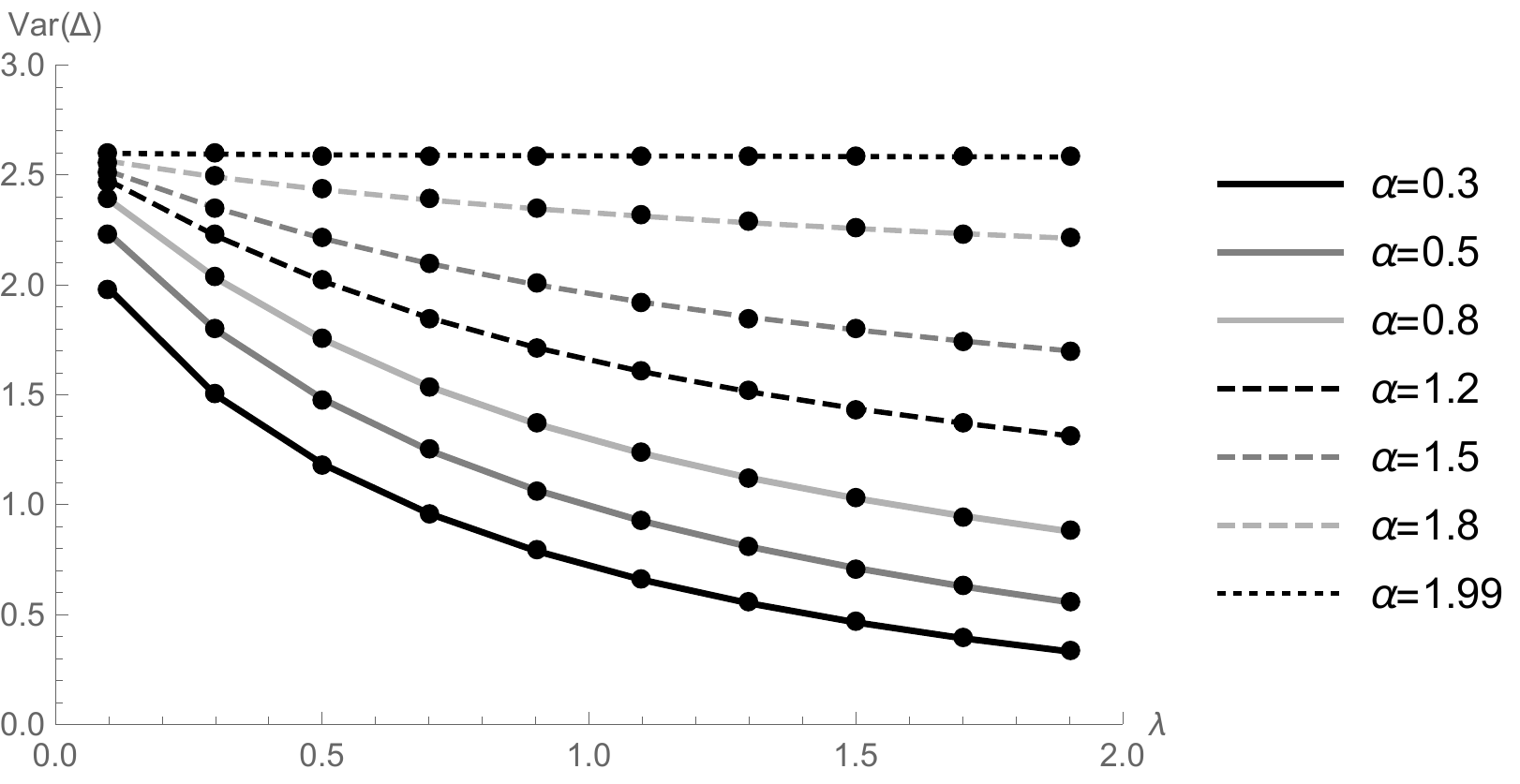}};
    \draw  (figure.north)  node[anchor=north,yshift=-1,black]{\textbf{\small{(d)}}};
\end{tikzpicture}
\end{subfigure}\\
\begin{subfigure}[b]{0.5\linewidth}
\centering
\begin{tikzpicture}
    \draw node[name=figure] {\includegraphics[width=8cm]{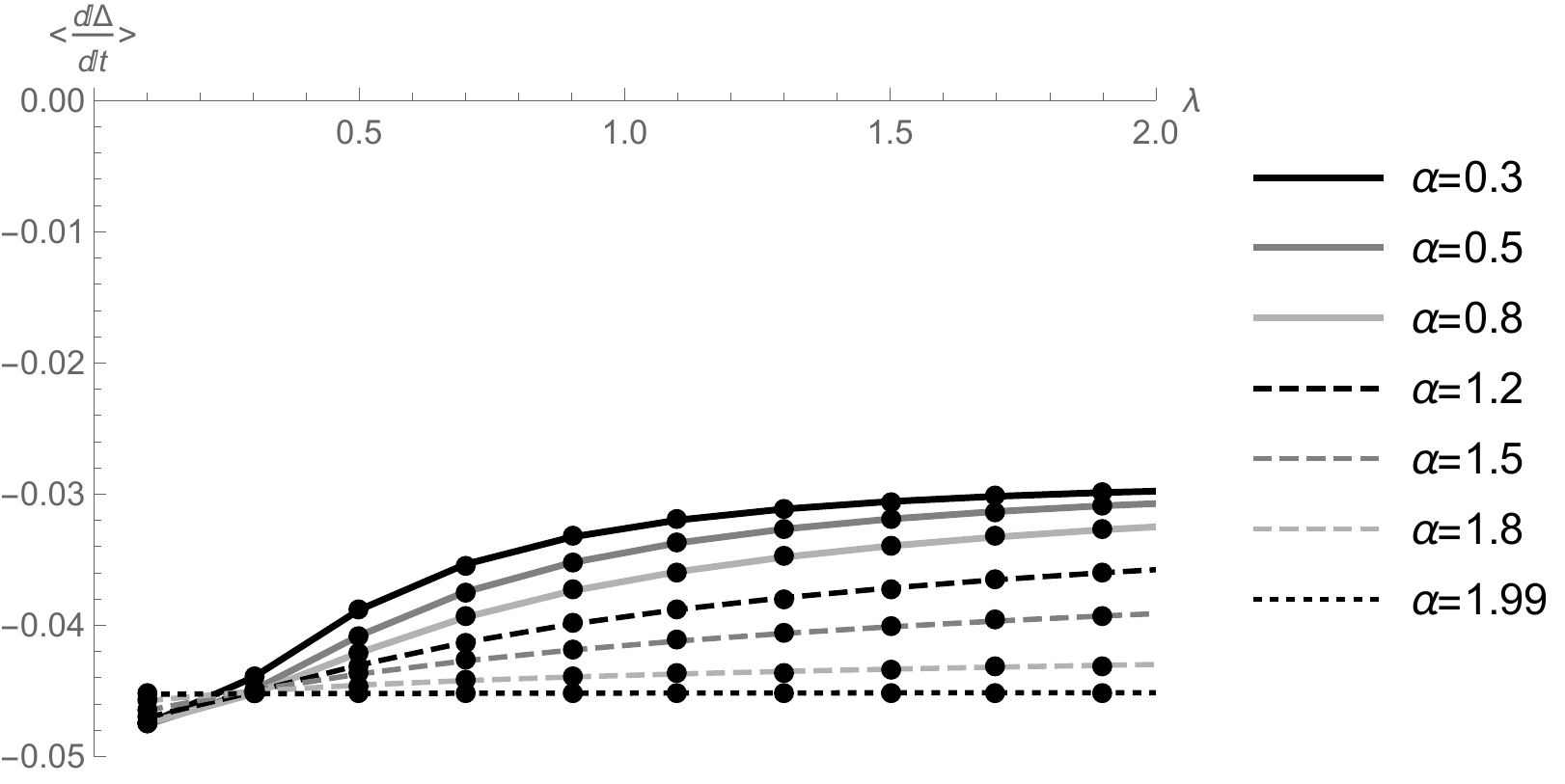}};
    \draw  (figure.north)  node[anchor=north,yshift=-1,black]{\textbf{\small{(e)}}};
\end{tikzpicture}
\end{subfigure}\hfill
\begin{subfigure}[b]{0.5\linewidth}
\centering
\begin{tikzpicture}
    \draw node[name=figure] {\includegraphics[width=8cm]{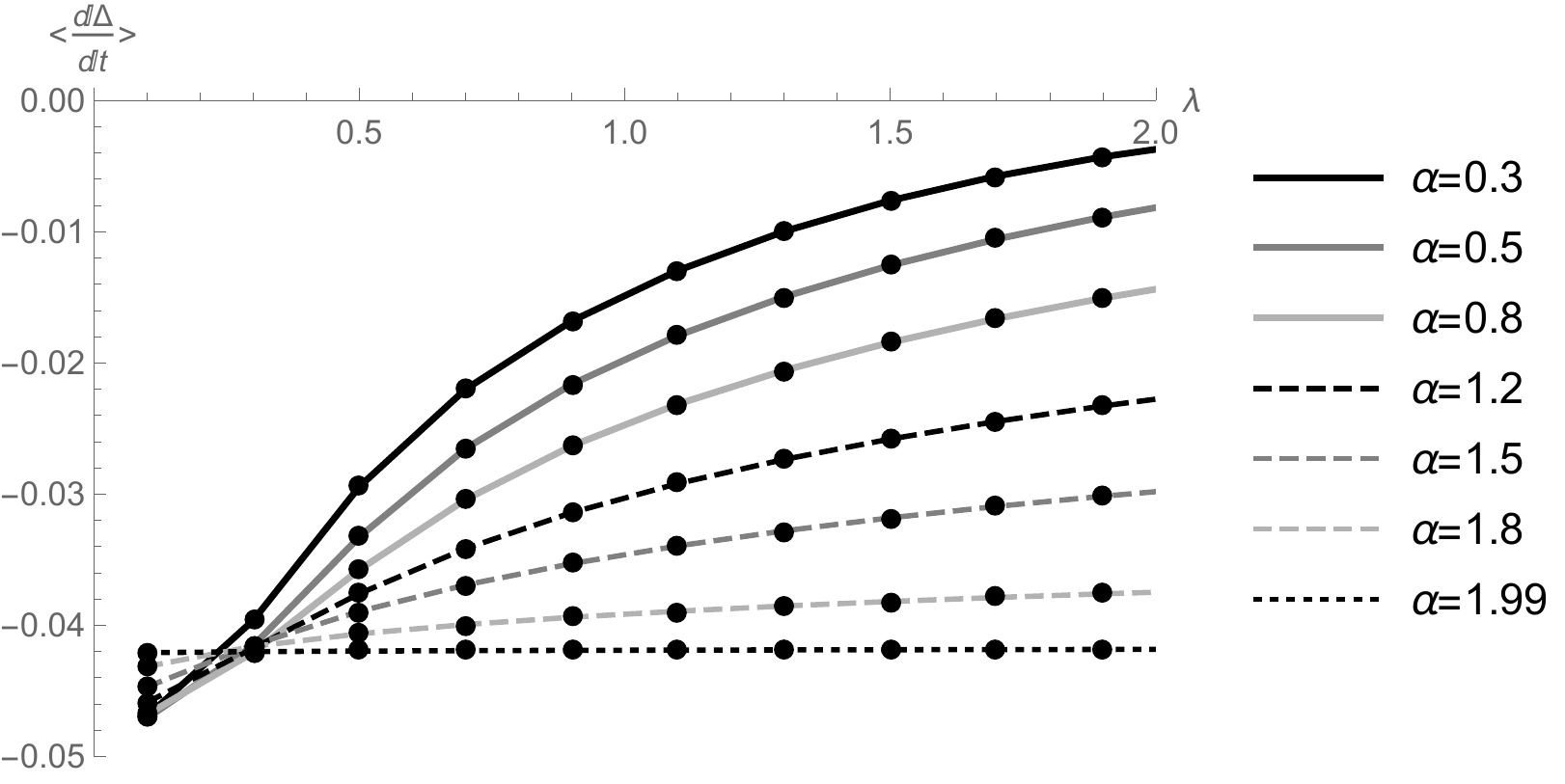}};
    \draw  (figure.north)  node[anchor=north,yshift=-1,black]{\textbf{\small{(f)}}};
\end{tikzpicture}
\end{subfigure}

\caption{Plots of the average ((a) and (b)) and variance ((c) and (d)) 
of $\Delta$ and the 
velocity $\langle \dot{\Delta} \rangle$ ((e) and (f)) given by Eq.(\ref{Levyvel}) as a function of the tempering parameter $\lambda$ for different values of the stability parameter $\alpha$. The couplings are as in Eqs.(\ref{coupls}) and the noise strength $\sigma=0.1$. The left hand plots show when the frustration parameters are $\phi=0.96\pi,\psi=0$ (when Blue and
Red deterministically drift)
and the right hand the case for $\phi=0.94\pi,\psi=0$
(where Blue and Red deterministically lock).}
\label{zeromoderatchet}
\end{figure*}

It is straightforward using the probability density to
compute the mean and variance of $\Delta$
which are shown in Fig.\ref{zeromoderatchet} for 
$\phi=0.96\pi$ and $\phi=0.94\pi$, across the 
threshold between Blue and Red deterministically locking
or drifting.
However we may also derive dynamical properties such
as the average velocity $\dot{\Delta}$ drawing upon \cite{ZupKallRob2017}.
From Eq.(\ref{defining-FP}) we obtain that each population density can be rewritten in terms of the \emph{probability current} ${\cal J}(x,t)$ as
\begin{align}
&\frac{\partial}{\partial t} {\cal P}(x,t) +\frac{\partial}{\partial x} {\cal J}(x,t) = 0\\
&\Rightarrow {\cal J}(x,t) = - \left\{  \sigma^2 \partial^{\alpha,\theta,\lambda}_{x} + \frac{\partial}{\partial x} V'(x)  \right\} {\cal P}(x,t).
\label{currentGauss}
\end{align}
We now restrict the support of $x$ to $\mathbb{S}^1$ by constructing the so-called \textit{reduced density} ${\cal P}^{(reduc)}(x,t)$ and \textit{reduced probability current} ${\cal J}^{(reduc)}(x,t)$ through
\begin{equation}
\begin{split}
{\cal P}^{(reduc)}(x,t) \equiv& \sum^{\infty}_{n=-\infty} {\cal P}(x+2 \pi n,t), \\
{\cal J}^{(reduc)}(x,t) \equiv& \sum^{\infty}_{n=-\infty} {\cal J}(x+2 \pi n,t).
\end{split}
\label{levyreduced}
\end{equation}
Due to the linearity of the Fokker-Planck equation, the reduced density and reduced probability current also obey Eq.(\ref{currentGauss}) but with the new boundary and normalisation conditions
\begin{equation}
{\cal P}^{(reduc)}(-\pi,t) = {\cal P}^{(reduc)}(\pi,t), \;\; \int^{\pi}_{-\pi} {\cal P}^{(reduc)}(x,t) \dx = 1.
\label{levyboundandnorm}
\end{equation}
Just like the Gaussian case~\cite{Reimann2002}, we expect that for tempered stable L{\'e}vy noise the steady state equivalent of Eq.(\ref{defining-FP}) with vanishing boundary conditions at the natural boundaries $x \rightarrow \pm \infty$ will be non-normalisable due to metastability. The steady state density ${\cal P}^{(reduc)}_{st}(x)$ on $\mathbb{S}^1$ satisfies
\begin{equation}
 \left\{ \sigma^2 \partial^{\alpha,\theta,\lambda}_{x} + \frac{\partial}{\partial x} V'(x)  \right\} {\cal P}^{(reduc)}_{st}(x) = 0
\label{stat-Levy}
\end{equation}
with boundary and normalisation conditions given by Eq.(\ref{levyboundandnorm}).

The solution is obtained by Fourier transformation of Eq.(\ref{stat-Levy}) revealing the expression,
\begin{eqnarray}
\begin{split}
\widehat{Q}(k+1) = - f_k \widehat{Q}(k)+\widehat{Q}(k-1),\\
\textrm{where}\;\; \widehat{Q}(k) \equiv e^{-i k \rho} \widehat{\cal{P}}^{(reduc)}_{st}(k),\\
\textrm{and}\;\; f_k \equiv -\frac{2}{k \sqrt{S^2+C^2}}[\Lambda(k)+i k \mu],
\end{split}
\label{threeterm}
\end{eqnarray}
and we note that because of periodicity, the Fourier variables $k$ are integer-valued.
We can then construct the corresponding reduced probability density via the discrete inverse Fourier transform
\begin{equation}
{\cal P}^{(reduc)}_{st}(x) = \frac{1}{2 \pi} \left\{ \widehat{Q}(0) + 2 \mathbb{R} \sum^{\infty}_{n=1} e^{-i n (x-\rho)} \widehat{Q}(n) \right\},
\label{PDF}
\end{equation}
where $\widehat{Q}(0) = 1$ from the normalisation condition in Eq.(\ref{levyboundandnorm}). The remaining  $\widehat{Q}(n)$ may be solved via the nonlinear transform
\begin{equation}
    S_{k+1}=\frac{\widehat{Q}(k+1)}{\widehat{Q}(k)},
\end{equation}
which transforms the linear three-term recurrence relation given by Eq.(\ref{threeterm}) into the following nonlinear two-term recurrence relation,
\begin{equation}
    S_k = \frac{1}{f_k + S_{k+1}}.
    \label{twoterm}
\end{equation}
Eq.(\ref{twoterm}) can be solved iteratively with the application of continued fractions, 
\begin{eqnarray*}
S_k = \frac{1}{f_k + \frac{1}{f_{k+1} + S_{k+2}}} =  \frac{1}{f_k + \frac{1}{f_{k+1} + \frac{1}{f_{k+2} +\frac{1}{ \ddots}}}}
\end{eqnarray*}
whose values can be used to reconstruct the corresponding $\widehat{Q}(k)$ expressions using,
\begin{equation}
\widehat{Q}(k) = S_k S_{k-1} \dots S_2 S_1  \widehat{Q}(0)
\label{SolN}
\end{equation}
for insertion into Eq.(\ref{PDF}).

In this paper we mainly focus on the average velocity $\langle \dot{\Delta} \rangle$, which is defined through the reduced steady state probability current ${\cal J}^{(reduc)}_{st}(x)$ with $x$ replaced by $\Delta$ as follows
 \begin{equation}
 \langle \dot{\Delta} \rangle = \frac{d}{dt}\langle \Delta \rangle = \int^{\pi}_{-\pi} {\cal J}^{(reduc)}_{st}(\Delta) d\Delta.
 \label{aveGauss}
 \end{equation}
 
Non-zero values of $\langle \dot{\Delta} \rangle$ indicate that through a combination of drift and noise there is a rotation of the centroid of Blue with respect to that of Red; positive values mean that Blue advances and negative values mean that Red advances. The average velocity $\langle \dot{\Delta} \rangle$ is solved for giving
\begin{equation}
\langle \dot{\Delta} \rangle  =  \left\{  \begin{array}{cl}
-\frac{\sigma^2 \alpha \theta \lambda^{\alpha -1}}{\cos \left( \frac{\pi \alpha}{2} \right)} + \mu -  2 \pi \gamma \mathbb{I} \left\{ \widehat{Q}(1) \right\} & 0 <\alpha < 1 \\
\mu -  2 \pi \gamma \mathbb{I} \left\{ \widehat{Q}(1) \right\}  & 1 < \alpha < 2.
\end{array} \right.
\label{Levyvel}
\end{equation}

 We now need to take into account the Laplacian projection of the noise. It is straightforward to see that as the equation
 for $\Delta$ arises from the difference $B-P$, the $N+M$ components of $u_j$ in Eq.(\ref{projcharfunc}) for this purpose are:
 \begin{eqnarray}
 u_j = \left\{  \begin{array}{l}
\frac{1}{\sqrt{N}}, j=1,\dots,N \\
-\frac{1}{\sqrt{M}}, j=N+1, \dots, N+M
\end{array}
\right. 
.
 \end{eqnarray}
 Eq.(\ref{projcharfunc}) thus gives
\begin{align}
\Lambda_0(k) =& \sigma^2\left\{ (1+\theta) \left[ N \left( \lambda + \frac{i k}{\sqrt{N}} \right)^{\alpha} + M \left( \lambda - \frac{i k }{\sqrt{M}} \right)^{\alpha} \right] 
\right. \nonumber \\
 & + (1-\theta) \left[ N \left( \lambda - \frac{i k}{\sqrt{N}} \right)^{\alpha} + M \left( \lambda + \frac{i k}{\sqrt{M}} \right)^{\alpha} \right]\nonumber \\
& - \left. 
 i k \left(\sqrt{N}-\sqrt{M}\right) v^{\alpha,\theta,\lambda} + 
 (N+M) \nu^{\alpha,\lambda} \right\}.
\end{align}
Noting that in the explicit example we consider, $N=M$, this leads to
\begin{align}
\Lambda_0(k) = 2 \sigma^2 N \left[ \left( \lambda + \frac{i k}{\sqrt{N}} \right)^{\alpha} + \left(\lambda - \frac{i k}{\sqrt{N}}  \right)^{\alpha} +  \nu^{\alpha,\lambda} \right].
\end{align}
This can be further simplified to
\begin{align}
\Lambda_0(k) = & 2 \sigma^2 N^{1-\alpha/2} \left[ \left(\sqrt{N} \lambda + i k \right)^{\alpha}  \right. \nonumber \\
 & \left. + \left(\sqrt{N} \lambda - i k\right)^{\alpha} +\nu^{\alpha,\sqrt{N} \lambda} \right].
\end{align}
We see that for the mode $\Delta$, the parameters of the noise $H$ for the zero mode projected
system are a simple rescaling of those of the noise $\eta$ on the original system:
\begin{eqnarray}
\sigma^2 &\rightarrow & \sigma^2 N^{1-\alpha/2} \\
\lambda & \rightarrow & \sqrt{N} \lambda \\
\theta & \rightarrow & 0.
\end{eqnarray}
Significantly, both the tempering and noise strength are {\it enhanced}, and the skew is {\it eliminated}.
The interplay of these for various $\alpha$ and $\lambda$ are not easily read off from Eq.(\ref{Levyvel}).

In the bottom row of Fig.\ref{zeromoderatchet} we show 
the $\lambda$-dependence of $\langle \dot{\Delta} \rangle$
for various $\alpha$ from Eq.(\ref{Levyvel}) for the couplings in Eqs.(\ref{coupls}) and with
 $\phi=0.96\pi,\psi=0$ (left, for which ${\cal K}<0$,
 o there is drift) and 
 $\phi=0.94\pi,\psi=0$ (right, for which ${\cal K}>0$, or the system
 locks).
 Note that for all of these plots we use $\sigma=0.1$,
 a slightly larger value than used for simulations as this 
makes the effects visible for all the properties across a common scale. We stress that these are all computed for
steady-state, {\it not static}, reduced distributions in the ratchet potential.

In all cases, we see several common broad features.
Whereas $\langle \Delta \rangle$ is positive, 
$\langle \dot{\Delta} \rangle$ is non-zero and negative.
This is why we emphasised earlier that a non-zero steady-state
average does not imply static behaviour,
so that the average of $\Delta$ needs to
be interpreted cautiously. It
suggests that though Blue succeeds on average in locking
ahead of Red, there is drift with Red advancing in relation to Blue.
This appears to have a knock-on effect that 
the average $\Delta$ is generally smaller
for the deterministically drifting case (left hand panels)
than for the deterministically locked case (right hand panels),
even though the average velocity is generally larger
for the former. In other words, when there is more drift,
$\Delta$ does not settle into a value that recurs
more frequently, and therefore a larger average value
does not develop.

We see in all cases that the near Gaussian case (red curve) is correctly $\lambda$ independent. The curves on the top, for $\phi=0.96\pi$, represent generally larger 
absolute values of $\langle \dot{\Delta} \rangle$,
consistent with ${\cal K}<0$ so that deterministic dynamics already imply $\dot{\Delta}\neq 0$. 
Tempering in general reduces the variance and average velocity.

There are key differences across the plots.
Significantly, for the average $\Delta$ the impact
of tempering is strongest for values of $\alpha<1$.
For both values of $\phi$ we observe that
tempering leads to more suppression of variance for
$\alpha<1$.

For small tempering $\lambda$ the average velocity is larger in absolute magnitude for smaller $\alpha$ (though difficult to discern on this scale), but there is a critical value of $\lambda$ at which this behaviour reverses, with small $\alpha$ values seeing more suppression. However the
effect eventually flattens out as $\lambda$ increases. Thus, even
when deterministically there is drift between clusters, as
in $\phi=0.94\pi$ on the left, tempered stable noise with small $\alpha$ and large tempering
$\lambda$ can cancel some of the diffusive drift.  For
$\phi=0.96\pi$, on the right, the black curve here for $\alpha=0.3$ genuinely approaches
zero. Thus,  for small $\alpha$ and sufficient tempering $\lambda$, the noise induced drift between Blue and Red may be cancelled entirely.

To summarise the key insights from this section: for 
small $\alpha$ in the presence of tempering, there is
a greater suppression of the average phase difference
between Blue and Red, and its variance. The noise
may also diminish the drift between Blue and Red clusters 
even when deterministically they might otherwise
be in relative motion. For the
deterministically locked situation, noise may be
cancelled entirely. 
Though we have not yet drawn attention to it, this behaviour is independent of the graph structure of
Blue and Red since it arises purely from the structure of
the fixed point as represented in the potential $V(\Delta)$
in Eq.~(\ref{potential}).

\section{Noisy Behaviour in the Full Dynamics}\label{sec:full}

We now focus on investigating the full nonlinear dynamics of the stochastic Blue vs Red model using numerical simulations. The aim of this numerical investigation is to identify the optimal internal synchronisation and phase lag of the Blue network against the Red network and to understand how the noise parameters $\alpha$, $\sigma$, and $\lambda$ lead to either steady, periodic, or noisy pathwise behaviour.

\subsection{Implementation}

To simulate the full tempered fractional stochastic Blue-v-Red model we implemented code in \textsf{C++} for maximum performance and uses the Messaging Passing Interface (\textsf{MPI}) to parallelise the simulation across an arbitrary number of CPU cores. The simulation is performed using an Euler method to simulate the dynamics at the $N$ vertices. This algorithm was chosen as it allows the code to easily switch between the cases of (1) no noise, (2) Gaussian noise, (3) stable noise, or (4) tempered stable noise without changing the underlying algorithm. The code takes the following parameters: the noise distribution parameters; the number of paths to simulate; the number of time steps in each path; and the filename containing the graph $\G$ stored in \textsf{graph6} format \cite{McKay1981}. The code also includes a number of hand-implemented random sampling algorithms for stable and tempered stable distributions. The evaluation of the expected values of the order parameter is approximated through a simple Monte-Carlo procedure, that is we approximate the expected value of the order parameter upon the discretisations $\mathbb{E}[(O_B)_i] = \mathbb{E}[(O_B(t_i)], t_i \in [0,T], i = 1,\dots,N$ through
the mean $(\bar{O}_B)_i = \frac{1}{M} \sum_{m=1}^M(O_B)_i ^{(m)}$ where $(O_B)_i ^{(m)}$ is the $i$'th discretised value of the $m$'th simulated path of the order parameter $O_B$. From the CLT we have that
\begin{equation} \label{eq:MCdist}
 (\bar{O}_B)_i \sim \mathcal{N} \left( \mathbb{E}[(O_B)_i], \frac{s}{\sqrt{M}} \right), 
\end{equation}
where $\mathcal{N}$ is the normal distribution and the standard deviation of the order parameter estimate is given as $s = \frac{1}{M-1} \sum_{m=1}^M((O_B)_i ^{(m)}-(\bar{O}_B)_i)^2 $.
This is similarly done for $(O_B)_i$ and $\Delta_i$.

\subsection{Pathwise dynamics}
In the following we show the pathwise dynamics,
namely individual but characteristic instances,
for tempered stable L{\'e}vy noise choices
for cases where we have previously examined
the deterministic \cite{KallZup2015} and Gaussian noise
\cite{HolZupKall2017} systems.
We first focus on the dynamics for specific values of $\sigma,\alpha,\lambda$ while always using the one-sided case $\theta=1$.
We generate time-series of the order parameters $O_B, O_R$ and $\Delta$. We choose values where both deterministic and stochastic
properties are evident, in other words where noise is not dominant. In most cases, this limits us to choices $\alpha>1$ so we use $\alpha=1.5$ as a value sufficiently far from Gaussian. Our choice of points in the $(\phi,\psi)$ landscape will be both below and above the threshold ${\cal K}=0$.
These should be compared to corresponding plots for the Gaussian case in \cite{HolZupKall2017}.
We note that whereas in that work we plotted box-whisker charts for the data at each time-step, here we simply plot the
mean and the standard deviation on a dual axis. Thus the fluctuations in $O_B$ and $O_R$ may appear to exceed unity - this is an artefact of how we represent the deviations.

\begin{figure}[hp]

\begin{subfigure}[b]{0.5\linewidth}
\centering
\begin{tikzpicture}
    \draw node[name=figure] {\includegraphics[width=4cm]{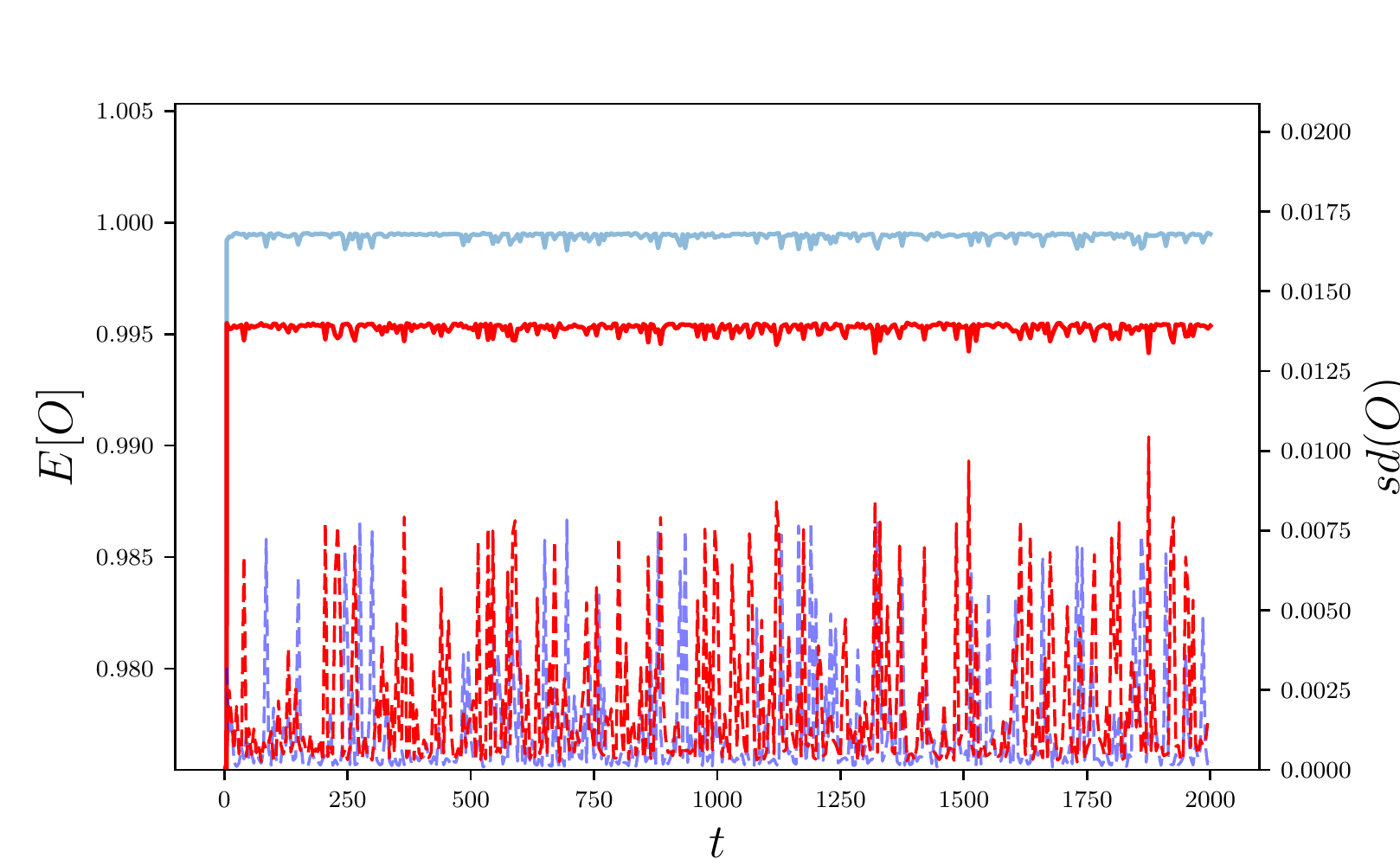}};
    \draw  (figure.north)  node[anchor=north,black]{\textbf{\small{(a)}}};
\end{tikzpicture}
\end{subfigure}\hfill
\begin{subfigure}[b]{0.5\linewidth}
\centering
\begin{tikzpicture}
    \draw node[name=figure] {\includegraphics[width=4cm]{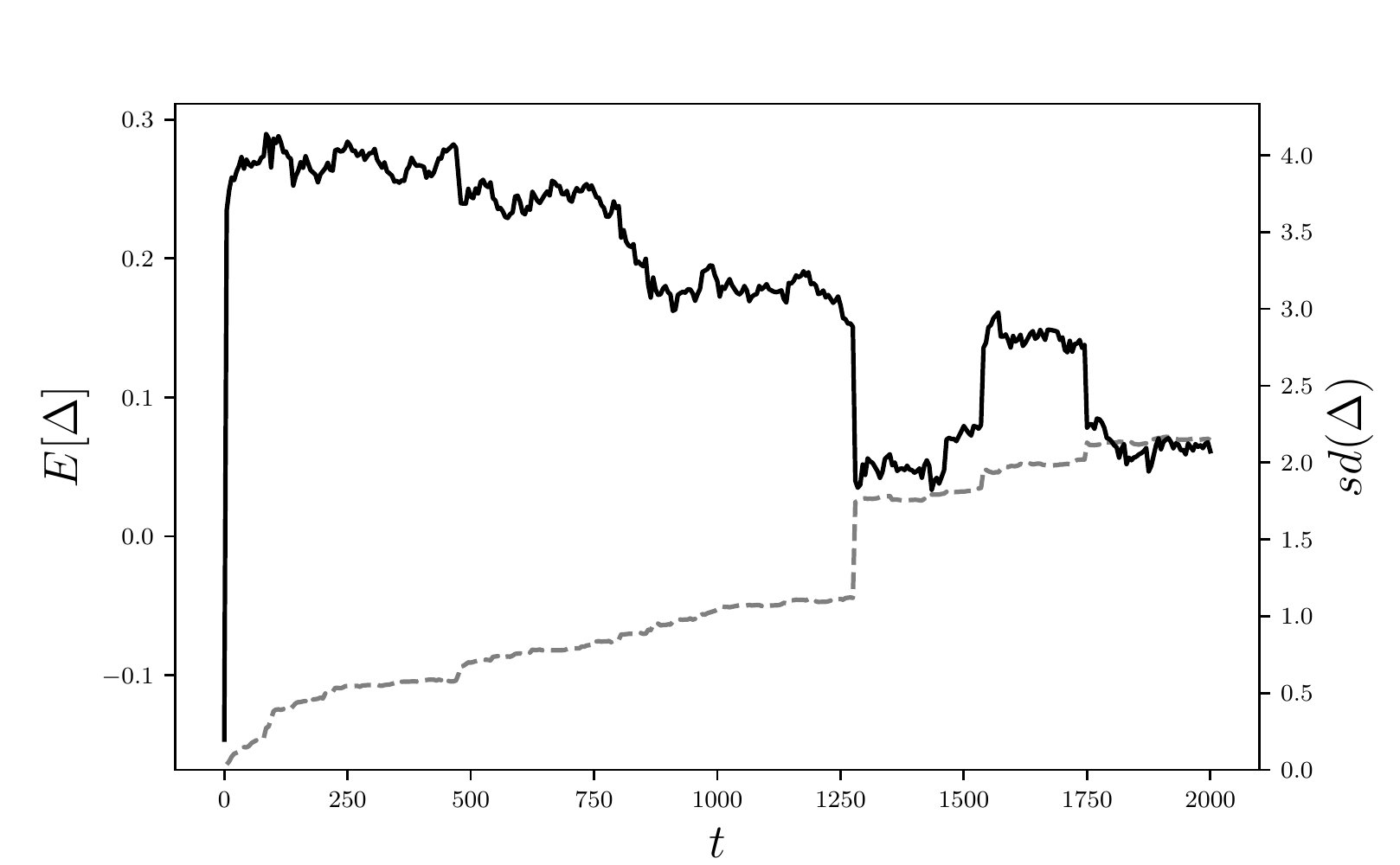}};
    \draw  (figure.north)  node[anchor=north,black]{\textbf{\small{(b)}}};
\end{tikzpicture}
\end{subfigure}\\
\begin{subfigure}[b]{0.5\linewidth}
\centering
\begin{tikzpicture}
    \draw node[name=figure] {\includegraphics[width=4cm]{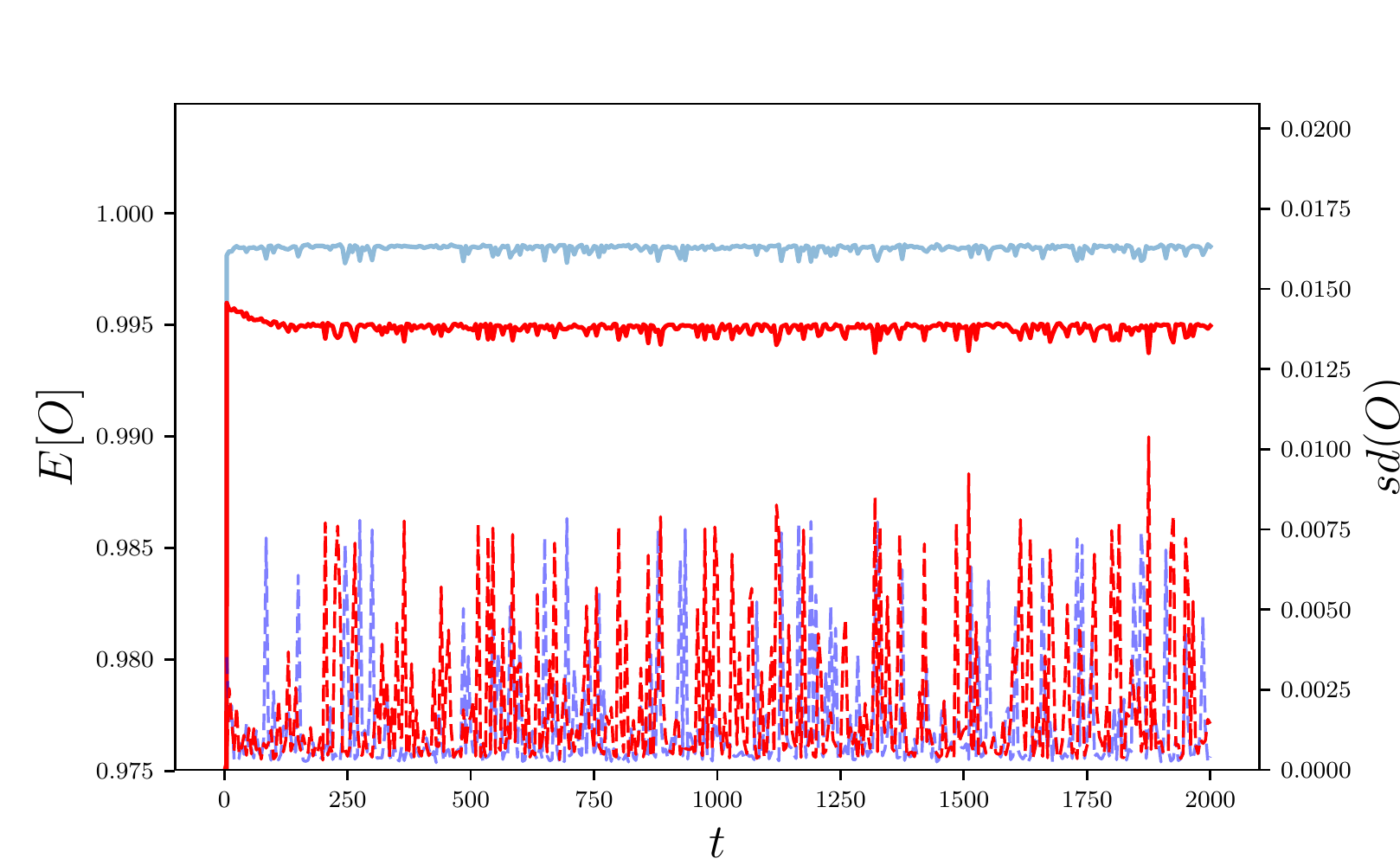}};
    \draw  (figure.north)  node[anchor=north,black]{\textbf{\small{(c)}}};
\end{tikzpicture}
\end{subfigure}\hfill
\begin{subfigure}[b]{0.5\linewidth}
\centering
\begin{tikzpicture}
    \draw node[name=figure] {\includegraphics[width=4cm]{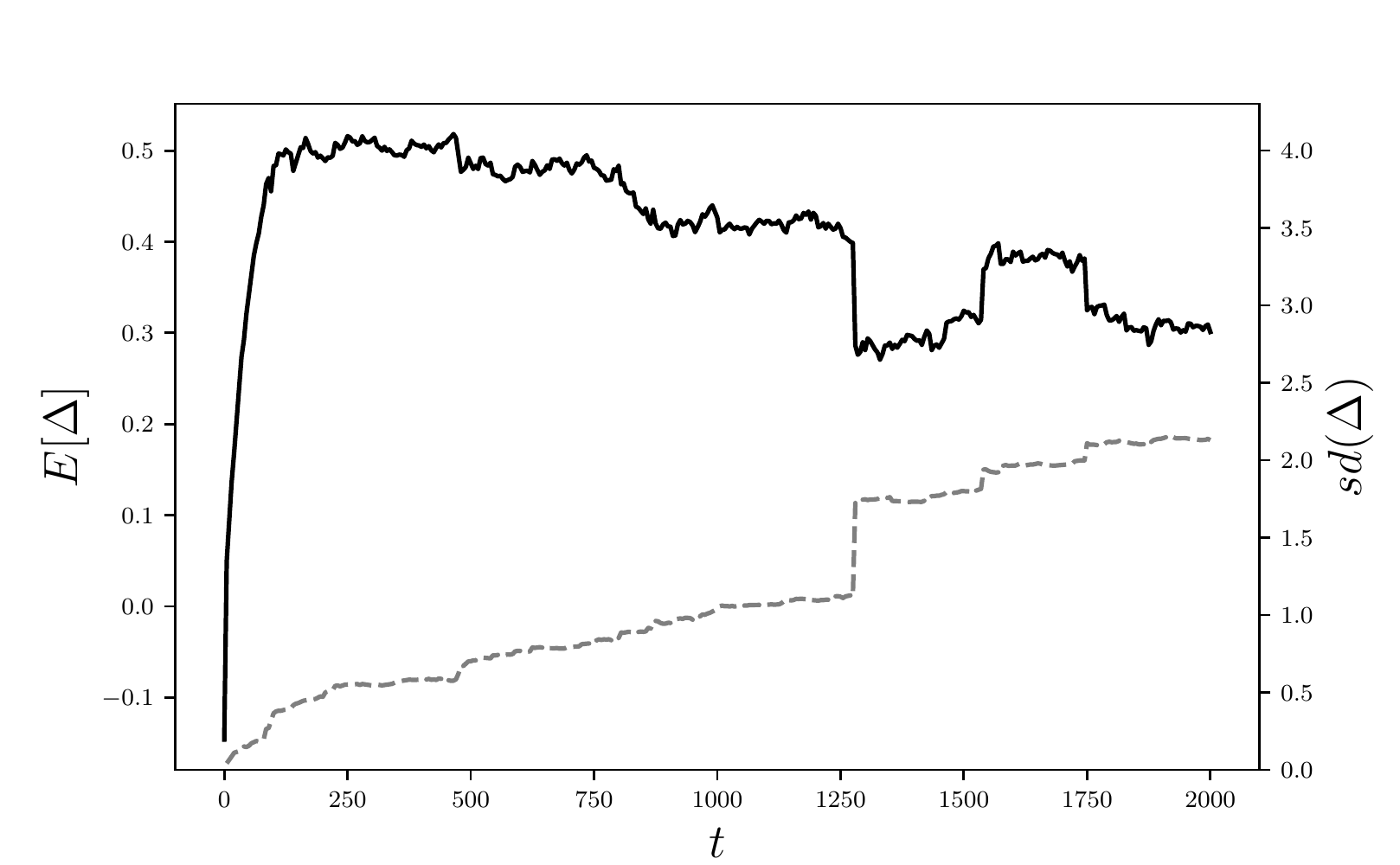}};
    \draw  (figure.north)  node[anchor=north,black]{\textbf{\small{(d)}}};
\end{tikzpicture}
\end{subfigure}\\
\begin{subfigure}[b]{0.5\linewidth}
\centering
\begin{tikzpicture}
    \draw node[name=figure] {\includegraphics[width=4cm]{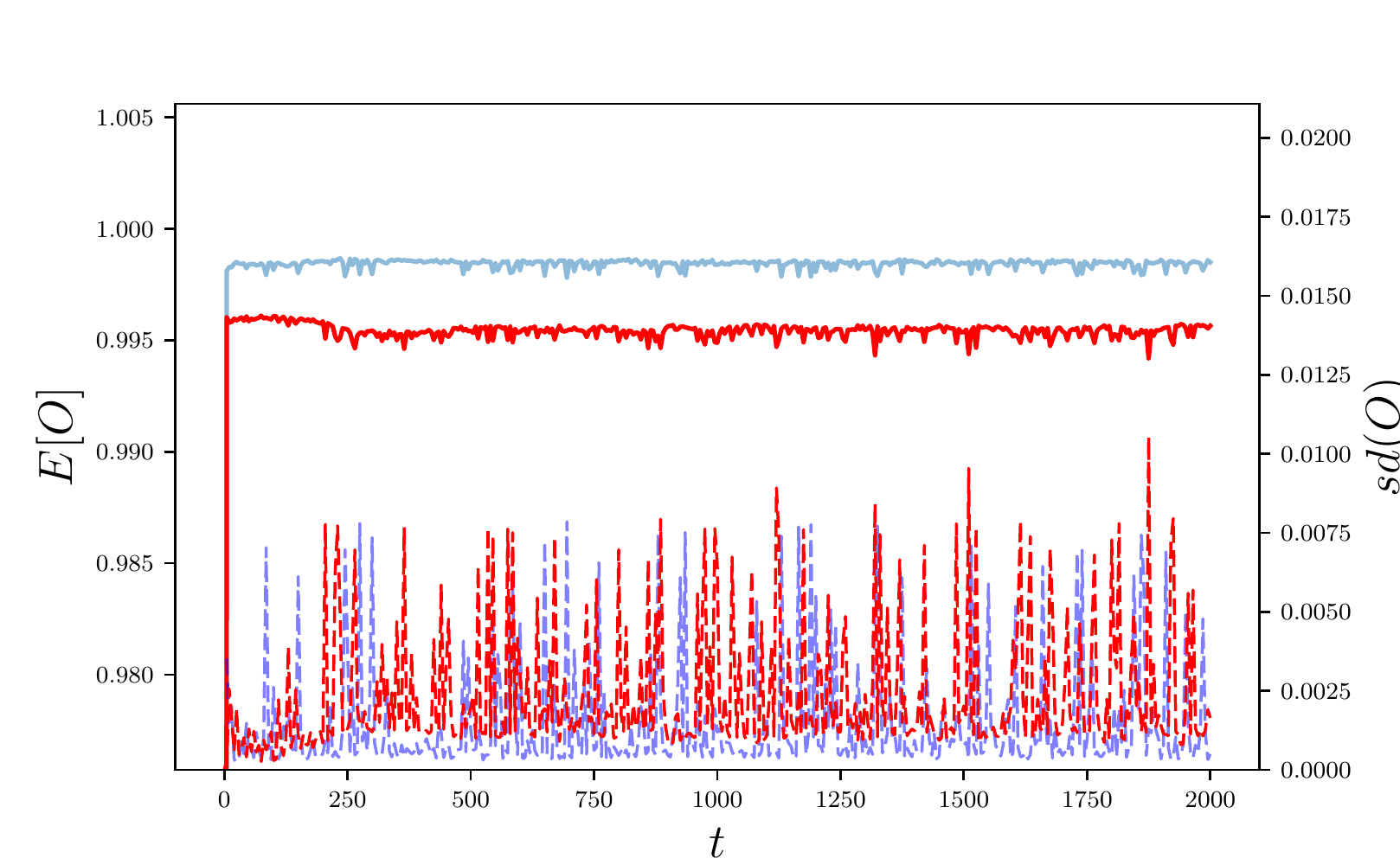}};
    \draw  (figure.north)  node[anchor=north,black]{\textbf{\small{(e)}}};
\end{tikzpicture}
\end{subfigure}\hfill
\begin{subfigure}[b]{0.5\linewidth}
\centering
\begin{tikzpicture}
    \draw node[name=figure] {\includegraphics[width=4cm]{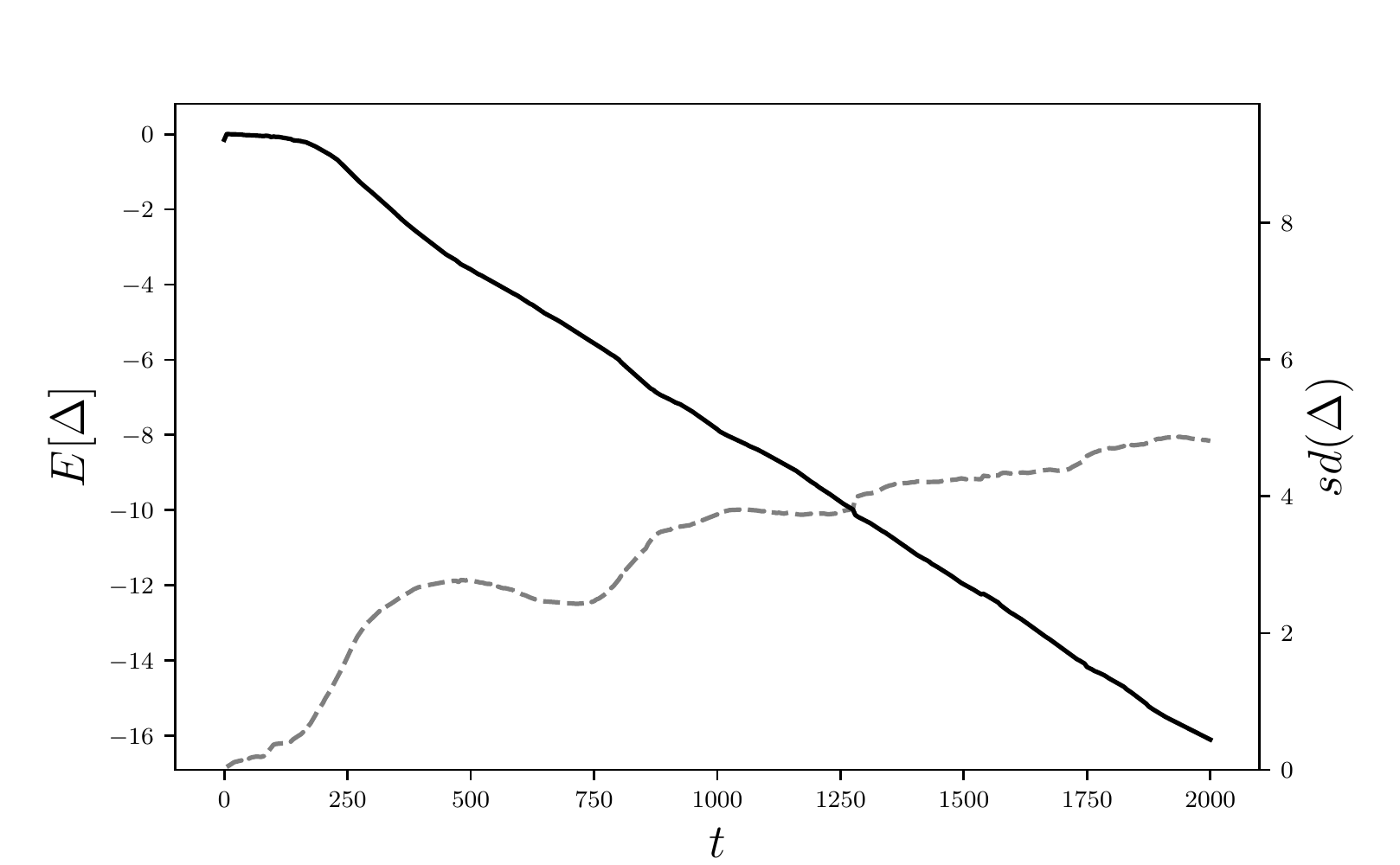}};
    \draw  (figure.north)  node[anchor=north,black]{\textbf{\small{(f)}}};
\end{tikzpicture}
\end{subfigure}\\
\begin{subfigure}[b]{0.5\linewidth}
\centering
\begin{tikzpicture}
    \draw node[name=figure] {\includegraphics[width=4cm]{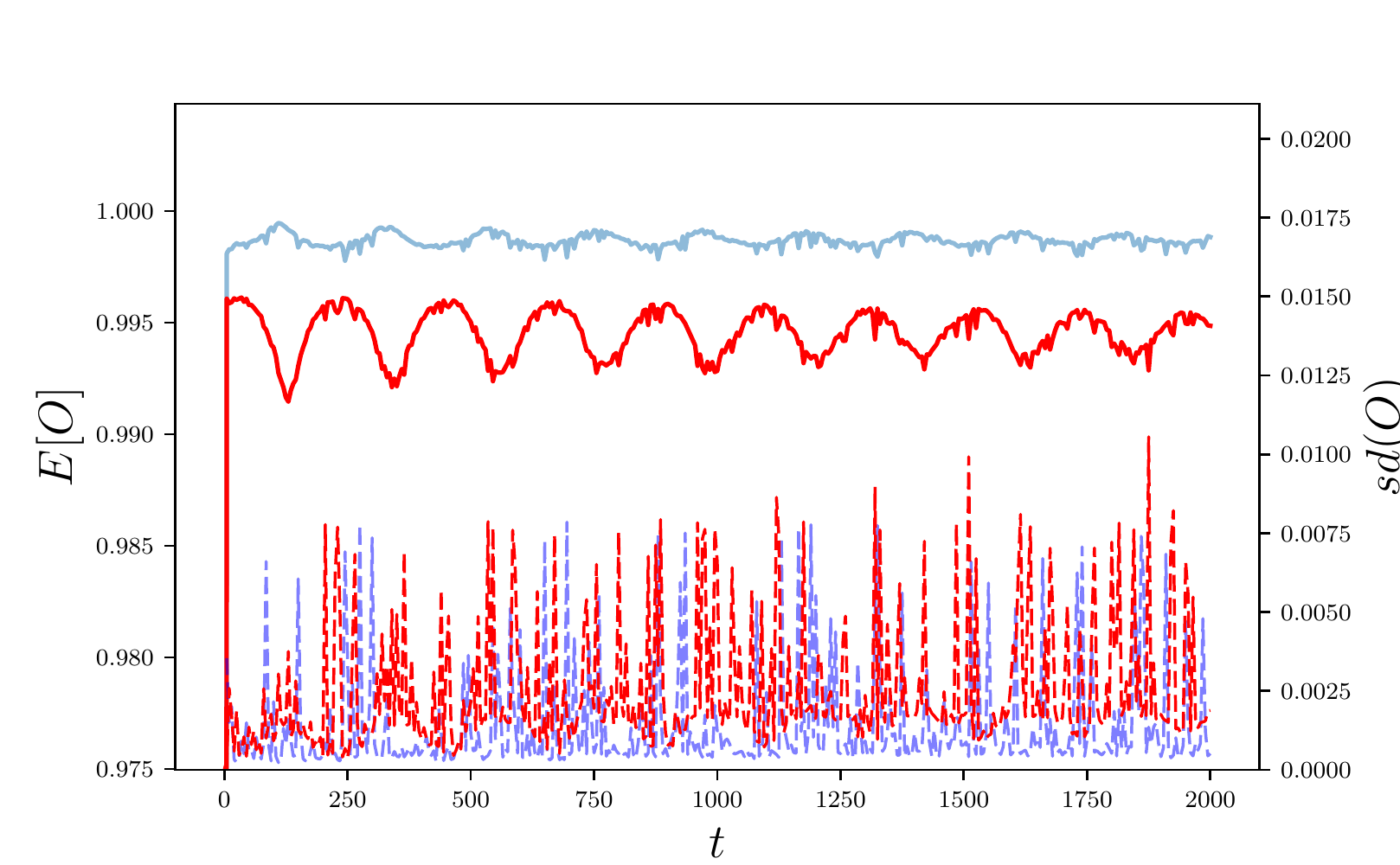}};
    \draw  (figure.north)  node[anchor=north,black]{\textbf{\small{(g)}}};
\end{tikzpicture}
\end{subfigure}\hfill
\begin{subfigure}[b]{0.5\linewidth}
\centering
\begin{tikzpicture}
    \draw node[name=figure] {\includegraphics[width=4cm]{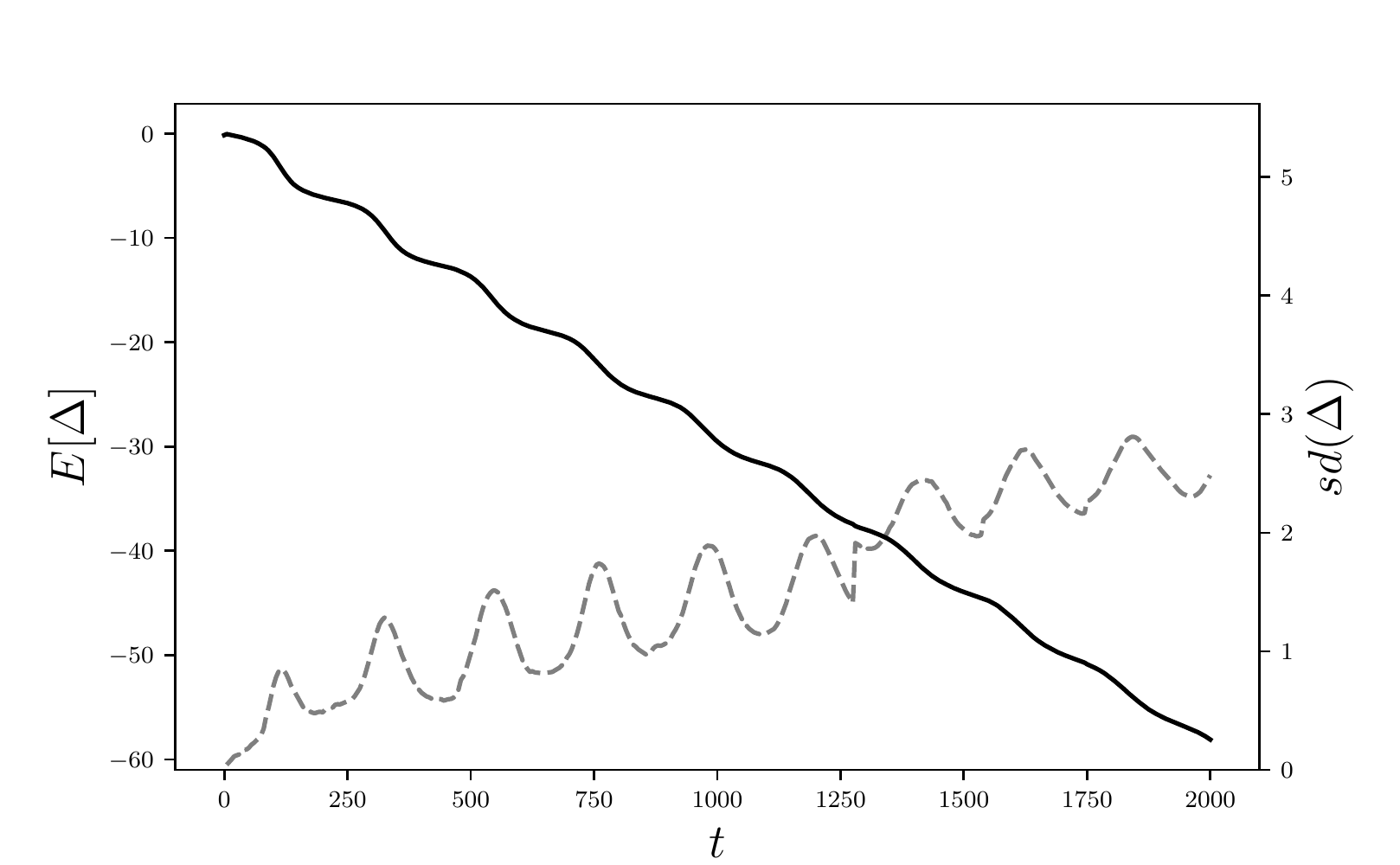}};
    \draw  (figure.north)  node[anchor=north,black]{\textbf{\small{(h)}}};
\end{tikzpicture}
\end{subfigure}
\caption{Time series plots in the stable L\'evy noise case with $\sigma = 0.05$ and $\alpha = 1.5$. Left axis: Time series plots for the average order parameters and $\Delta$. Right axis:  Time series plots (dashed lines) for the standard deviation of the Monte-Carlo estimates of the order parameters and $\Delta$. Left column: order parameters for the blue and red networks as given by the  blue (light grey) and  red (grey) lines respectively. Right column: $\Delta$ as denoted by the black lines. (a) and (b): $(\phi,\psi) = (0.2\pi,0)$, (c) and (d): $(\phi,\psi) = (0.94\pi,0)$,(e) and (f): $(\phi,\psi) = (0.95\pi,0)$,(g) and (h): $(\phi,\psi) = (0.96\pi,0)$ }
\label{fig:StableTS3}
\end{figure}

To begin with we consider stable noise for $\sigma = 0.05$ and $\alpha = 1.5$ in Fig.\ref{fig:StableTS3}.
We show examples for four points in the $(\phi,\psi)$ plane, two below, one at the threshold of ${\cal K}=0$ and the other above.
We observe, firstly, that levels of synchronisation in $O_B$ and $O_R$ are high in all cases, however the nature of the fluctuations
can be quite different. While ${\cal K}>0$, as for the top two cases, the fluctuations are quite random about a constant mean. The
heavy tail jumps are more evident in $\Delta$ in contrast to $O_B, O_R$.
Below ${\cal K}=0$, as for $\phi=0.96\pi,\psi=0$ a clear periodicity is evident in all measures, however, for $O_B$ and $O_R$ the
superimposed jumps are also marked. These should be contrasted with a far more incremental stochasticity in the Gaussian case shown in
\cite{HolZupKall2017}. The behaviour of $\Delta$ in the fourth instance is quite telling: the periodicity is evident, however after each cycle there
is a distortion in the period that incrementally increases. 

The important point to emphasise is that we are applying noise here to all nodes so that we are stimulating in essence both the
zero and normal modes according to the Laplacian.
Nevertheless, we can distinguish the impact on the two classes of mode quite cleanly. Firstly, the dominance of the periodicity
when ${\cal K}<0$ with noise superimposed is a consequence of the noise on {\it normal modes} but the
underlying deterministic behaviour of the zero mode, namely $\Delta(t)$, leaving its mark.
Secondly, the stochasticity in the periodicity itself of $\Delta$ when ${\cal K}<0$, is a consequence of the noise on the
zero mode itself. Specifically, the L{\'e}vy ratchet potential manifests itself in the probabilistic nature of how long the system
stays in the metastable well when subject to noise. We recall from \cite{HolZupKall2017} that when Gaussian noise is only applied to normal modes
the behaviour of $\Delta(t)$ in the ${\cal K}<0$ regime is strictly periodic.

\begin{figure}[hp] 
\begin{subfigure}[b]{0.5\linewidth}
\centering
\begin{tikzpicture}
    \draw node[name=figure] {\includegraphics[width=4cm]{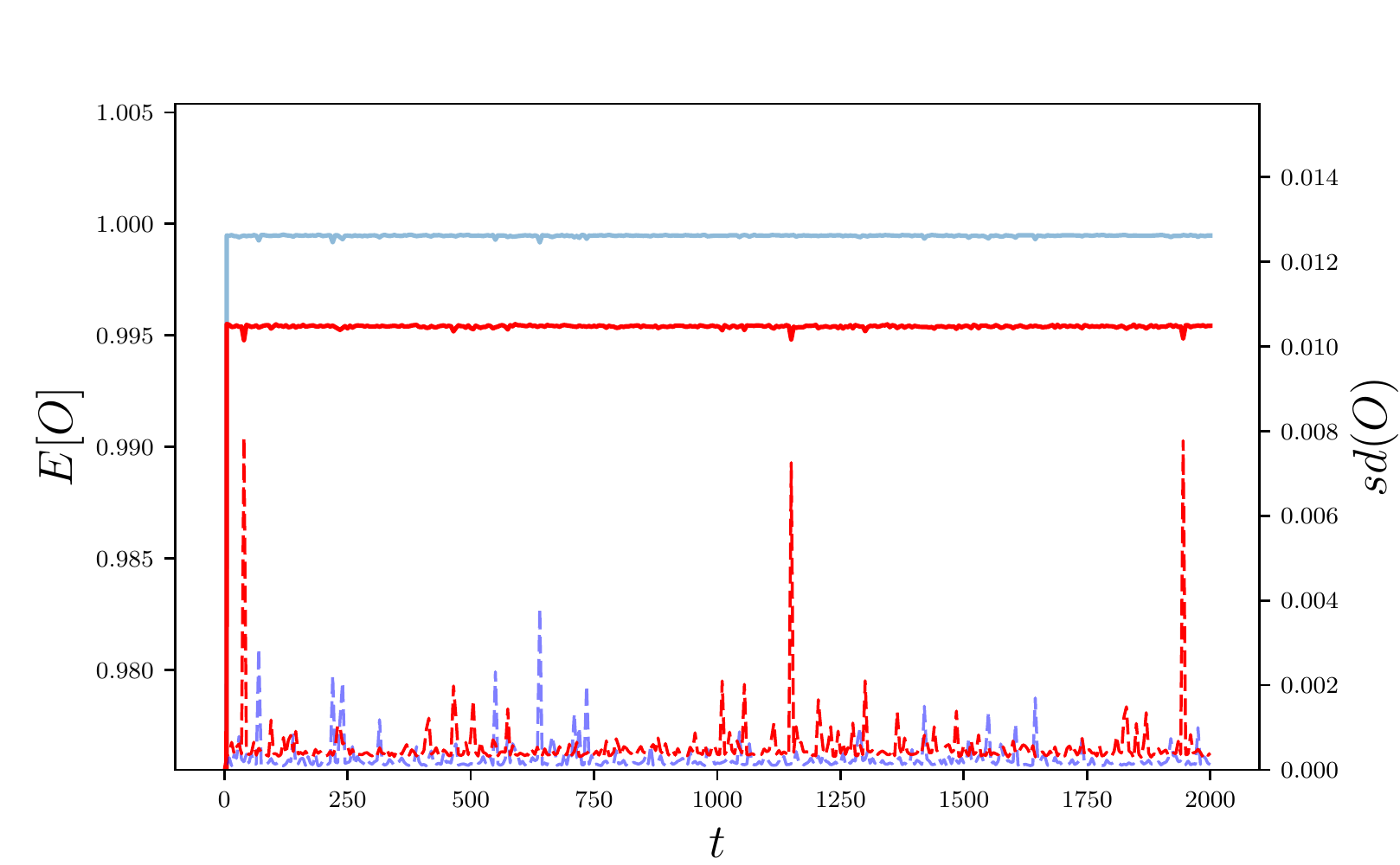}};
    \draw  (figure.north)  node[anchor=north,black]{\textbf{\small{(a)}}};
\end{tikzpicture}
\end{subfigure}\hfill
\begin{subfigure}[b]{0.5\linewidth}
\centering
\begin{tikzpicture}
    \draw node[name=figure] {\includegraphics[width=4cm]{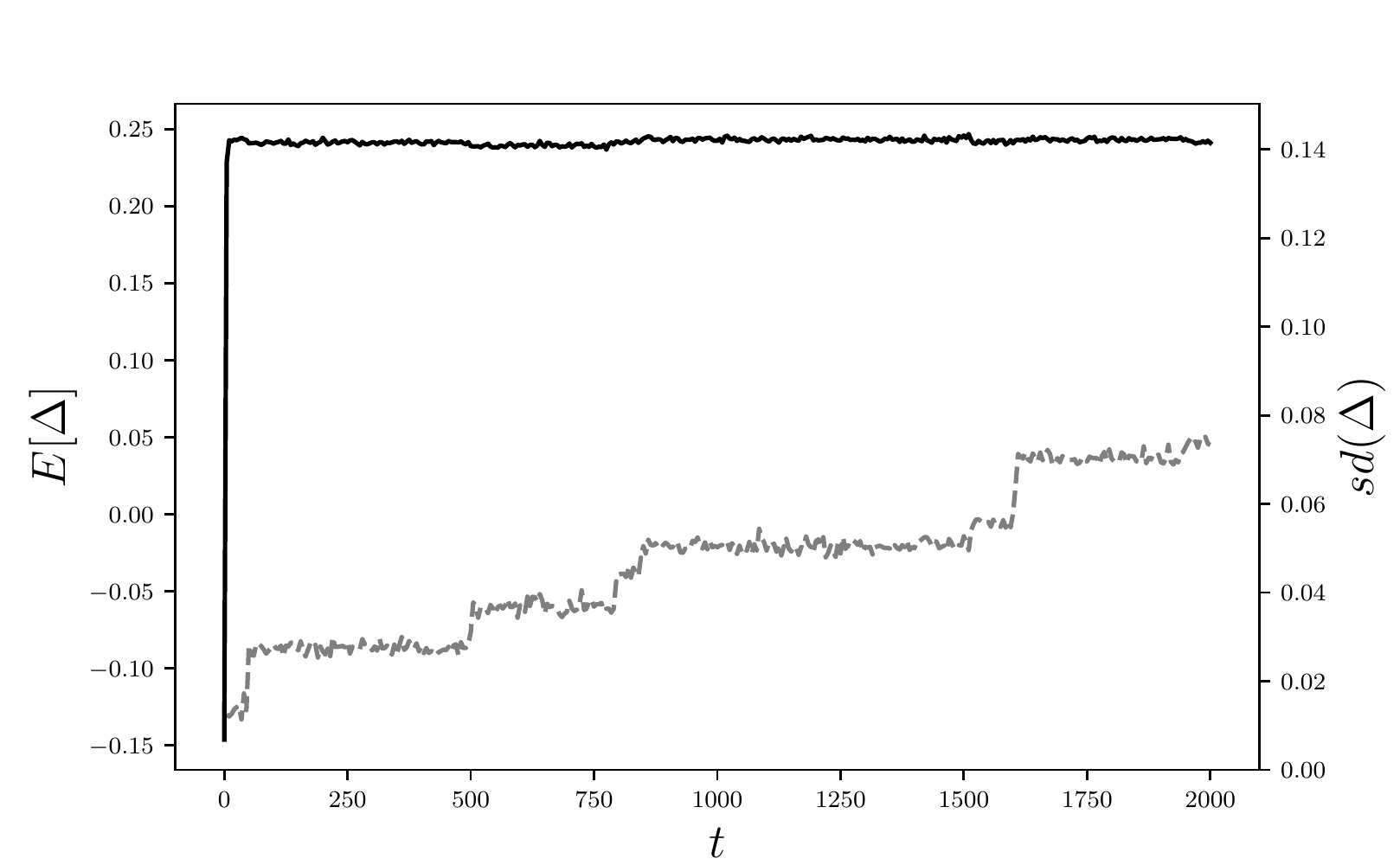}};
    \draw  (figure.north)  node[anchor=north,black]{\textbf{\small{(b)}}};
\end{tikzpicture}
\end{subfigure}\\
\begin{subfigure}[b]{0.5\linewidth}
\centering
\begin{tikzpicture}
    \draw node[name=figure] {\includegraphics[width=4cm]{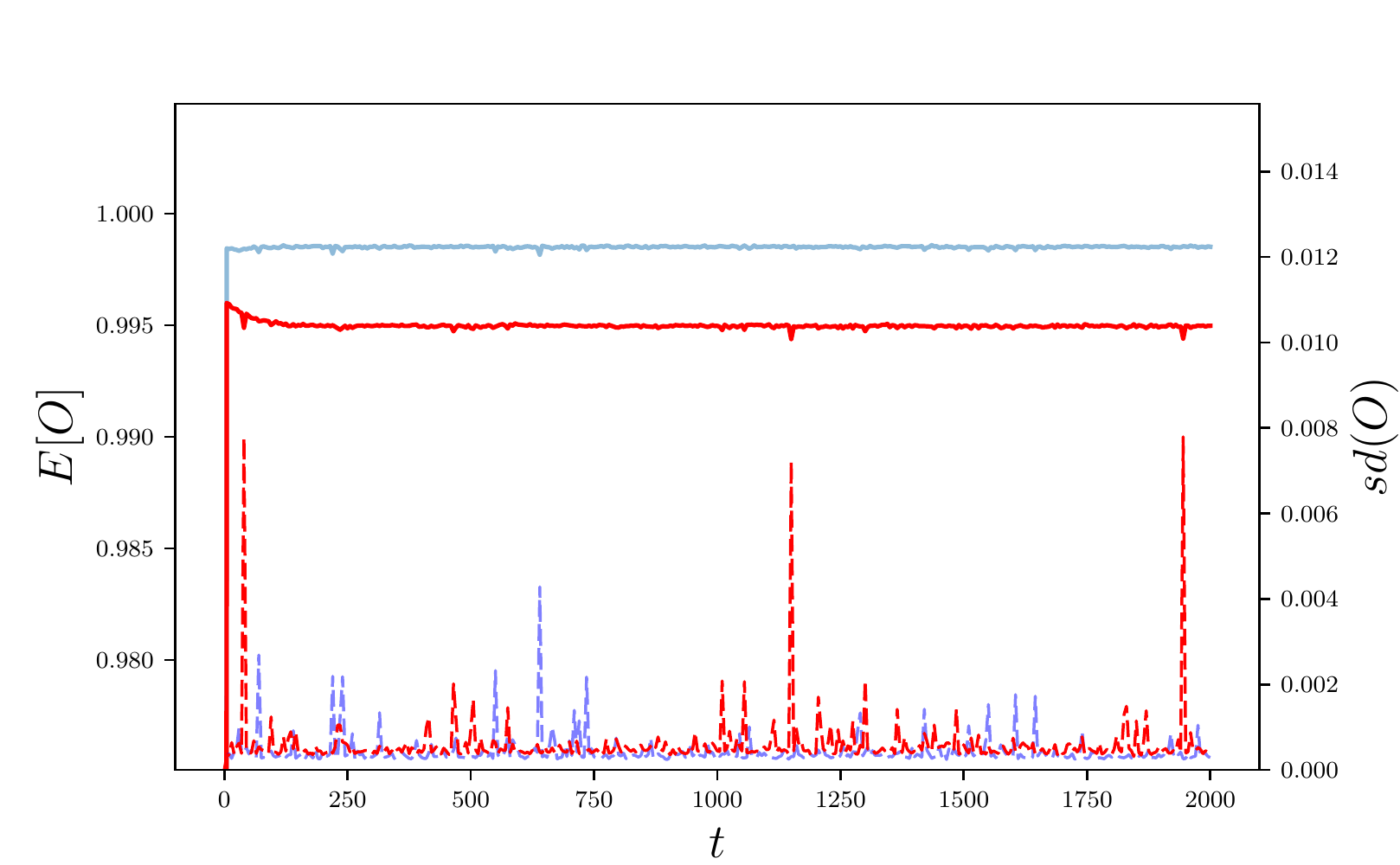}};
    \draw  (figure.north)  node[anchor=north,black]{\textbf{\small{(c)}}};
\end{tikzpicture}
\end{subfigure}\hfill
\begin{subfigure}[b]{0.5\linewidth}
\centering
\begin{tikzpicture}
    \draw node[name=figure] {\includegraphics[width=4cm]{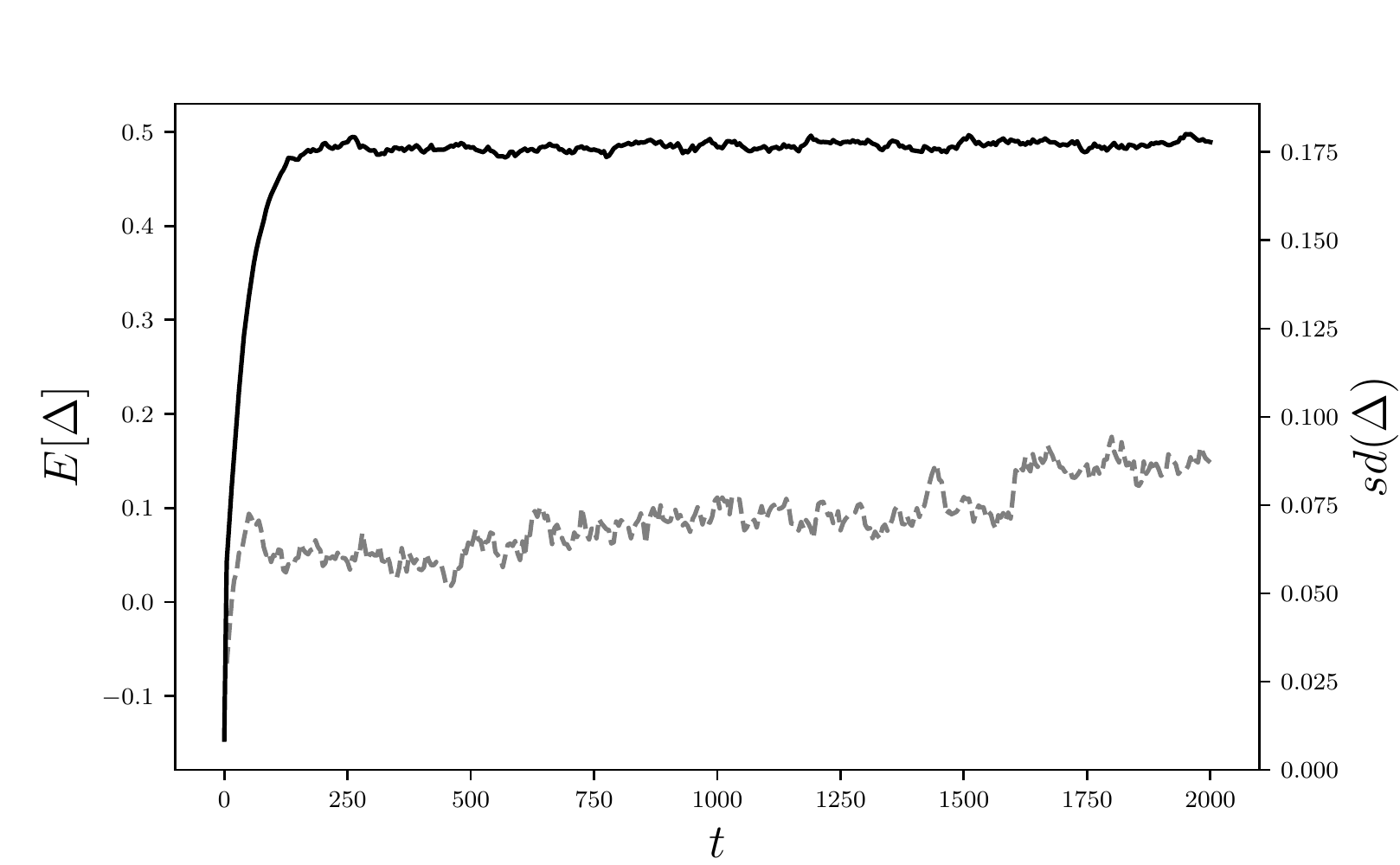}};
    \draw  (figure.north)  node[anchor=north,black]{\textbf{\small{(d)}}};
\end{tikzpicture}
\end{subfigure}\\
\begin{subfigure}[b]{0.5\linewidth}
\centering
\begin{tikzpicture}
    \draw node[name=figure] {\includegraphics[width=4cm]{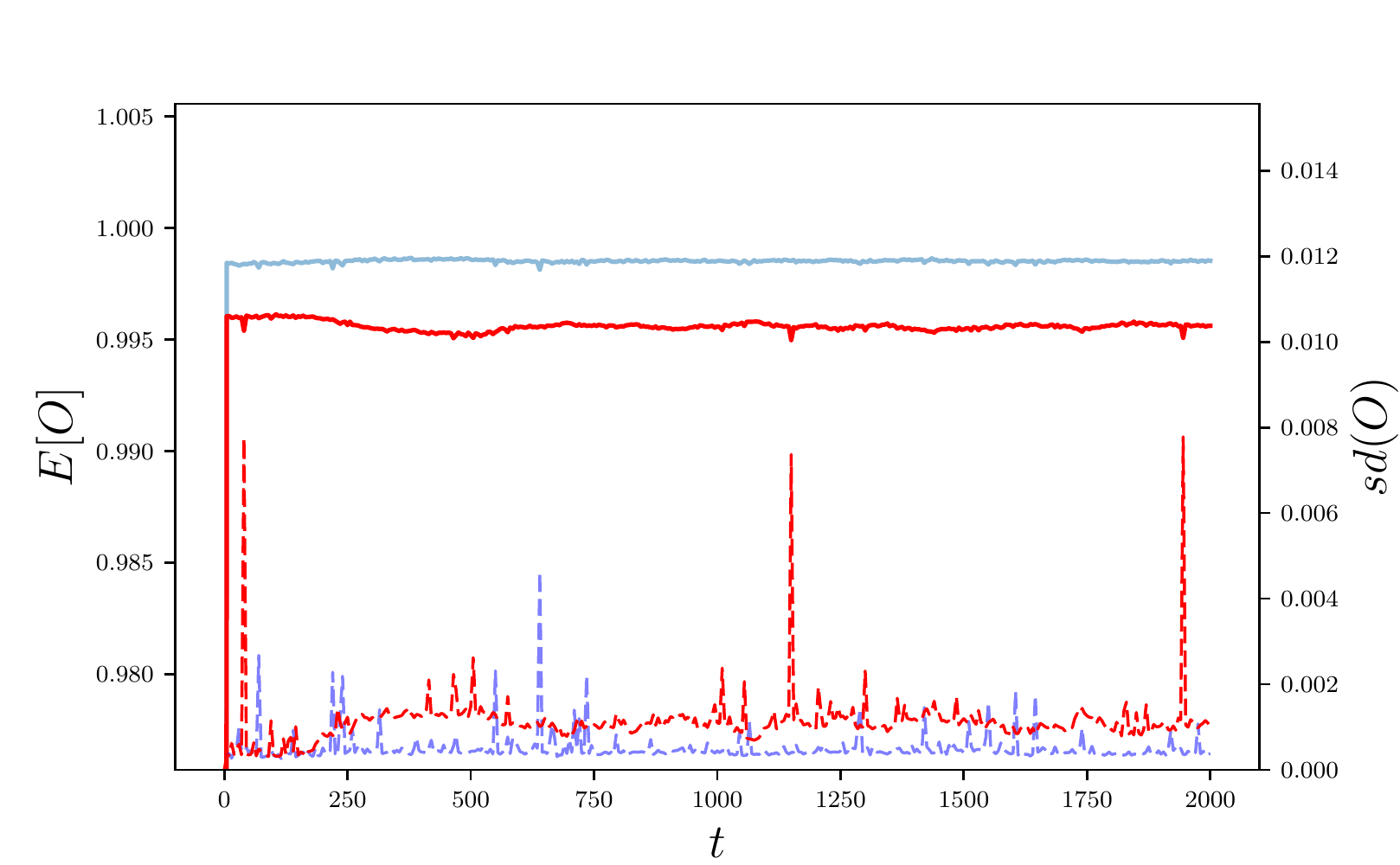}};
    \draw  (figure.north)  node[anchor=north,black]{\textbf{\small{(e)}}};
\end{tikzpicture}
\end{subfigure}\hfill
\begin{subfigure}[b]{0.5\linewidth}
\centering
\begin{tikzpicture}
    \draw node[name=figure] {\includegraphics[width=4cm]{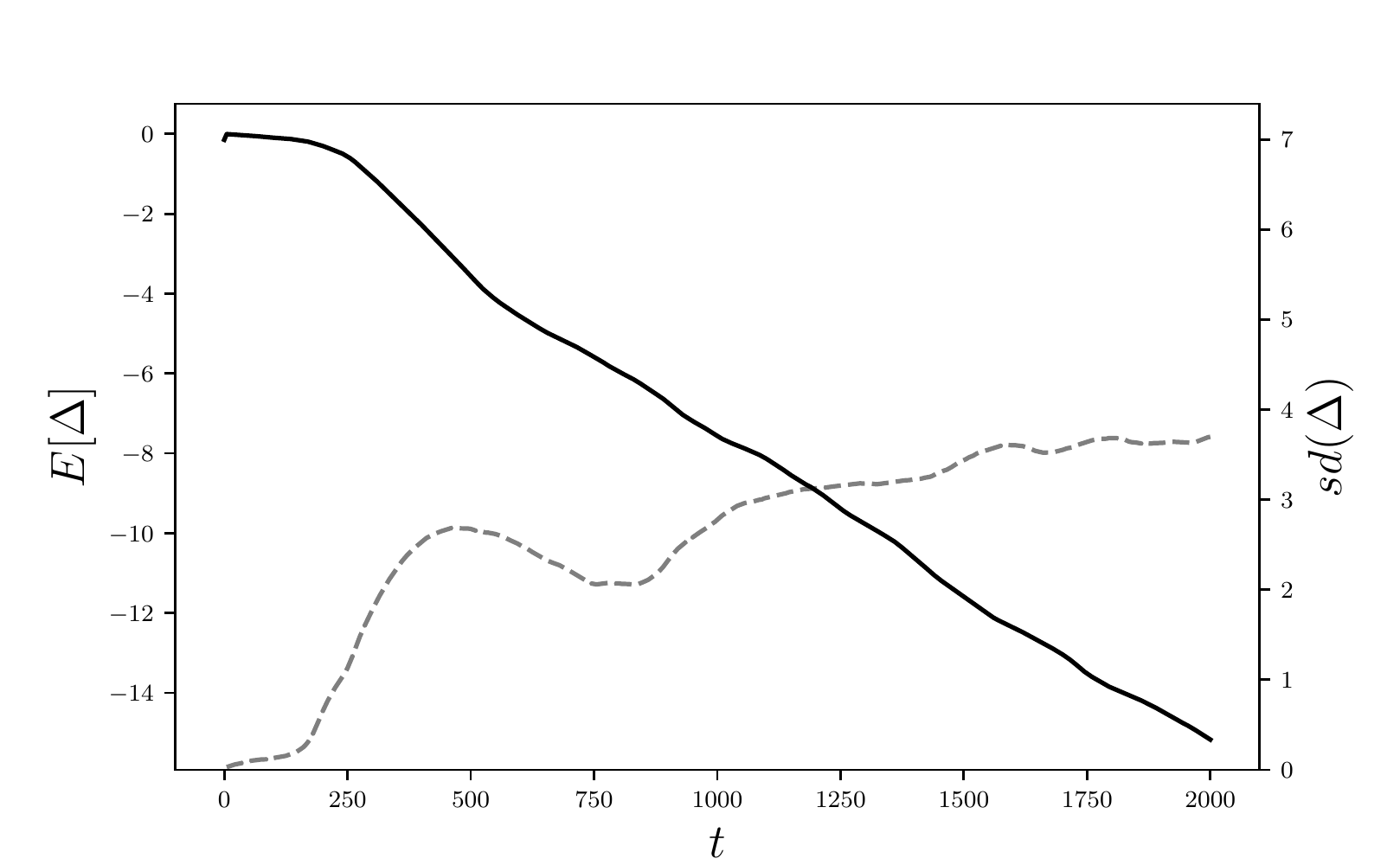}};
    \draw  (figure.north)  node[anchor=north,black]{\textbf{\small{(f)}}};
\end{tikzpicture}
\end{subfigure}\\
\begin{subfigure}[b]{0.5\linewidth}
\centering
\begin{tikzpicture}
    \draw node[name=figure] {\includegraphics[width=4cm]{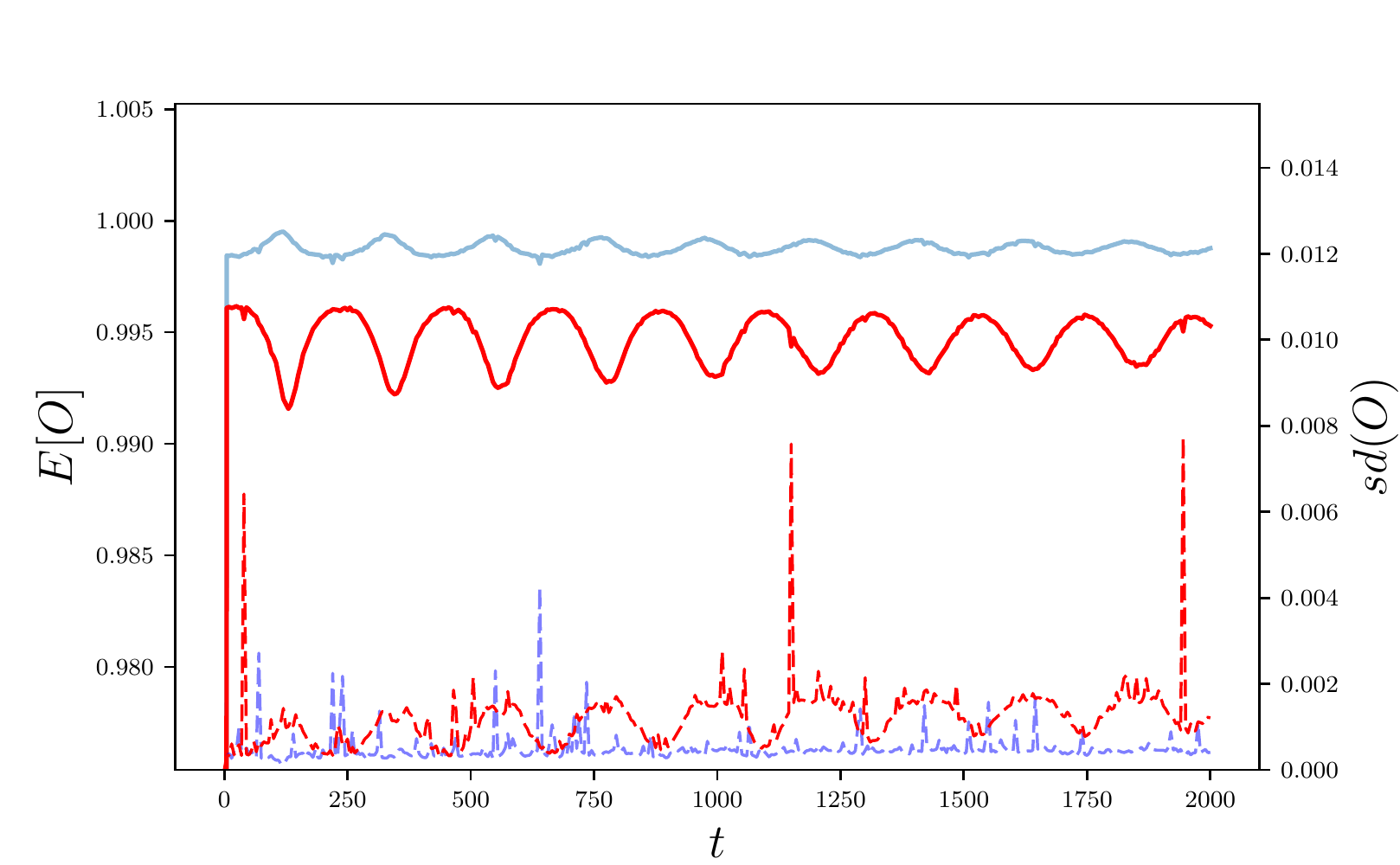}};
    \draw  (figure.north)  node[anchor=north,black]{\textbf{\small{(g)}}};
\end{tikzpicture}
\end{subfigure}\hfill
\begin{subfigure}[b]{0.5\linewidth}
\centering
\begin{tikzpicture}
    \draw node[name=figure] {\includegraphics[width=4cm]{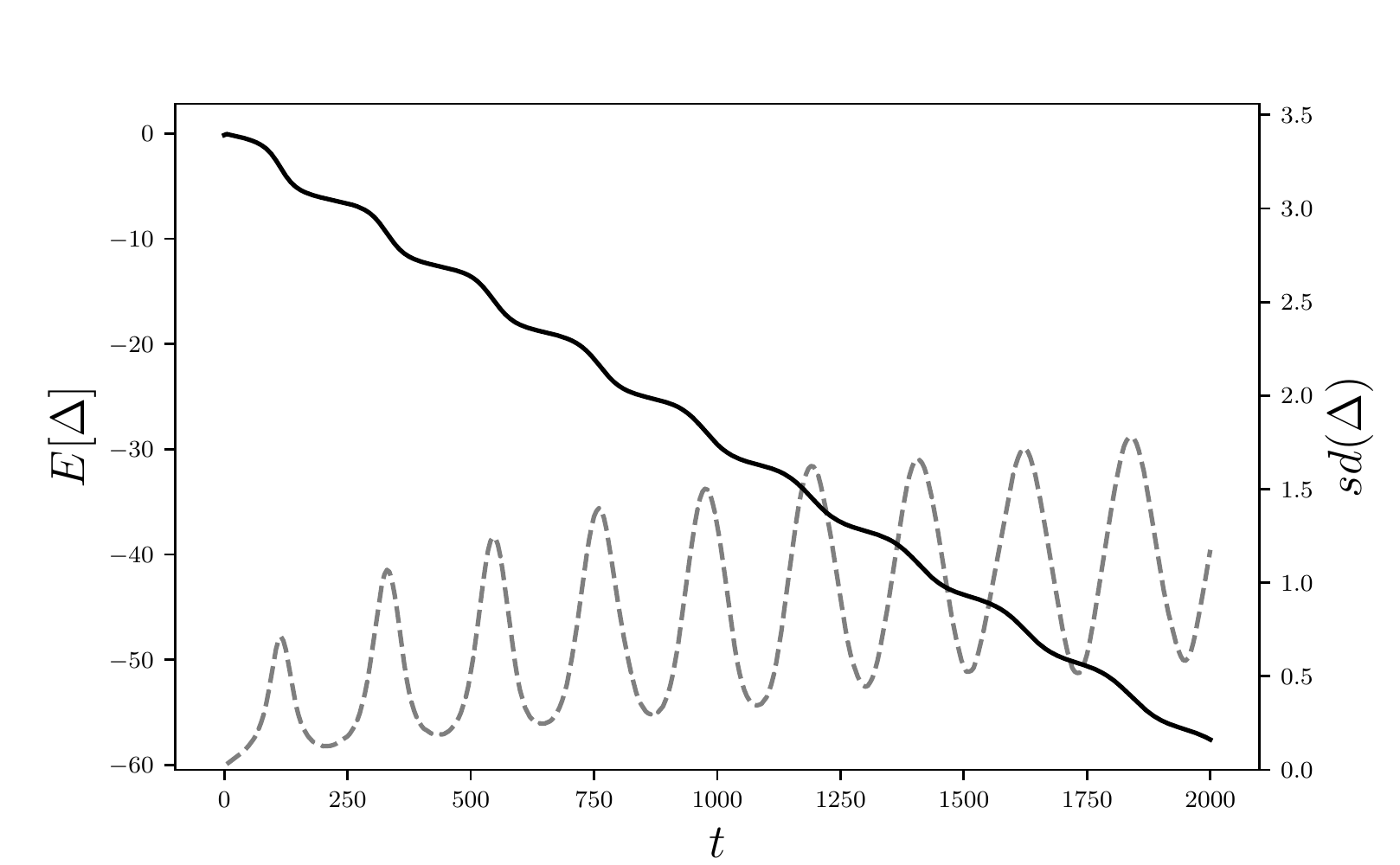}};
    \draw  (figure.north)  node[anchor=north,black]{\textbf{\small{(h)}}};
\end{tikzpicture}
\end{subfigure}
\caption{Time series plots in the stable L\'evy noise case with $\sigma = 0.05$ and $\alpha = 1.5$ and $\lambda = 1$. Left axis: Time series plots for the average order parameters and $\Delta$. Right axis:  Time series plots (dashed lines) for the standard deviation of the Monte-Carlo estimates of the order parameters and $\Delta$. Left column: order parameters for the blue and red networks as given by the  blue (light grey) and  red (grey) lines respectively. Right column: $\Delta$ as denoted by the black lines. (a) and (b): $(\phi,\psi) = (0.2\pi,0)$, (c) and (d): $(\phi,\psi) = (0.94\pi,0)$,(e) and (f): $(\phi,\psi) = (0.95\pi,0)$,(g) and (h): $(\phi,\psi) = (0.96\pi,0)$ }
\label{fig:TStableTS2}
\end{figure}

In light of this discussion, understanding the tempered case is straightforward. Shown in Fig.\ref{fig:TStableTS2},
we observe essentially the same characteristics but with moderation of the stochasticity. Tempering here is at $\lambda=0.75$ so
that the jumps in $O_B$ and $O_R$ are significantly smaller and sparser, both for constant and periodic behaviours. In the
behaviour of $\Delta$ the broadening for ${\cal K}>0$ is smoother. For the periodic regime, ${\cal K}<0$, the period is far more well-defined
as a consequence of the stabilisation of the system in the ratchet potential.

\begin{figure}[hp] 
\begin{subfigure}[b]{0.5\linewidth}
\centering
\begin{tikzpicture}
    \draw node[name=figure] {\includegraphics[width=4cm]{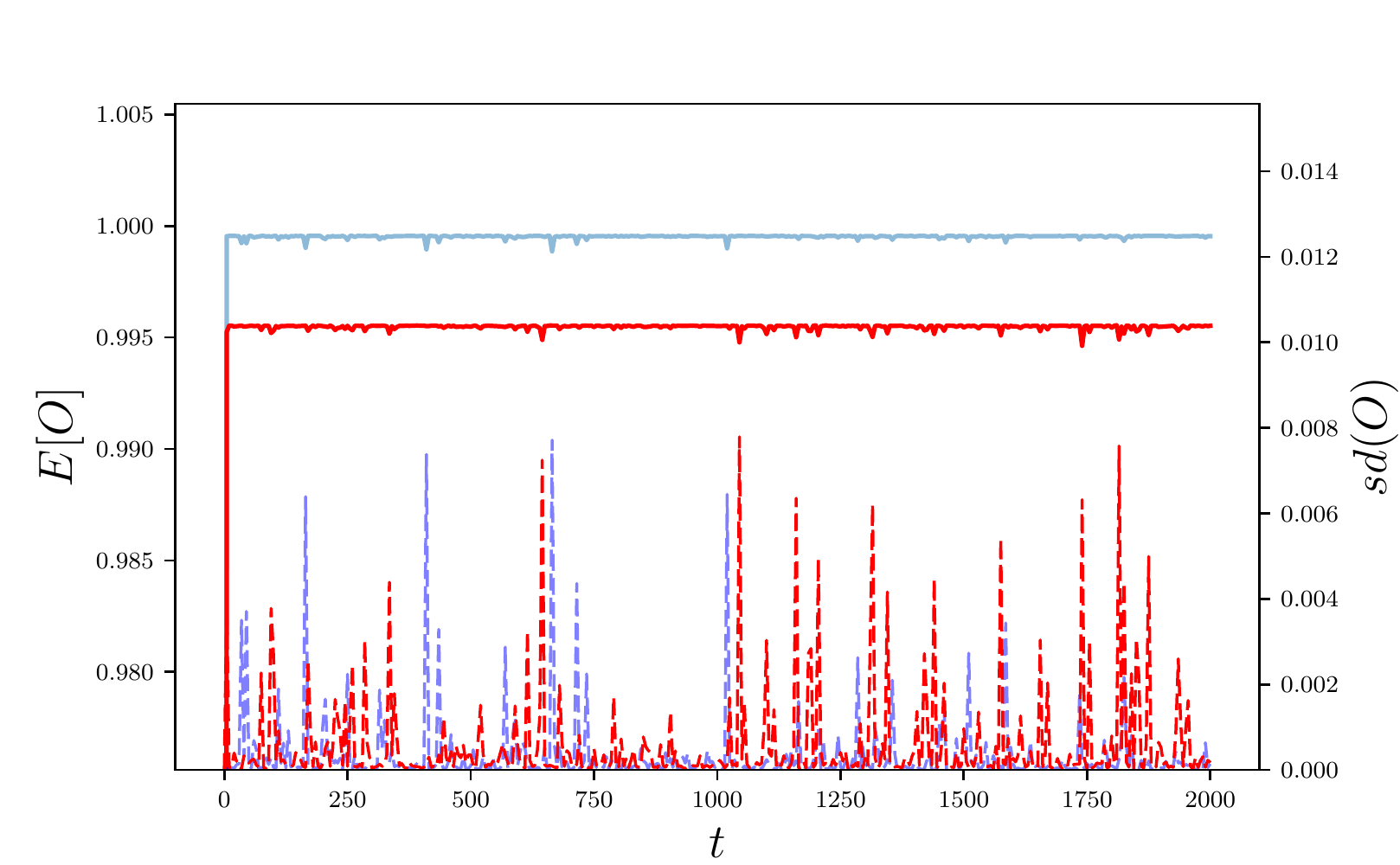}};
    \draw  (figure.north)  node[anchor=north,black]{\textbf{\small{(a)}}};
\end{tikzpicture}
\end{subfigure}\hfill
\begin{subfigure}[b]{0.5\linewidth}
\centering
\begin{tikzpicture}
    \draw node[name=figure] {\includegraphics[width=4cm]{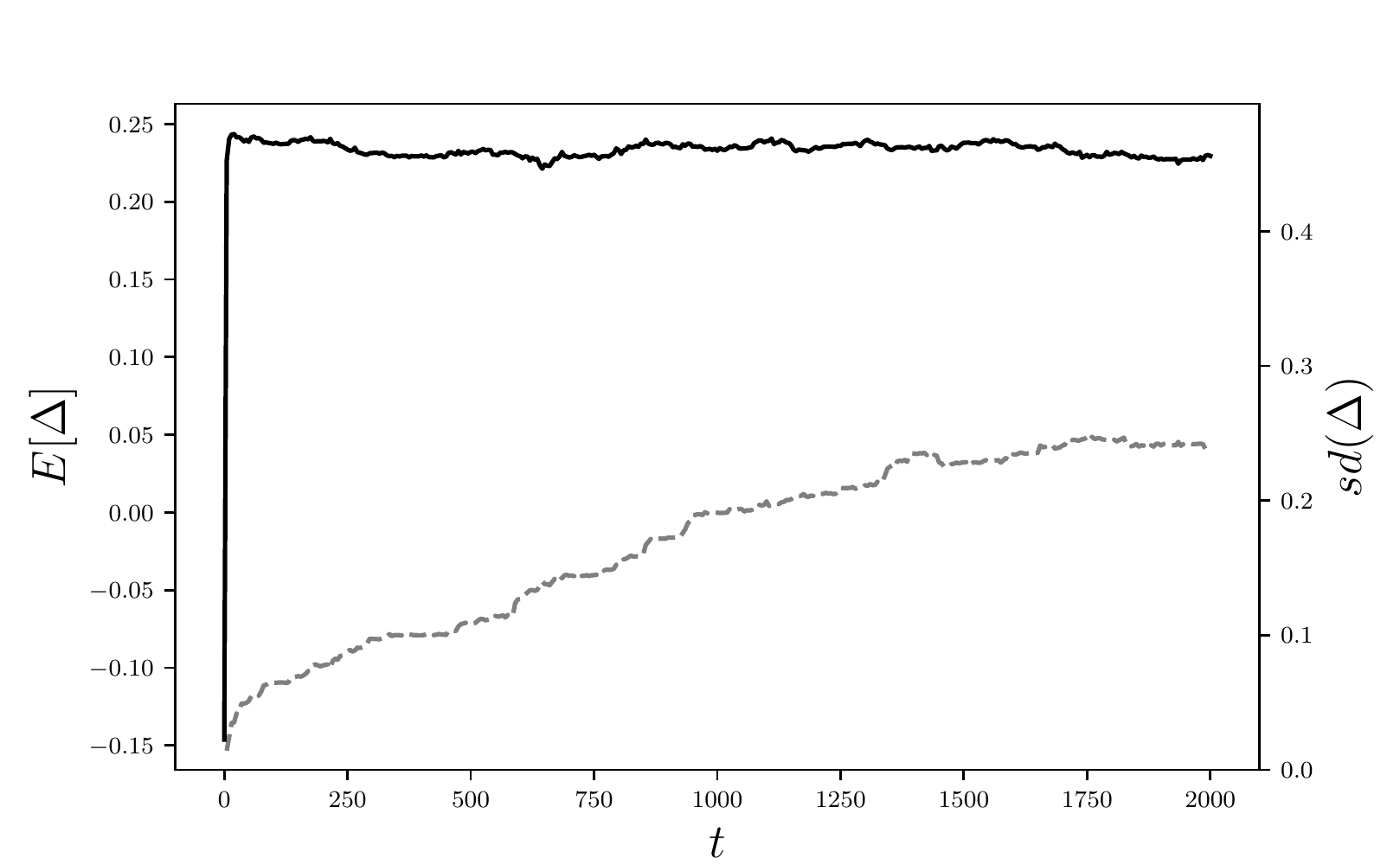}};
    \draw  (figure.north)  node[anchor=north,black]{\textbf{\small{(b)}}};
\end{tikzpicture}
\end{subfigure}\\
\begin{subfigure}[b]{0.5\linewidth}
\centering
\begin{tikzpicture}
    \draw node[name=figure] {\includegraphics[width=4cm]{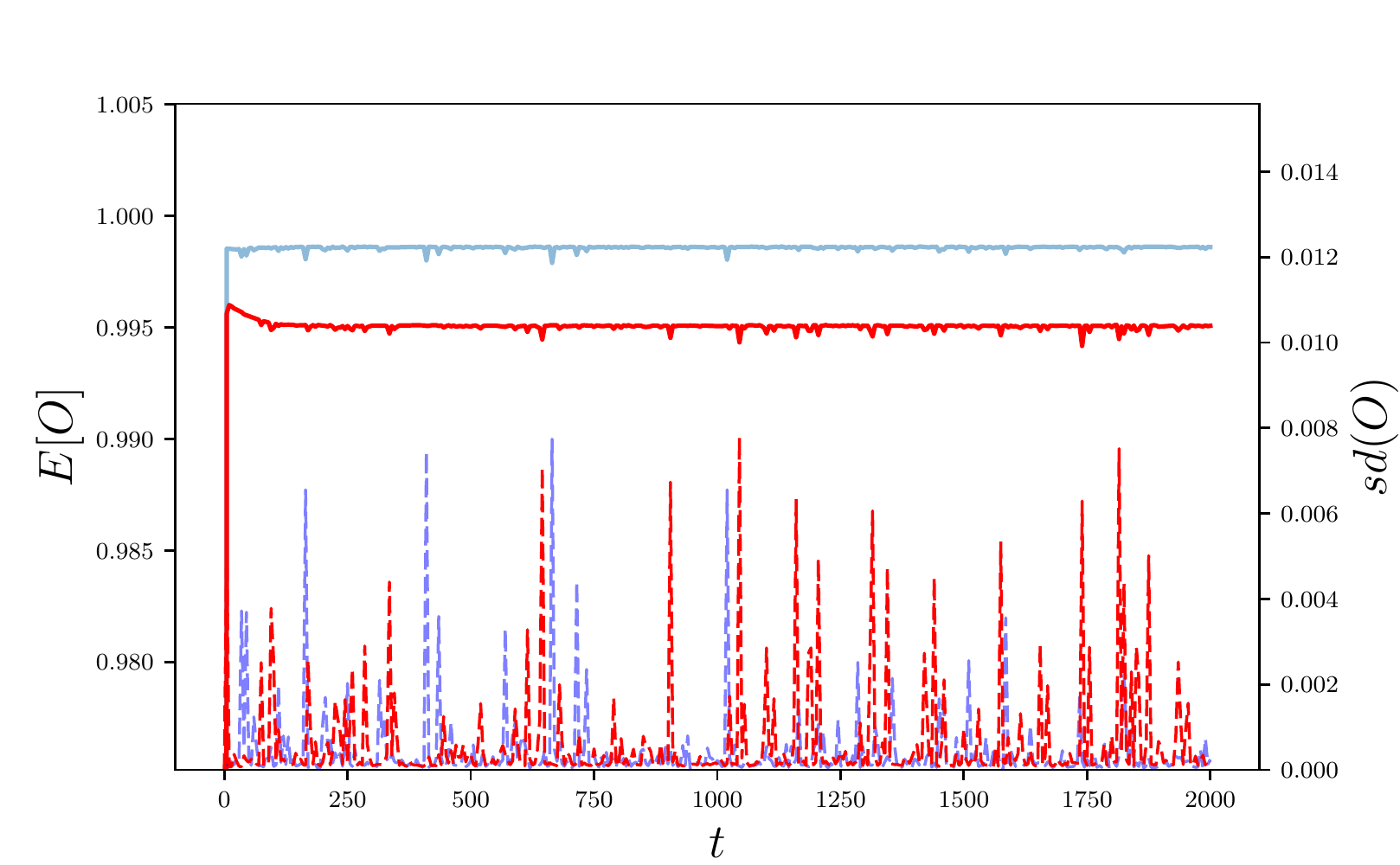}};
    \draw  (figure.north)  node[anchor=north,black]{\textbf{\small{(c)}}};
\end{tikzpicture}
\end{subfigure}\hfill
\begin{subfigure}[b]{0.5\linewidth}
\centering
\begin{tikzpicture}
    \draw node[name=figure] {\includegraphics[width=4cm]{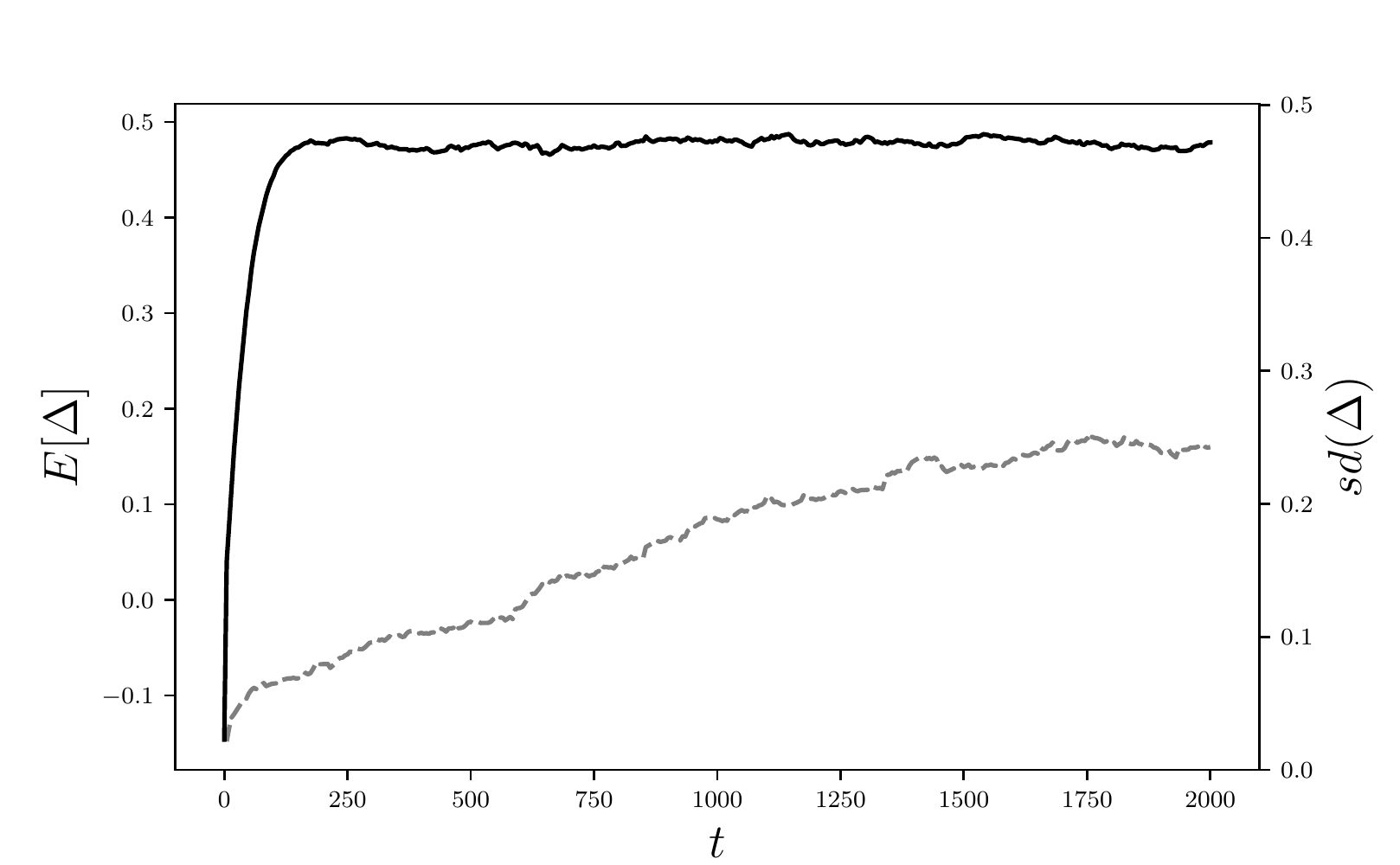}};
    \draw  (figure.north)  node[anchor=north,black]{\textbf{\small{(d)}}};
\end{tikzpicture}
\end{subfigure}\\
\begin{subfigure}[b]{0.5\linewidth}
\centering
\begin{tikzpicture}
    \draw node[name=figure] {\includegraphics[width=4cm]{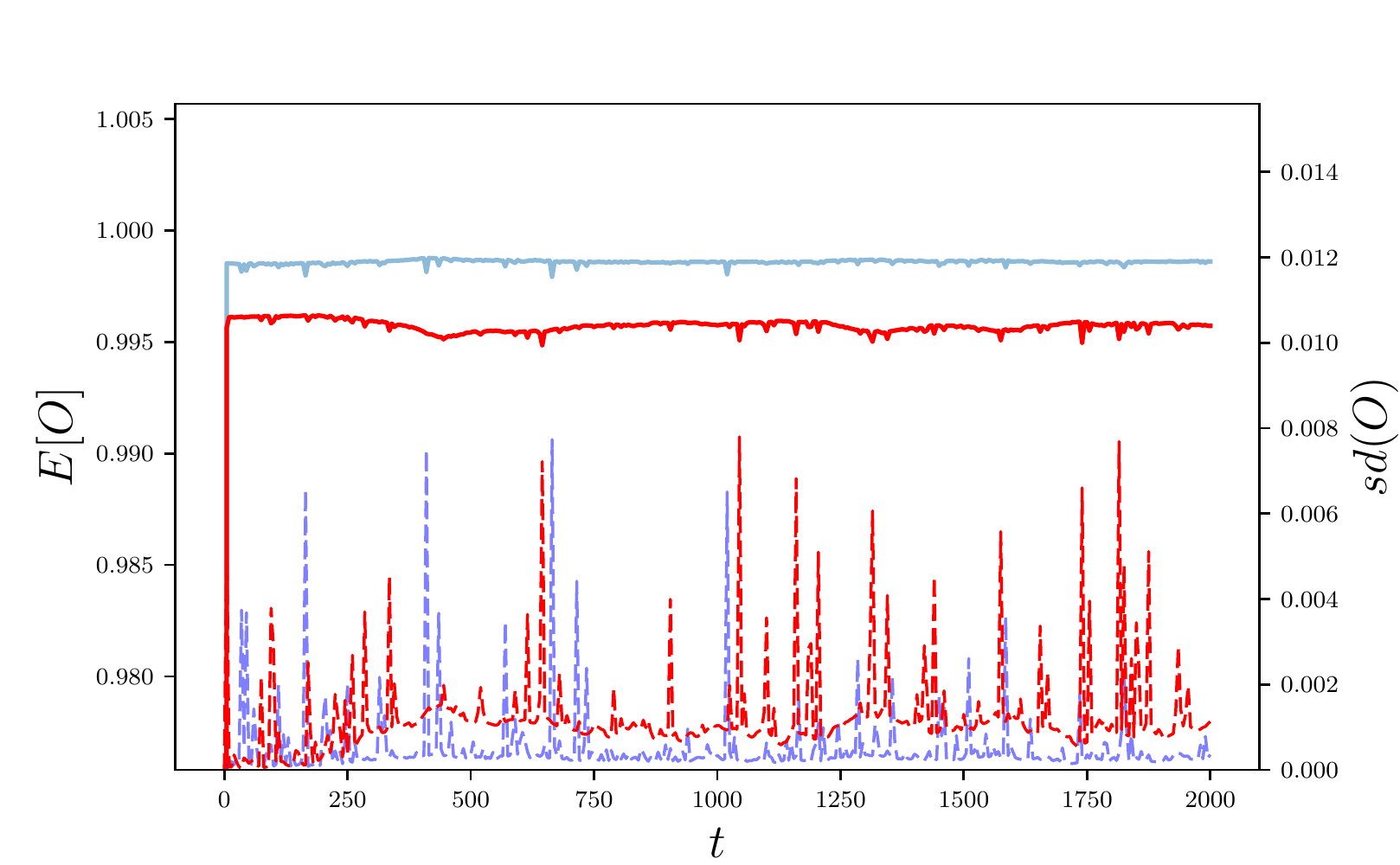}};
    \draw  (figure.north)  node[anchor=north,black]{\textbf{\small{(e)}}};
\end{tikzpicture}
\end{subfigure}\hfill
\begin{subfigure}[b]{0.5\linewidth}
\centering
\begin{tikzpicture}
    \draw node[name=figure] {\includegraphics[width=4cm]{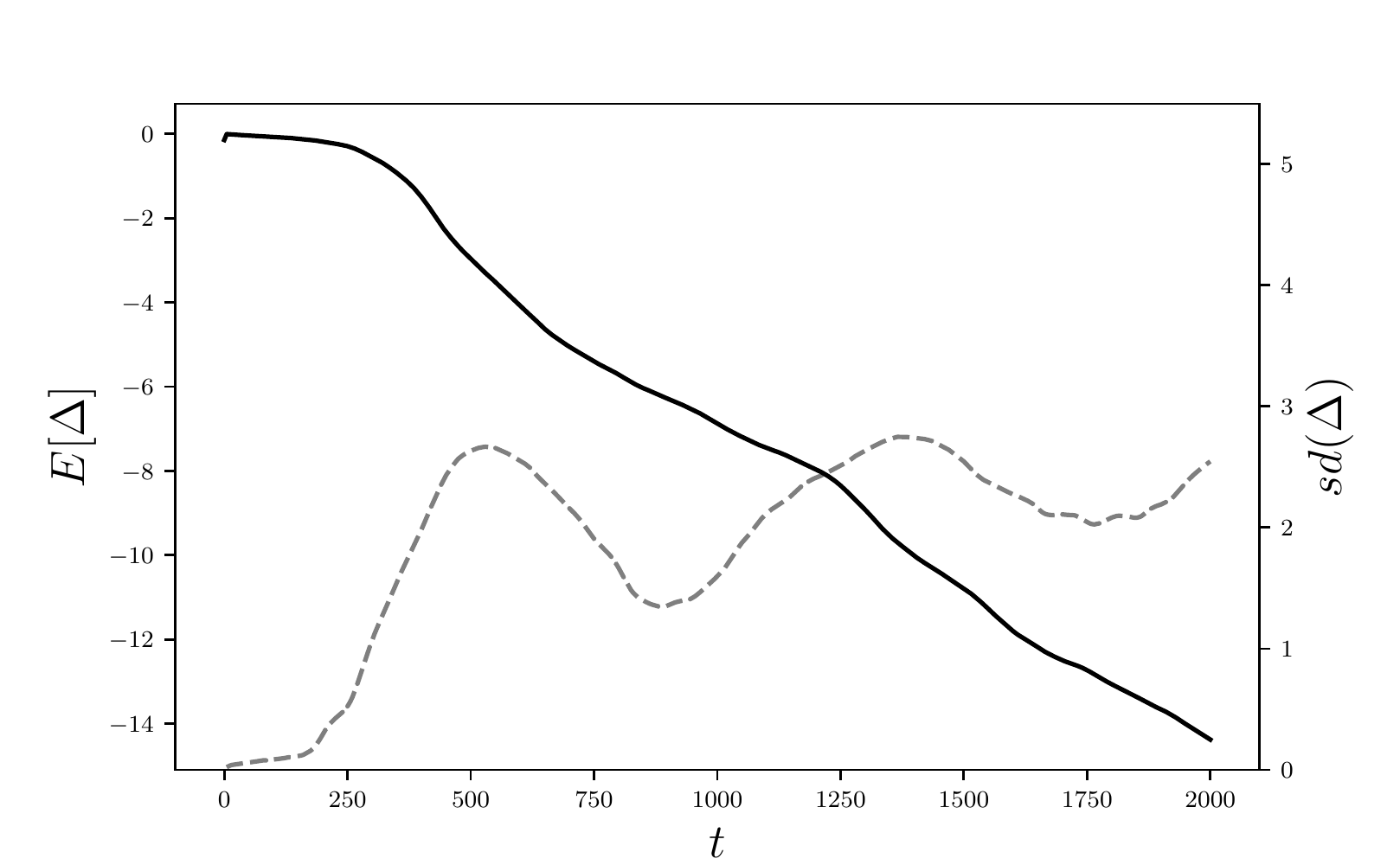}};
    \draw  (figure.north)  node[anchor=north,black]{\textbf{\small{(f)}}};
\end{tikzpicture}
\end{subfigure}\\
\begin{subfigure}[b]{0.5\linewidth}
\centering
\begin{tikzpicture}
    \draw node[name=figure] {\includegraphics[width=4cm]{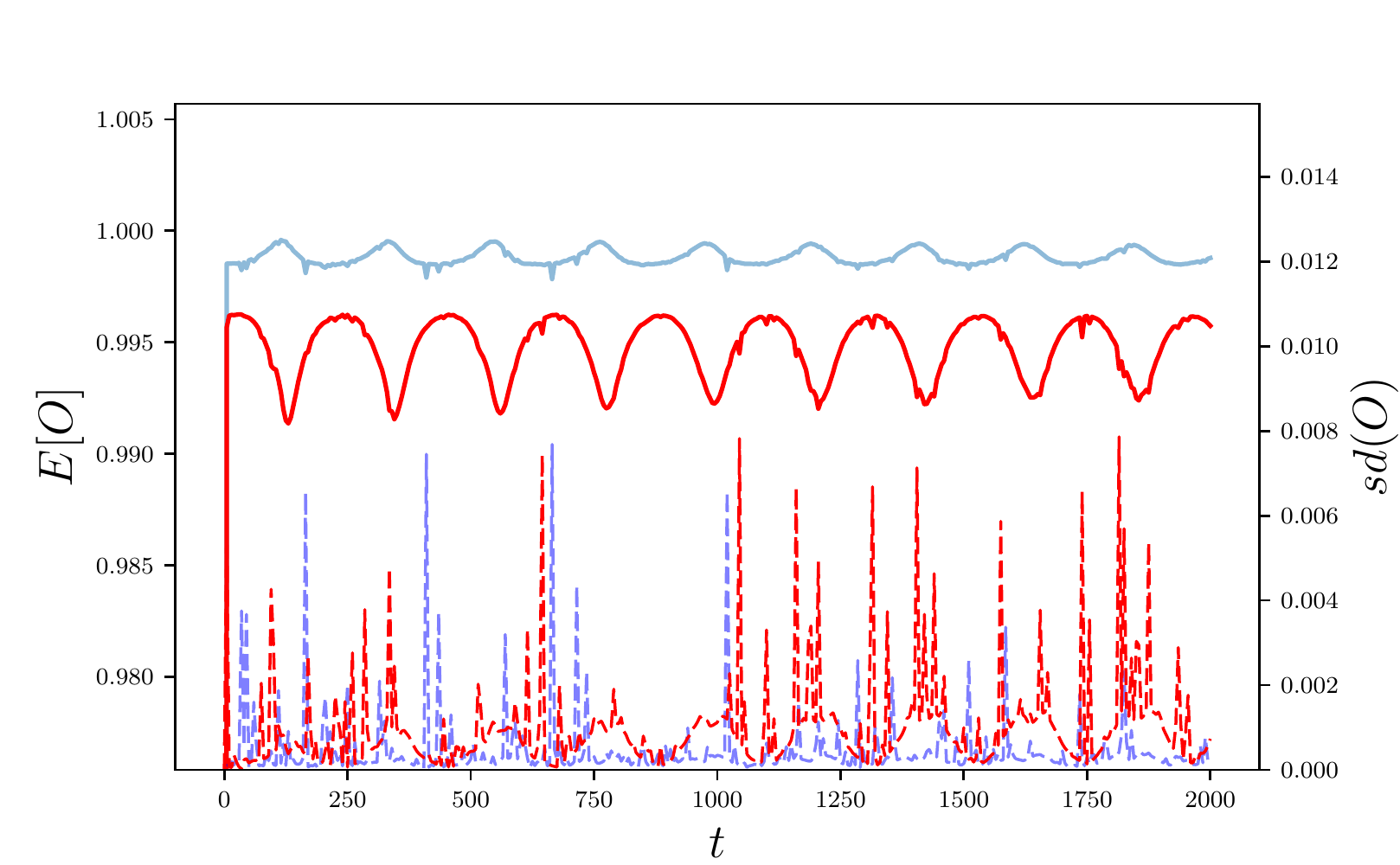}};
    \draw  (figure.north)  node[anchor=north,black]{\textbf{\small{(g)}}};
\end{tikzpicture}
\end{subfigure}\hfill
\begin{subfigure}[b]{0.5\linewidth}
\centering
\begin{tikzpicture}
    \draw node[name=figure] {\includegraphics[width=4cm]{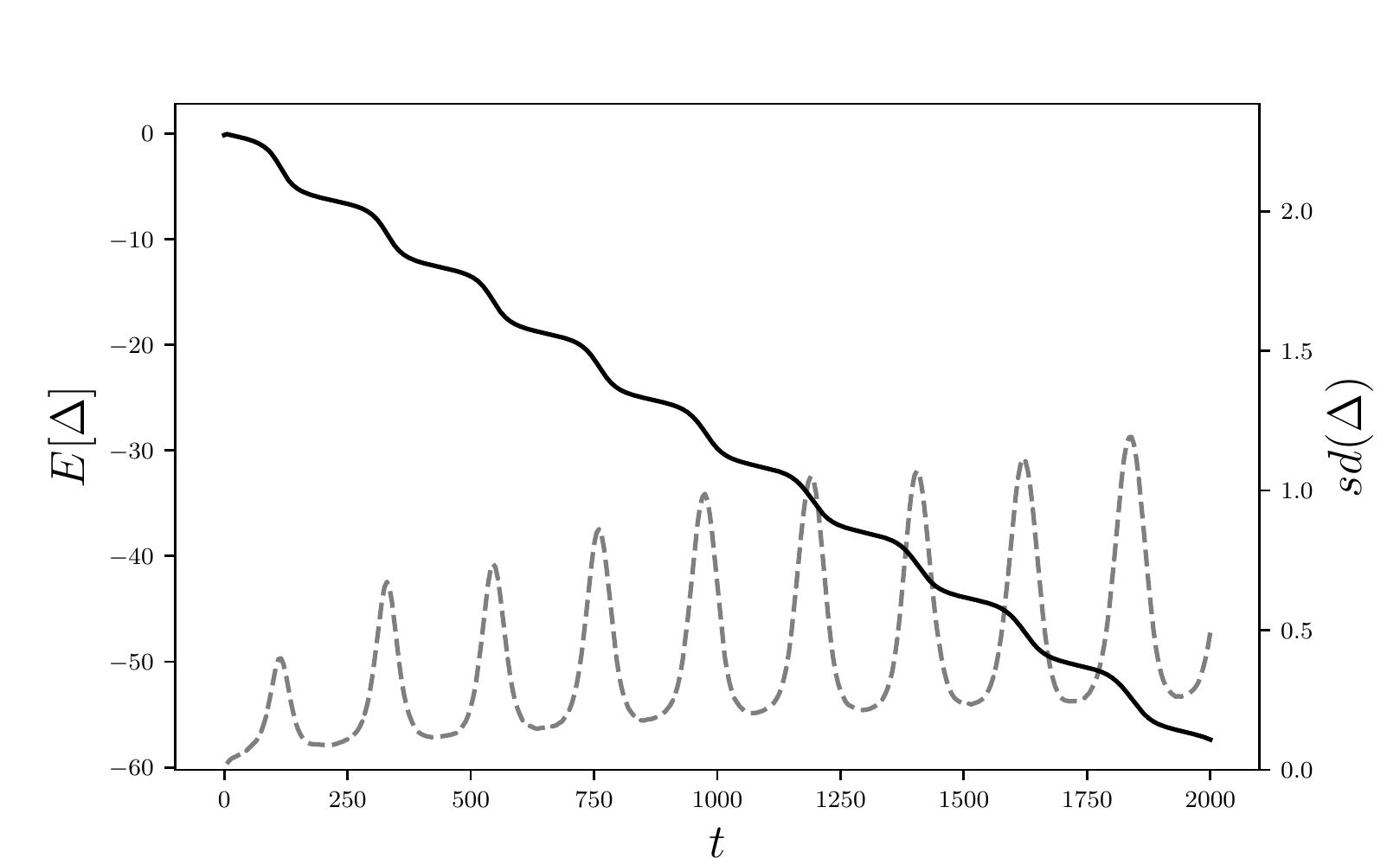}};
    \draw  (figure.north)  node[anchor=north,black]{\textbf{\small{(h)}}};
\end{tikzpicture}
\end{subfigure}
\caption{Time series plots in the stable L\'evy noise case with $\sigma = 0.05$ and $\alpha = 0.5$ and $\lambda = 0.75$. Left axis: Time series plots for the average order parameters and $\Delta$. Right axis:  Time series plots (dashed lines) for the standard deviation of the Monte-Carlo estimates of the order parameters and $\Delta$. Left column: order parameters for the blue and red networks as given by the  blue (light grey) and  red (grey) lines respectively. Right column: $\Delta$ as denoted by the black lines. (a) and (b): $(\phi,\psi) = (0.2\pi,0)$, (c) and (d): $(\phi,\psi) = (0.94\pi,0)$,(e) and (f): $(\phi,\psi) = (0.95\pi,0)$,(g) and (h): $(\phi,\psi) = (0.96\pi,0)$ }

\label{fig:TStableTS4}
\end{figure}

We have avoided thus far to show results for $0.5<\alpha<1$ as there is little structure distinguishable due to the heaviness of the tails.
However, in the presence of tempering with
$\alpha=0.5$ recognisable structures emerge as shown in Fig.\ref{fig:TStableTS4}. We see here the behaviour
predicted analytically, that with tempering and small values of $\alpha$ a smoother dynamics is restored.

\subsection{Objective function for deterministic system}
We now focus on the fitness landscape, where fitness refers to optimal {\it internal synchronisation and phase lag}
for one population with respect to the other; we choose Blue for this purpose. 
To this end we choose an objective function given by 
\begin{equation}\label{eq:objfunc}
U(\phi,\psi) = \left(\sum_{i=1}^3 w_{i}U_{i}\right)^{-\frac{1}{4}},
\end{equation}
where $w_i$ are the weights of sub-objective functions $U_i$. These sub-objective functions, upon the discretisations $(O_B)_i = O_B(t_i), t_i \in [0,T], i = 1,\dots,N$ , and similarly for $(O_R)_i$ and $\alpha_i$, are given by
\begin{align}\label{eq:subobjfunc}
U_1 =& \sqrt{\sum_{i=1}^N (  ( O_B)_i - 1  )^2    } +         \sqrt{\sum_{i=1}^N (  \frac{d}{dt}( O_B)_i - 0  )^2    } ,\\ 
U_2 =& \sqrt{\sum_{i=1}^N (  \Delta_i - \phi  )^2    }, \\
U_3 =& \sqrt{\sum_{i=1}^N (  \frac{d}{dt} ( O_R)_i - 0  )^2 },
\end{align}
where $U_1$,$U_2$ and $U_3$ corresponds to $O_B$ being constant and close to 1, $\Delta$ being close to $\phi$ and $O_R$ being constant respectively.
We indicate the value of $0$ in Eq.(\ref{eq:subobjfunc}) to emphasise that objective function seeks constant $O_B$ and $O_R$.
This choice of combination of $U_1, U_2$ and $U_3$ captures the trade-offs required by the Blue agents to maintain both stable internal synchronisation $O_B=1, \dot{O}_B=0$, and 
maintain a lead in the cycle $\phi$ over the competitor, Red, namely $\Delta=\phi$. We also include a requirement that Red have static
value of the order parameter $O_R$, without seeking it to be close to 1. In other words, Blue prefers to be internally well-synchronised, and Red
at best frequency synchronised but to the extent that Blue agents interacting with Red may achieve a phase advantage of $\phi$. The choice
of the power $-\frac{1}{4}$ in Eq.(\ref{eq:objfunc}) is for visualisation purposes to enable better resolution of peaks and troughs in the landscape,
as will be seen shortly.

\subsection{Bayesian optimisation}\label{bayesopt}

Instead of performing an exhaustive grid search across the parameters $\phi$ and $\psi$ over the region $[0,2\pi] \times [0, 2 \pi]$ we approach the problem of finding a global optimum of the various objective functions using the theory of constrained global optimisation built upon the ideas of Bayesian inference and Gaussian processes
\cite{Ras2006,Snoek2012}. In other words, we seek
\begin{equation}
\max_{\phi,\phi}\left\{ U(\phi,\psi): \phi,\psi \in [0, 2\pi] \right\},
\end{equation}
using the minimal number of probes of $U(\phi, \psi)$. One of the consequences of this approach is that we may obtain a view of the fitness landscape by interpolating over the intermediate values obtained by the algorithm as it searches for the global maximum.

Bayesian optimisation is different to other kinds of optimisation in that it constructs a probabilistic model for 
seeking the maximum of a function $f(x)$ on some bounded set $\sX \subset \R^d$ and then exploits this model to make decisions about where in $\sX$ to evaluate next. It uses all of the information available from previous evaluations of $f(x)$ and not simply local gradient and Hessian approximations. 
This gives an algorithm that finds the maximum of non-convex functions with relatively few evaluations at the cost of performing more computation to determine the next point to try. This approach is particularly well suited when the evaluation of $f(x)$ is expensive to perform (as in our case).
When performing this approach in practice there are two major implementation choices that need to be made. The first is a model for the prior over functions that express assumptions about the function being optimised. We assume here a Gaussian process prior for the influence of $\phi$ and $\psi$ on ${\cal K}$. Second, the implementation requires a choice of an `acquisition function' used to construct a utility function from the model posterior and allows the algorithm to determine the next point to evaluate. We used the implementation of Bayesian optimisation provided by the \texttt{BayesOptimisation} package~\cite{BayesOpt}.

To improve the performance of the Bayesian optimisation algorithm and, in the case of the stochastic system, to account for the standard error of our estimate in Eq.(\ref{eq:MCdist}), we perform data smoothing on a coarser version of the original data set through the application of a symmetric moving average filtering algorithm. The symmetric moving average filter that we used is of window length $2n +1$ and is given by
\begin{equation}\label{eq:smoothed}
(\hat{O}_B)_i = \sum_{j =-n}^{n} {\varpi}_j \mathbb{E}\left[(O_B)_i\right], \quad n < i < N - n ,
\end{equation}
where the weights are chosen such that $\sum_{j} \varpi_j = 1$ and the expectations are approximated through Eq.(\eqref{eq:MCdist}). In our application of this algorithm we make the simple selection of $\varpi_j = \frac{1}{2n+1}$. For the $n$ observations at the beginning and end of the time series, we use an asymmetric moving average filter to preserve all observations. Unless otherwise stated, we use $n=10$ in our window length. 

\subsection{Fitness landscape for deterministic system}
\begin{figure}[hp]
\begin{subfigure}[b]{\linewidth}
\centering
\begin{tikzpicture}
    \draw node[name=figure] {\includegraphics[width=\columnwidth, trim={2.5cm 0 1.5cm 1cm}]{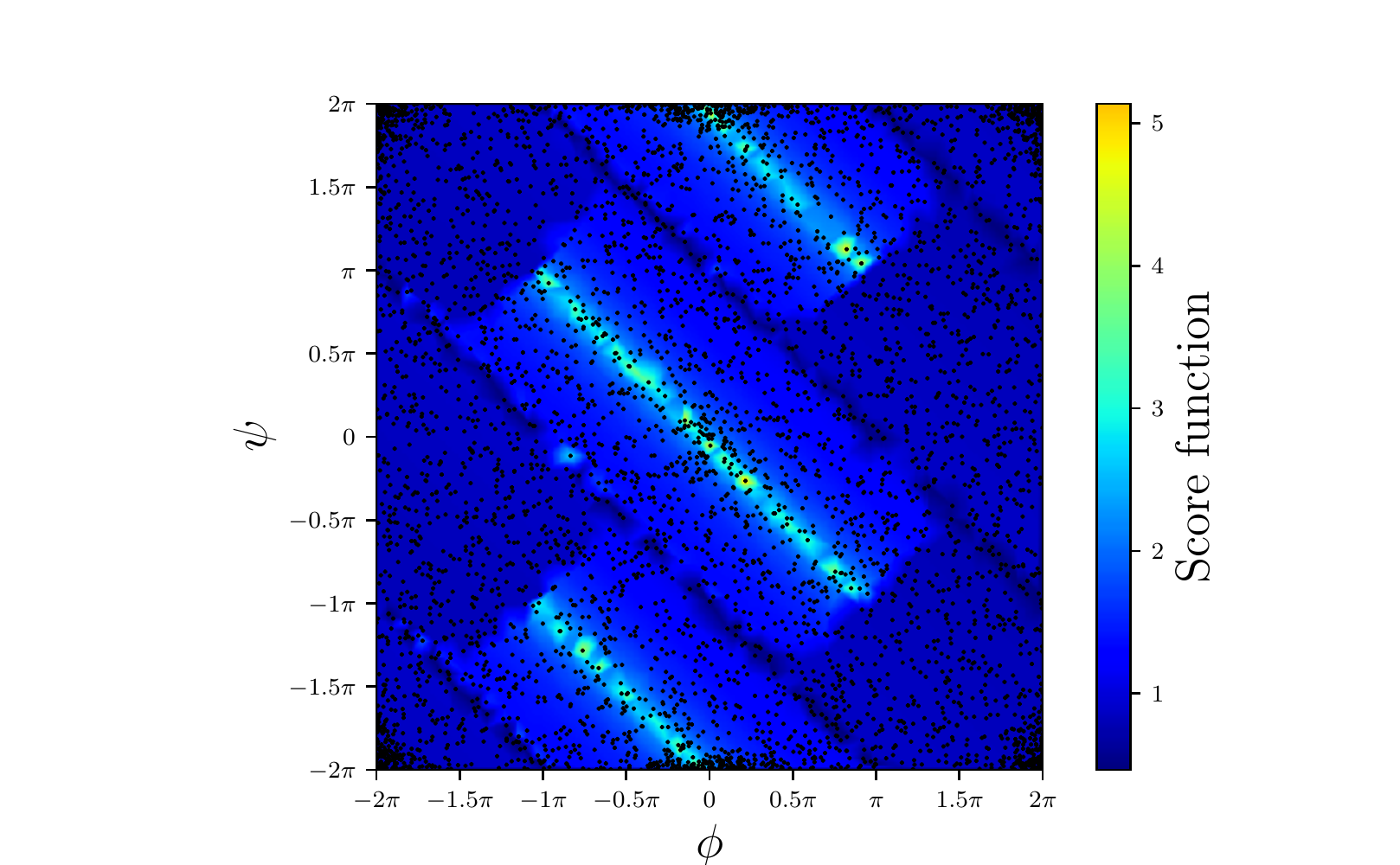}};
    \draw  (figure.north west)  node[anchor=north west,black]{\textbf{\small{(a)}}};
\end{tikzpicture}
\end{subfigure}
\begin{subfigure}[b]{\linewidth}
\centering
\begin{tikzpicture}
    \draw node[name=figure] {\includegraphics[width=0.83\columnwidth]{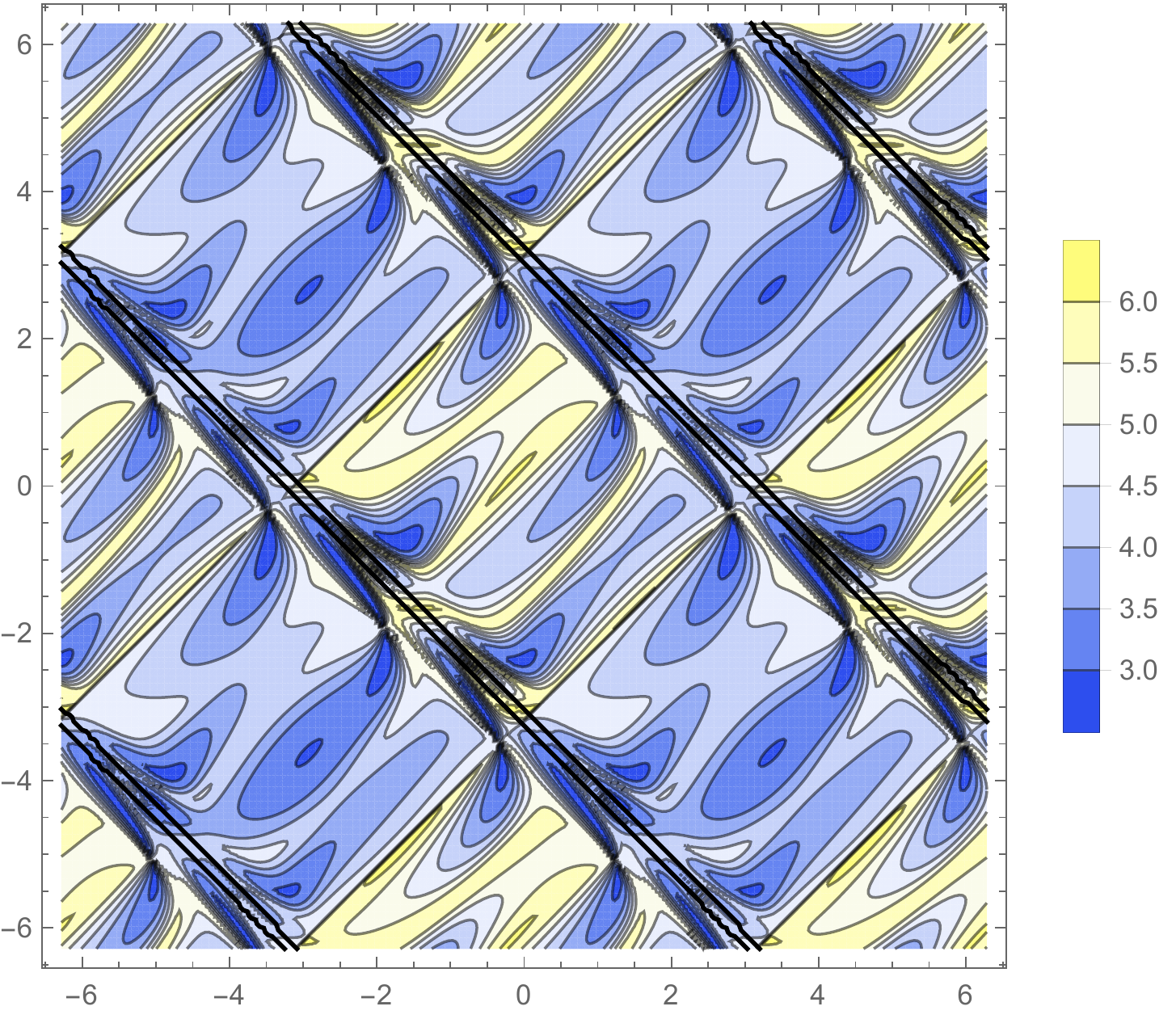}};
    \draw  (figure.north west)  node[anchor=north west,black,xshift=-10]{\textbf{\small{(b)}}};
\end{tikzpicture}
\end{subfigure}
\caption{(a): Objective score function for $U(\phi,\psi)$ in the deterministic case with 500 iterations and T = 1000. The objective weights are given by $w_1 = 1$, $w_2 = 0$ and $w_3 = 1$. High scoring (red) values correspond to $O_B$ constant and close to 1 and $O_R$ constant whilst low scoring values (blue) correspond to either periodicity or noisy behaviour in the times series or $O_B$ far away from 1. Note that the optimal value near $(1.1 \pi,1.85 \pi)$ in this fitness landscape corresponds to a periodic time series with a very long periodicity, beyond that of T=1000. The subsequent plot repeats the construction of the fitness landscape for T=2000.
(b): the objective function
including only the analytical evaluation of $O_B$ truncating to leading order in the fixed point.
This is overlaid in black with the lines where ${\cal K}=0$.
}
\label{ObjFuncPlots}
\end{figure}

In the top panel of Fig.\ref{ObjFuncPlots} we show the objective function for the deterministic system in the $\phi,\psi$  plane. 
We compare it below with a plot of 
the objective function but with only the term
corresponding to $w_1$ included, where
we evaluate $O_B(\phi,\psi)$ based on the truncation to lowest order in the fixed point approximation,
Eq.(\ref{O_Btrunc}), as discussed earlier. In addition, we overlay on this latter plot the lines (in black) where ${\cal K}=0$. 

In the top panel of Fig.\ref{ObjFuncPlots} we observe an overall `ridge' of optimality in the bright strip on the diagonal, and, with same alignment, dark narrow troughs far
from optimality. This ridge however sits at the top of a similarly aligned region that drops off
orthogonally to the bright ridge down to the darkly coloured troughs.
Correspondingly, the overall symmetry of the
objective function on the top is displayed in the truncated form of $O_B$ below. 
We may infer that the negatively sloped diagonal of maxima and minima
in the top panel 
are a consequence completely of the behaviour of ${\cal K}$ shown in the lower plot. Because of the specific small choice of $\mu$ in our example,
the region where ${\cal K}<0$ (where the angle between Blue and Red centroids, $\Delta$,
becomes time-dependent) is very narrow - in the narrow gap between the black lines in the bottom plot. This is, therefore, responsible
for the narrow troughs
in the objective function on the top. Similarly, where the objective function takes its maximum coincides 
at points with where ${\cal K}$ achieves its maximum
positive value, as well as $O_B$ itself as seen
periodically along the central diagonal of the lower plot.

However, something difficult to detect in a single quantity such as used in $O_B$ on the bottom but evident in the full objective function on the top is that 
it is impossible for Blue to achieve the {\it combined objective} of optimal internal synchronisation {\it and} large frustration in relation to Red, $\phi$, for a fixed value of Red's frustration in relation to Blue. This is visible in the top plot as follows. On the one hand, following the bright ridge of the plot, positive values of $\phi$ require negative values of $\psi$ to maintain high optimality. This ridge corresponds to
Red being `compliant' with Blue's intent: large positive phase $\phi$ for Blue is consistent with large negative phase, $-\phi$, for Red. 
However these bands do not extend indefinitely along the diagonals of the
$(\phi,\psi)$ plane but are truncated. 
This is a consequence of the presence of $U_2$ in the objective function Eq.(\ref{eq:subobjfunc})
where optimality requires a sign of $\Delta$: Blue seeks to be a certain phase {\it ahead} of Red.
This truncation of the
diagonal band is only seen partially in the bottom plot
in that the central cell is a square rather than a rectangle
in the upper plot.

Once values of $(\phi,\psi)$ off the diagonal are chosen, the degree of optimality drops off into
rectangular bands that align with the diagonal ridge. This drop off is a consequence of
the combination of internal synchronisation {\it and} achievement by Blue of maintaining close to the phase difference $\phi$ ahead of Red,
but where Red chooses `non-compliant' values $\psi$.
The larger is $\psi>0$, the smaller the range of $\phi>0$
available before instability occurs, namely a dark ridge in the plot is encountered. 
Thus, this landscape representation summarises the trade-offs in internal synchronisation and ability to synchronise in
a stable manner ahead of the competitor available to one side or the other given the choices of the other.
The question we address next is how much of this structure is maintained in the presence of stable and tempered stable noise.

\subsection{Fitness landscape for stochastic system}

To study the fitness landscape in the presence of noise, we use the objective function of Eq.(\ref{eq:objfunc}) 
but with sub-objective functions using the discretisations $(\hat{O}_B)_i = \hat{O}_B(t_i), t_i \in [0,T], i = 1,\dots,N$ , and similarly for $(\hat{O}_R)_i$ and $(\hat{\Delta})_i$, are given in terms of the smoothed expected values as defined in Eq.(\ref{eq:smoothed}). The sub-objective functions then are as follows
\begin{align}\label{eq:subobjfunc3}
U_1 =& \sqrt{\sum_{i=1}^N (  (\hat{O}_B)_i - 1  )^2    } +   \sqrt{\sum_{i=1}^N (  \frac{d}{dt} (\hat{O}_B)_i - 0  )^2    }, \\ 
U_2 =& \sqrt{\sum_{i=1}^N (  (\hat{\Delta})_i - \phi  )^2    }, \\
U_3 =& \sqrt{\sum_{i=1}^N (  \frac{d}{dt} (\hat{O}_R)_i - 0  )^2 },
\end{align}
where $U_1$,$U_2$ and $U_3$ corresponds to $\hat{O}_B$ being constant and close to 1, $\hat{\Delta}$ being close to $\phi$ and $\hat{O}_R$ being constant respectively. 

\begin{figure*}[tp]
\begin{subfigure}[b]{0.5\linewidth}
\centering
\begin{tikzpicture}
    \draw node[name=figure] {\includegraphics[width=\columnwidth, trim={2cm 0 0 0}]{Figures/deterministic-BvR-landscape.pdf}};
    \draw  (figure.north)  node[anchor=north,black]{\textbf{\small{(a)}}};
\end{tikzpicture}
\end{subfigure}\hfill
\begin{subfigure}[b]{0.5\linewidth}
\centering
\begin{tikzpicture}
    \draw node[name=figure] {\includegraphics[width=\columnwidth, trim={2cm 0 0 0}]{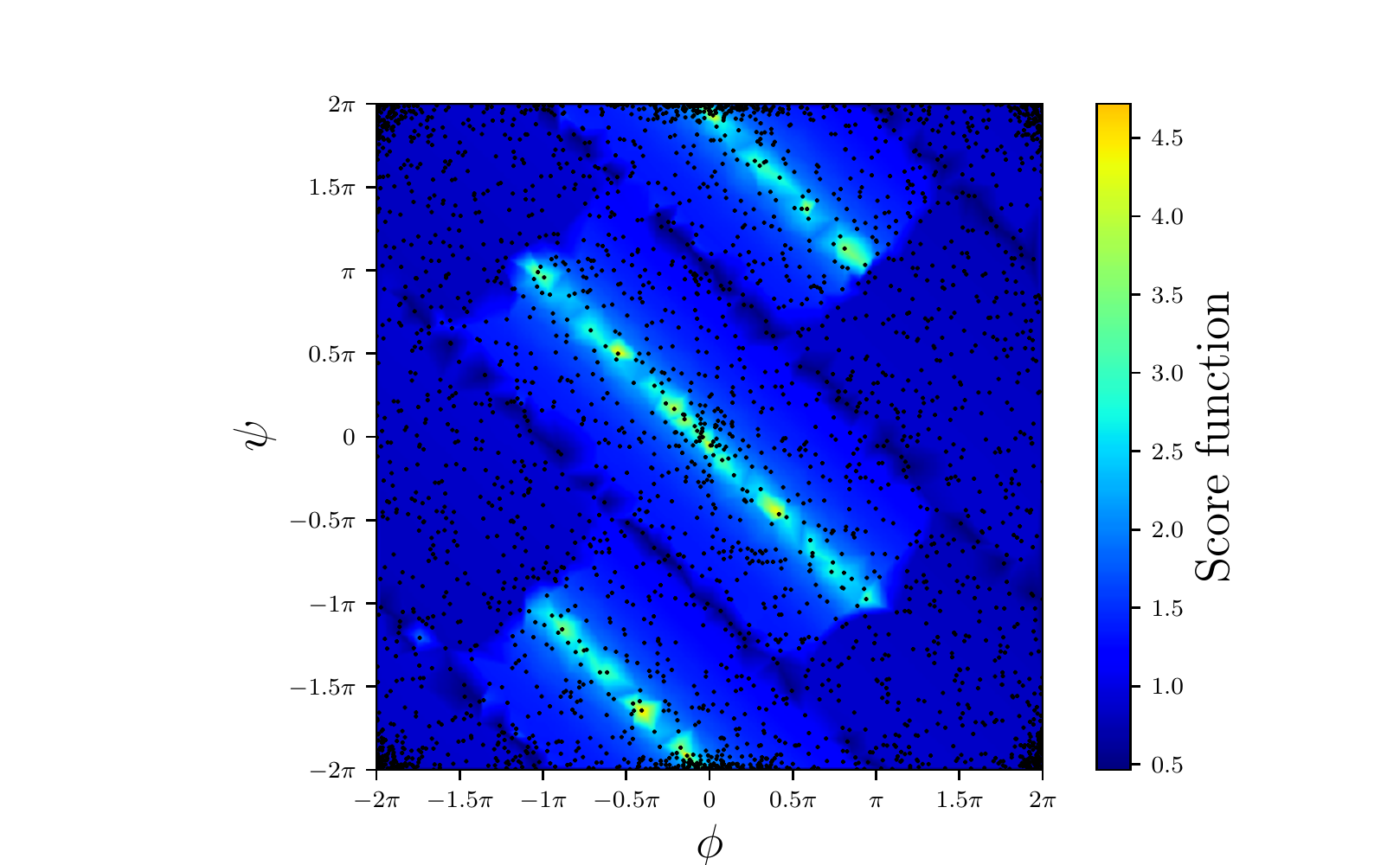}};
    \draw  (figure.north)  node[anchor=north,black]{\textbf{\small{(b)}}};
\end{tikzpicture}
\end{subfigure}\\
\begin{subfigure}[b]{0.5\linewidth}
\centering
\begin{tikzpicture}
    \draw node[name=figure] {\includegraphics[width=\columnwidth, trim={2cm 0 0 0}]{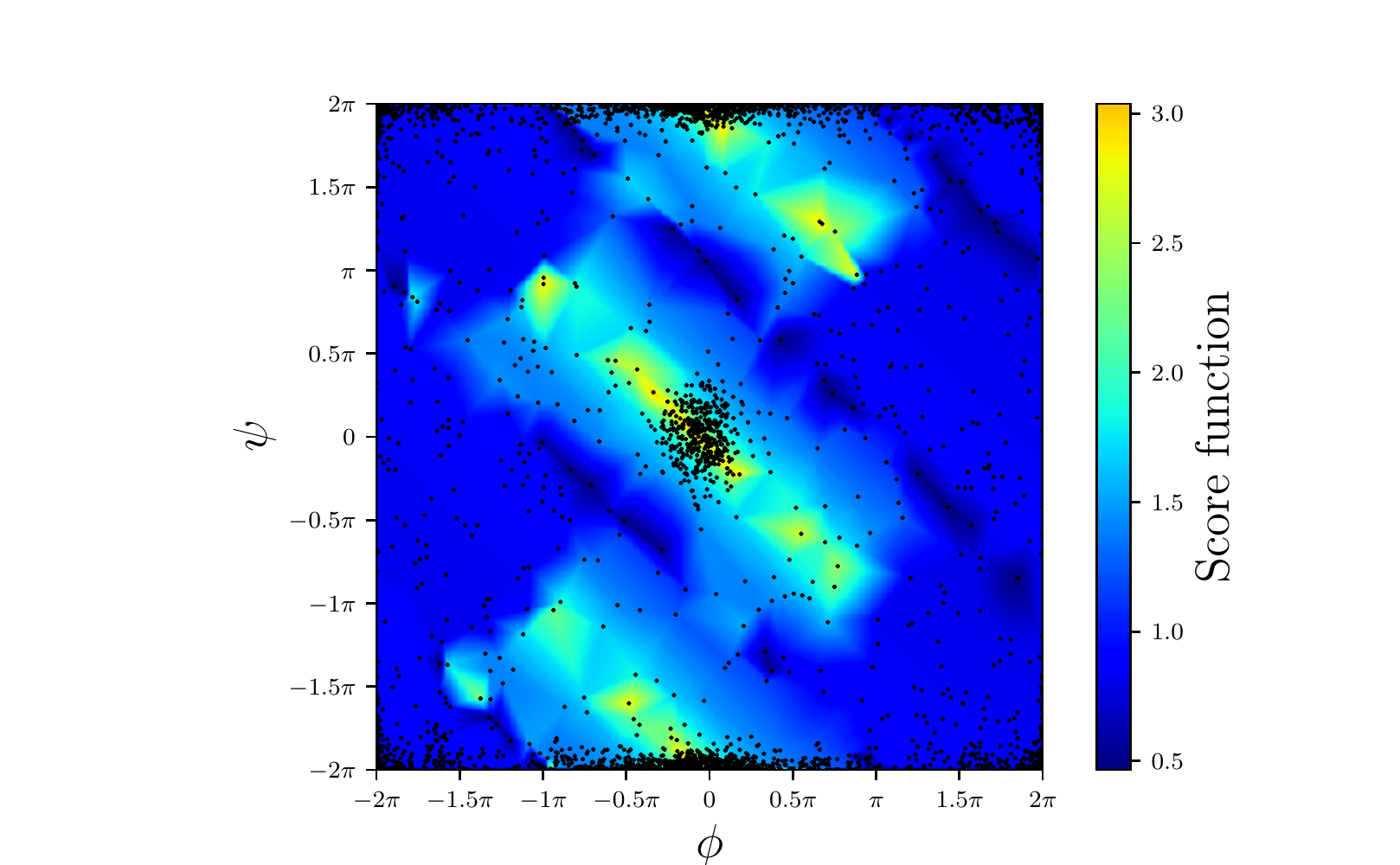}};
    \draw  (figure.north)  node[anchor=north,black]{\textbf{\small{(c)}}};
\end{tikzpicture}
\end{subfigure}\hfill
\begin{subfigure}[b]{0.5\linewidth}
\centering
\begin{tikzpicture}
    \draw node[name=figure] {\includegraphics[width=\columnwidth, trim={2cm 0 0 0}]{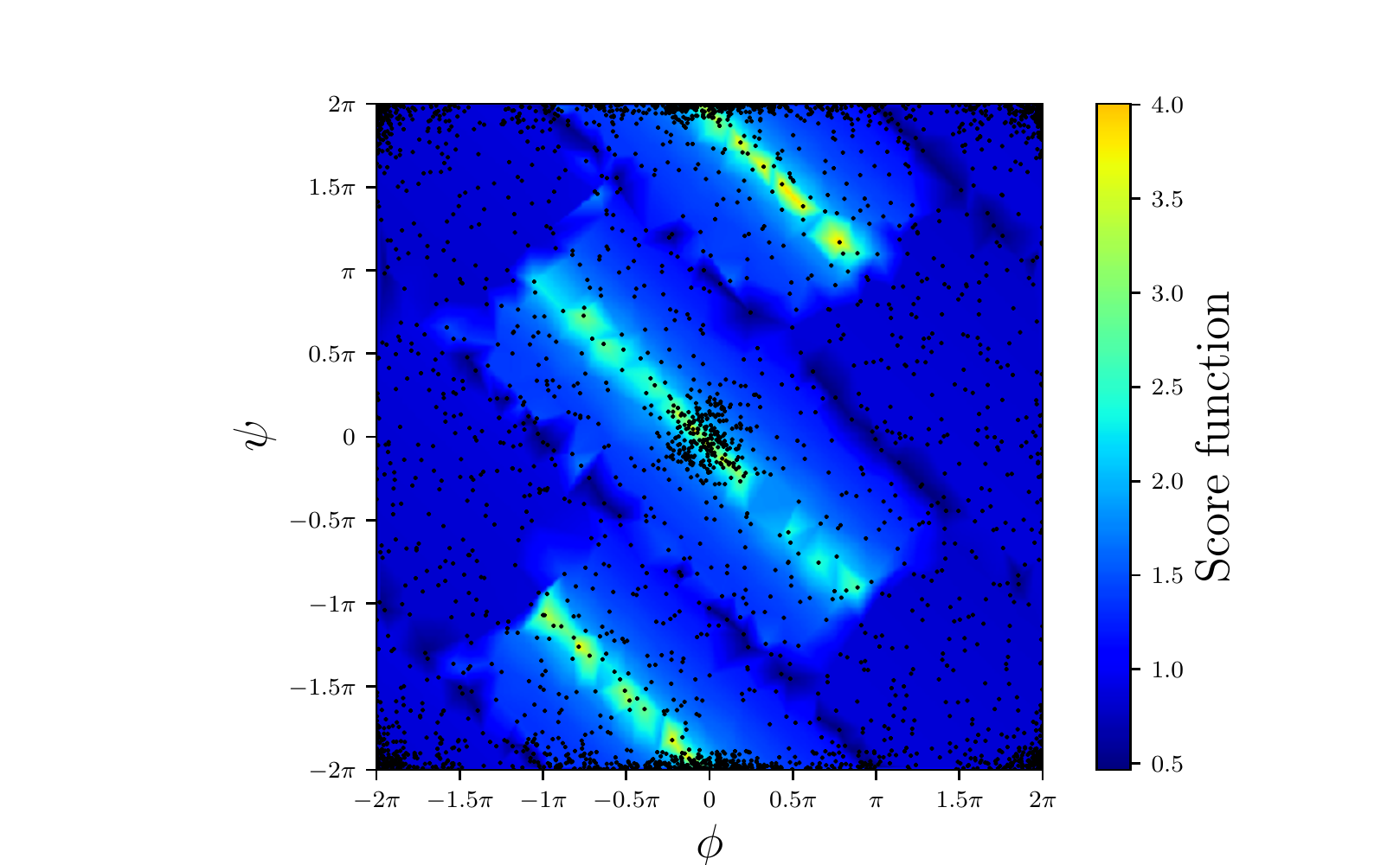}};
    \draw  (figure.north)  node[anchor=north,black]{\textbf{\small{(d)}}};
\end{tikzpicture}
\end{subfigure}
\caption{Objective score function for $U(\phi,\psi)$ for different noise cases: (A) deterministic , (b) Gaussian (top right), (c) stable with
$\sigma=0.05, \alpha=1.5$, and (d) tempered stable with $\sigma=0.05,\alpha=1.5,\lambda=0.5$, with 500 iterations and $T = 1000$. The objective weights are given by $w_1 = 1$, $w_2 = 0$ and $w_3 = 1$. High scoring (red) values correspond to $O_B$ constant and close to 1 and $O_R$ constant whilst low scoring values (blue) correspond to either periodicity or noisy behaviour in the times series or $O_B$ far away from 1. Note that the optimal value near $(1.1 \pi,1.85 \pi)$ in this fitness landscape corresponds to a periodic time series with a very long periodicity, beyond that of $T=1000$. The subsequent plot repeats the construction of the fitness landscape for $T=2000$.
}
\label{ObjFuncStochPlots}
\end{figure*}

In Fig.\ref{ObjFuncStochPlots} we show four cases of the objective function, comparing the deterministic, Gaussian (top left and right) and, importantly, the
stable case for $\alpha=1.5$ (bottom left) and tempered stable case with $\alpha=1.5,\lambda=0.5$. In particular, for the two non-Gaussian cases
we have used the Bayesian optimisation approach.

\begin{figure}[hp] 
\begin{subfigure}[b]{1\linewidth}
\centering
\begin{tikzpicture}
    \draw node[name=figure] {\includegraphics[width=\textwidth]{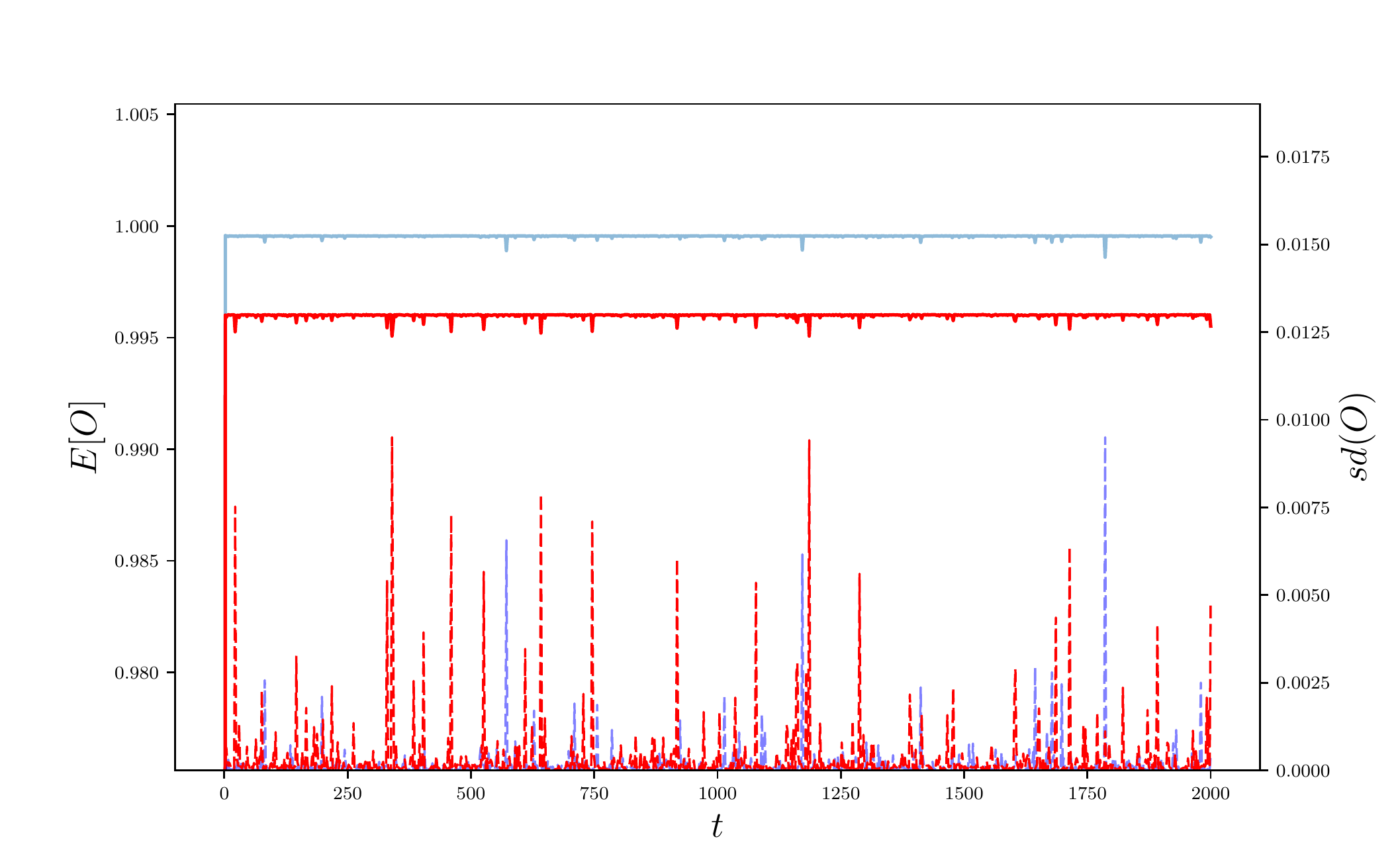}};
    \draw  (figure.north)  node[anchor=north,black]{\textbf{\small{(a)}}};
\end{tikzpicture}
\end{subfigure}
\begin{subfigure}[b]{1\linewidth}
\centering
\begin{tikzpicture}
    \draw node[name=figure] {\includegraphics[width=\textwidth]{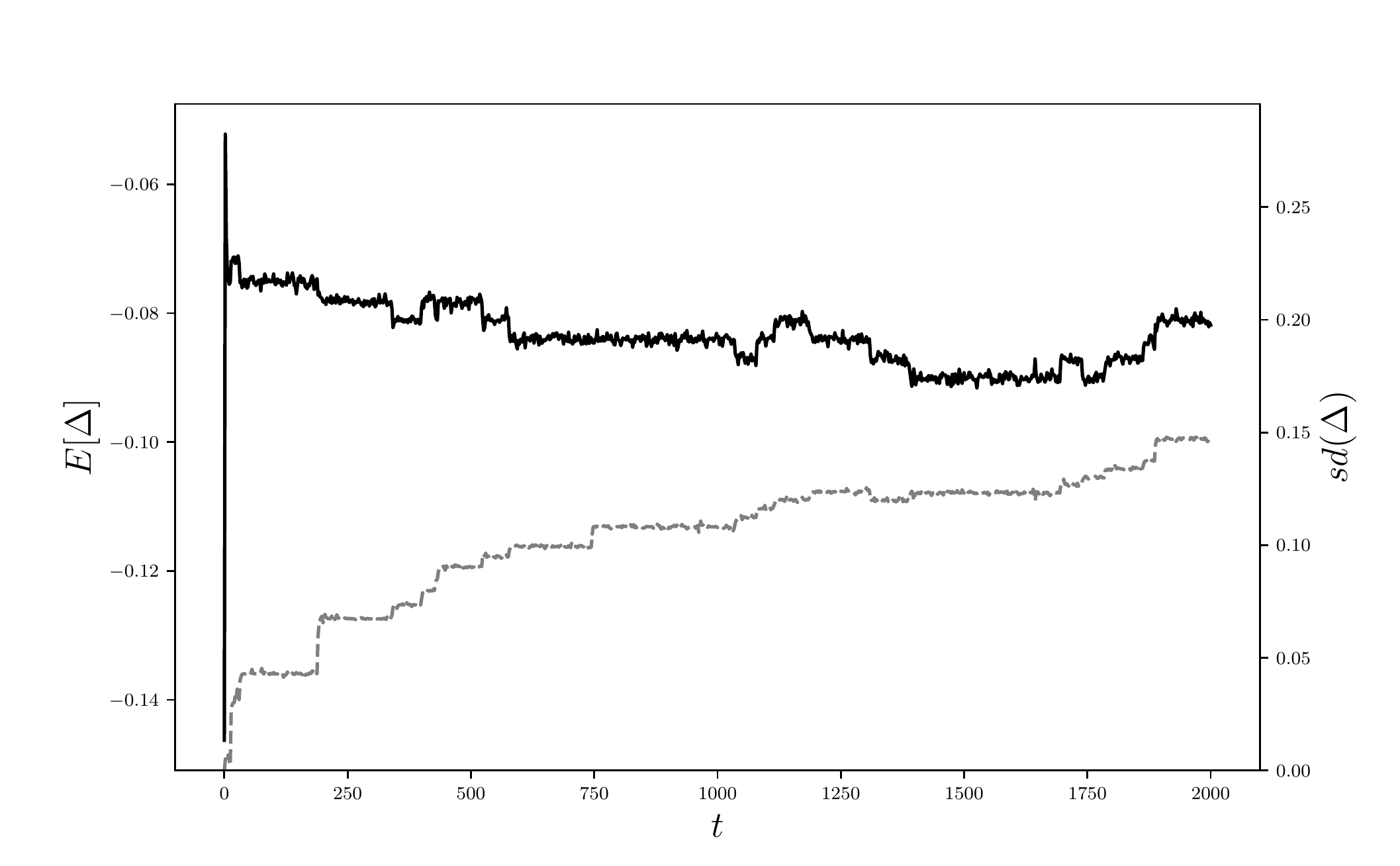}};
    \draw  (figure.north)  node[anchor=north,black]{\textbf{\small{(b)}}};
\end{tikzpicture}
\end{subfigure}
\caption{Time series plots in the optimal regime of $U(\phi,\psi)$ in the tempered stable L\'evy noise case with $\sigma = 0.05$, $\alpha = 1.5$, $\lambda = 0.5$ and $(\phi,\psi)=(0,0)$. Left axis (solid lines): Time series plots for the average order parameters and $\Delta$. Right axis (dashed lines): Time series plots for the standard deviation of the Monte-Carlo estimates of the order parameters and $\Delta$. (a) order parameters for the blue and red networks as given by the  blue (light grey) and  red (grey) lines respectively. (b) $\Delta$ as denoted by the black lines.}
\label{fig:TSoptimal}
\end{figure}

We observe firstly that the Gaussian case (top right) maintains to a high degree the integrity of the landscape from the deterministic result.
However, stable noise (bottom left), as a consequence of the heavy tails, significantly smears the landscape - we note to this end the
change in the colour gradations for this case showing that overall values of the objective function are suppressed in comparison to
deterministic and Gaussian cases. Also, the accumulation of points searched by the Bayesian optimisation
underscores the relative sub-optimality of $(\phi,\psi)$ values away from the origin. Nevertheless the structure of the landscape is maintained.
With suppression of the heavy tails by tempering (bottom right), the landscape is closer to the Gaussian case. In Fig.\ref{fig:TSoptimal} we plot numerical evidence that the optimal fitness landscape results picks out optimal values by plotting the time series of the various order parameters at a point within the optimal band.

In conclusion for this stage, the overall imprint of the deterministic dynamics is retained in non-Gaussian noise cases. In particular, the
partition of the $(\phi,\psi)$ plane according to the sign of ${\cal K}$ remains key for determining regions of stochastic optimality.

\subsection{Determining the presence of regular and noisy path behaviour }
In order to analyse for which values of $(\alpha,\sigma,\lambda)$ the property of periodicity holds we introduce a new objective function 
\begin{equation}
\label{U01test}
U^{(01)}(\alpha,\sigma) = 1 - K,
\end{equation}
where $K$ is the result of the application of the `0-1 test' upon the $\hat{O}_R$ time series as defined in Eq.(\ref{eq:smoothed}). The 0-1 test is the means for determining the presence of chaos in time series data that is a computationally efficient alternative to computing a Lyapunov exponent \cite{Gottwald2004,Gottwald2009}. In this work, we shall utilise the 0-1 test to differentiate between regular and noisy path behaviour of $\hat{O}_R$.
To apply the 0-1 test upon the time series $\hat{O}_R$, we select an observable from the solution $\psi(\hat{O}_R)$ and some parameter $c > 0$. The method used in this test is independent of the form of the function $\psi$. We now define
\begin{align}
\theta(t) &= ct + \int_0^t \psi ( \hat{O}_R ) ds, \\
p(t) &= \int_0^t \psi(\hat{O}_R) \cos(\theta(s)) ds.
\end{align}
If the underlying dynamics are regular then $p(t)$ is bounded otherwise $p(t)$ behaves asymptotically like Brownian motion. To identify the growth of $p(t)$, the mean-square displacement is used:
\begin{equation}
M(t) = \lim_{T\rightarrow \infty} \frac{1}{T} \int_0^T \left( p (t+ \tau)-p(\tau) \right) ^2 d\tau.
\end{equation}
If the behaviour $p(t)$ is Brownian, then $M(t)$ will grow linearly in time; otherwise if $p(t)$ is bounded, then $M(t)$ will be bounded. To distinguish between these two properties, the asymptotic growth rate of the mean-square displacement is calculated:
\begin{equation}
K = \lim_{t} \frac{\log(M(t))}{\log t}.
\end{equation}
This can be calculated through a linear regression of $\log(M(t))$ and $\log t$ and will be near 1 if the underlying dynamics are chaotic or noisy and 0 otherwise.
Note that the choice of $1-K$ in Eq.(\ref{U01test}) is the reverse of the typical test, used 
in this way here for consistency with the landscape plots wherein large values (periodic behaviour in the 0-1 test and desired dynamics in the previous objective functions) are `good', and low scoring values (noisy behaviour in the 0-1 test and undesired dynamics in the previous objective functions ) 
are 'bad'.

\begin{figure}[hp] 
\begin{subfigure}[b]{1\linewidth}
\centering
\begin{tikzpicture}
    \draw node[name=figure] {\includegraphics[height=5cm]{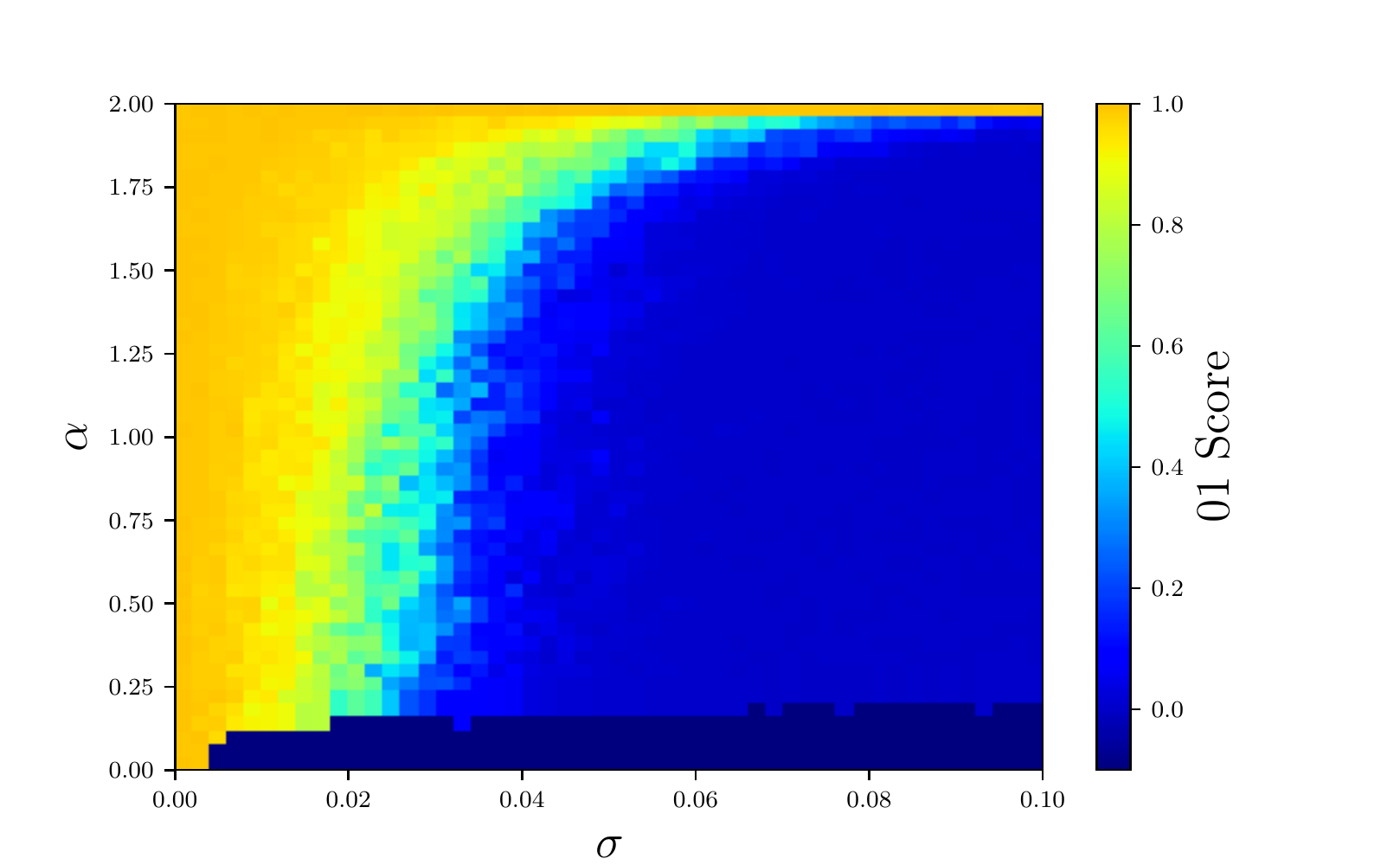}};
\draw  (figure.north)  node[anchor=north,black]{\textbf{\small{(a)}}};
\end{tikzpicture}
\end{subfigure}
\begin{subfigure}[b]{1\linewidth}
\centering
\begin{tikzpicture}
    \draw node[name=figure] {\includegraphics[height=5cm]{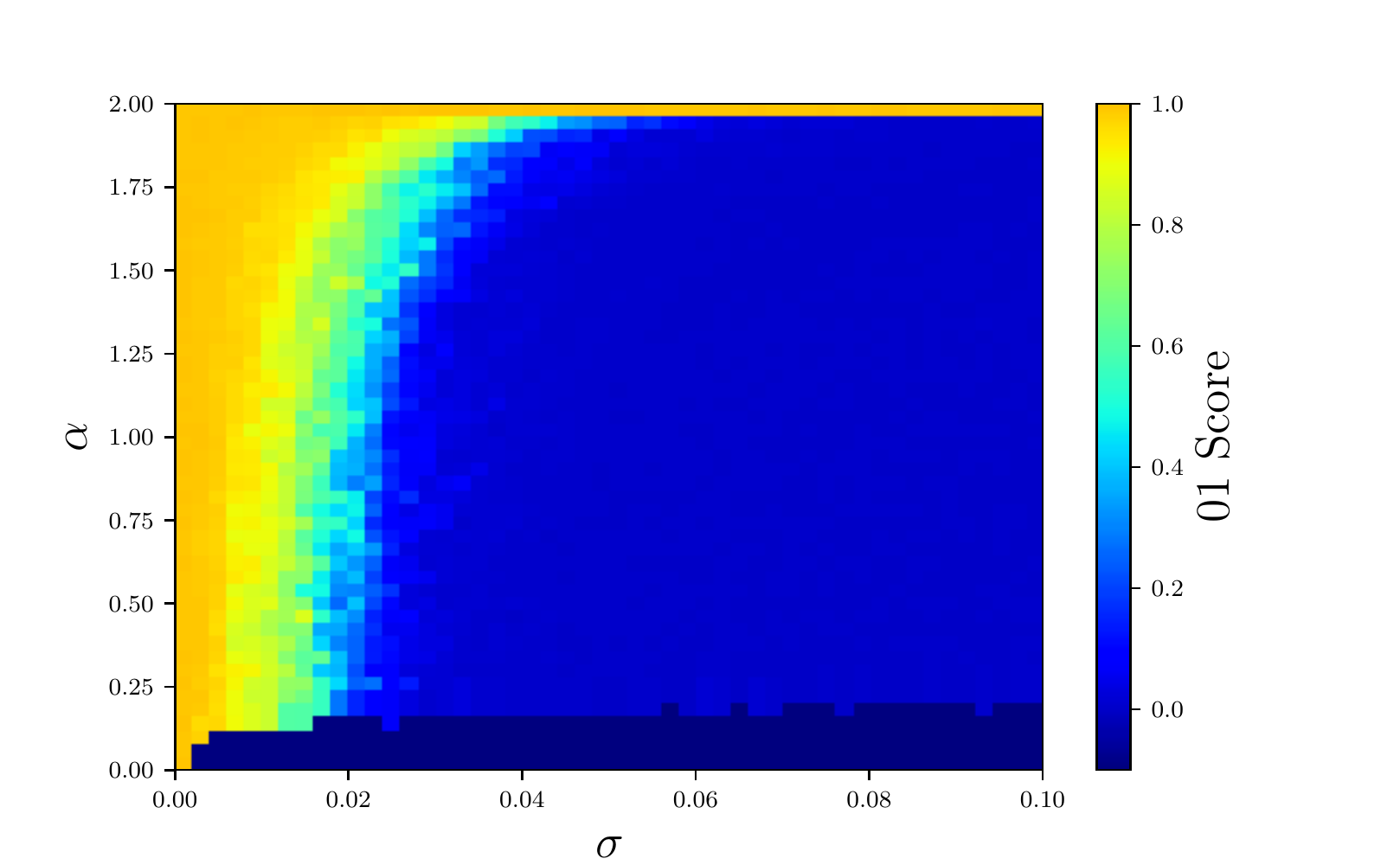}};
\draw  (figure.north)  node[anchor=north,black]{\textbf{\small{(b)}}};
\end{tikzpicture}
\end{subfigure}
\caption{Heat maps of the objective score function for $U^{(01)}(\alpha,\sigma)$ in the stable L\'evy noise case with $\phi = 0.96 \pi$, $\psi = 0.$ and 100 iterations for T = 6000; (a) is the one-sided and (b) is the two-sided case.  High scoring (dark yellow) values correspond to $O_R$ steady or periodic
behaviour, low scoring non-zero values (light blue) correspond to noise, and negative values (dark blue) correspond to
long transients in the times series. }
\label{fig:01teststable}
\end{figure}

\begin{figure}[hp] 
\begin{subfigure}[b]{1\linewidth}
\centering
\begin{tikzpicture}
    \draw node[name=figure] {\includegraphics[height=5cm]{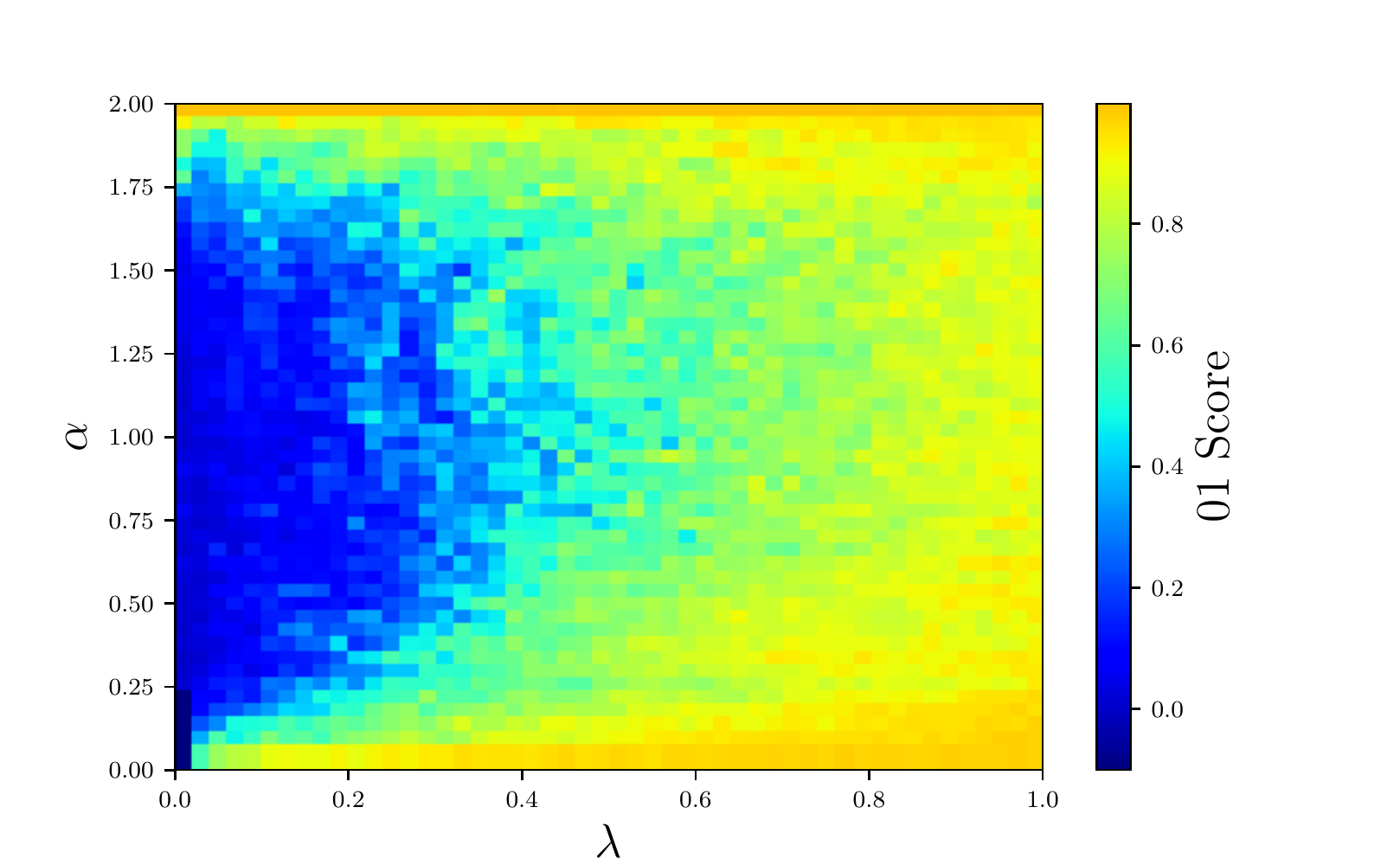}};
 \draw  (figure.north)  node[anchor=north,black]{\textbf{\small{(a)}}};
\end{tikzpicture}
\end{subfigure}
\begin{subfigure}[b]{1\linewidth}
\centering
\begin{tikzpicture}
    \draw node[name=figure] {\includegraphics[height=5cm]{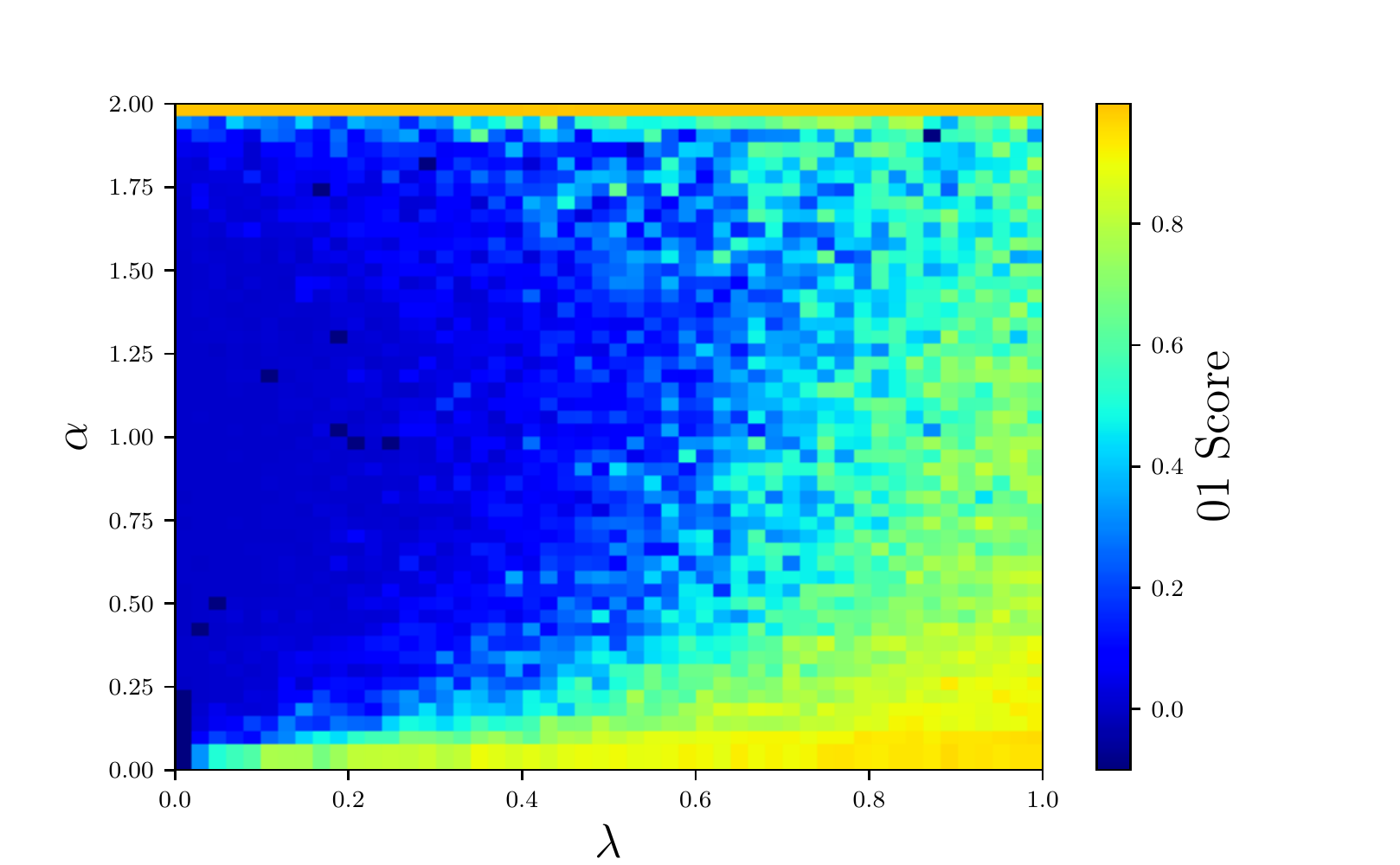}};
\draw  (figure.north)  node[anchor=north,black]{\textbf{\small{(b)}}};
\end{tikzpicture}
\end{subfigure}
\caption{Heat maps of the objective score function for $U^{(01)}(\alpha,\lambda)$ in the tempered stable L\'evy noise case,
with $\psi = 0.$, $\sigma = 0.05$ and 100 iterations for T = 6000;
top  for the one-sided $\phi=0.96\pi$ and bottom for the two-sided case with $\phi=0.96\pi$. High scoring (dark yellow) values correspond to $O_R$ steady or periodic
behaviour, low scoring non-zero values (light blue) correspond to noise, and negative values (dark blue) correspond to
long transients in the times series. }
\label{fig:01testTstable}
\end{figure}

We now conduct straightforward grid search through the range of parameters in presenting landscape plots.
We show two examples of computation of $U^{(01)}$ for stable ($\lambda=0$) and tempered stable (varying across $\lambda$), in 
Figs.\ref{fig:01teststable},\ref{fig:01testTstable}, where
yellow indicates ordered regions and blue noisy regions. We choose a case of $\phi=0.96\pi,\psi=0$ where ${\cal K}>0$ and therefore we know that
the deterministic dynamics
should be constant in quantities such as $O_B,O_R$ and $\Delta$. 

For the stable case, we show both one-sided (top panel) and two-sided (bottom panel) cases in  Fig.\ref{fig:01teststable}.
We observe that across all values of $\alpha$, for $\sigma>0.05$
there is an onset of noise with the strongest departures to noise at low $\alpha$ consistent with the property that tails in the noise
distribution become very heavy. Evidently the role of skew in the noise plays a marginal role.

Contrastingly, we now choose $\sigma=0.05$ and explore the tempered stable case in Fig.\ref{fig:01testTstable} for $\phi=0.96\pi$ for both the one-sided $(\theta=1)$ (top) and symmetric $(\theta=0)$ (bottom) cases. 
We see that close to the Gaussian regime
$\alpha\approx 2$, there is an absence of noisy behaviour across all values of $\lambda$, as to be expected. We also see that for $\lambda=0$,
going vertically down the left hand edge of the plot,
there is a transition from steady, to periodic behaviour, to noisy behaviour and then
long transients, reflecting the structure of the landscape for
the stable case in Fig.\ref{fig:01teststable}.
However, once $\lambda\neq 0$ this
monotonic trend from steady state to noisy behaviour changes.
For any fixed $0<\lambda<0.3$ as $\alpha$ decreases
there is a transition to noisy behaviour and then back to periodic
behaviour (yellow goes to blue which goes back to yellow).
Thus there is a {\it restoration} of ordered behaviour for all $\lambda$ for {\it small} values of $\alpha$,
typically $\alpha\leq 0.25$.

In other words,
Blue and Red will exhibit smoother periodic behaviour in
relation to each other. This is broadly consistent with the pattern observed analytically for the dynamics of $\Delta$ using the 
ratchet potential in the zero mode projection of the Blue-vs-Red system. Recall three significant
effects were seen to occur in that approach: 
the skew drops out, there
is a greater suppression of variance 
in $\Delta$ for small $\alpha$ for
non-zero $\lambda$, and there is a
reduction of the average drift between Blue and Red clusters. That
the skew does not play a role can be seen in the result for the landscape of 
for $U^{(01)}$ for zero skew (namely
symmetric two-sided noise) in the bottom panel, with $\phi=0.96\pi$ again. Once more we observe the suppression of noisy behaviour at small $\alpha$. Though the onset of noisy behaviour is more general
in the full system, applying between and within Blue and Red. However, the analytical result for the
variance of $\Delta$ and the average value
of $\dot{\Delta}$ shows that the generation, or
suppression, of variability in the zero mode of the system 
is {\it consistent} with this behaviour.
Of course there is need for caution, as the
analytic approach is not valid in the deep noisy
regime; the {\it onset} of drift and variation
from the noise is a precursor to the stronger
regime of noisy behaviour.
In this sense, we may say there is qualitative
agreement between numerical and analytical results
on the impact of small $\alpha$ in suppressing
the tendency to 
noisy behaviour of the nearly locked Blue-Red system
in the presence of tempering; increasing the
exponent of the tail of the distribution
does not disrupt Blue and Red locking
with monotonically increasing strength.
The observation of this,
and the broad consistency between analytical
and numerical approaches
are the main results of this paper.


\subsection{Optimisation across multiple parameters}

We have seen throughout the work that the non-linear nature of the interactions in the Blue-v-Red model mean that, even deterministically, it is not generally possible for one population, say Blue, to achieve the desired phase ahead of the other, for example $\phi$. We now use the Bayesian approach to search for optima {\it for Blue} across the four cross-population parameters $\phi,\psi,\zeta_{BR}, \zeta_{RB}$
given the Blue and Red networks and internal couplings $\zeta_B, \zeta_R$ as previously given. We conduct 500 iterations running the simulation out to $T=2000$ and consider four cases: deterministic, Gaussian (as used in
\cite{HolZupKall2017}), stable ($\alpha=1.5$) and tempered stable $(\alpha=1.5, \lambda=0.75)$. The results are shown in Table~\ref{tab:optimalvalues}. 
\begin{table}
{
\footnotesize
\centering
\begin{tabular}{| c | c | c c c c| }
  \hline	
  Noise & $U$ & \multicolumn{4}{|c|}{Parameters} \\
   & & $\psi$ & $ \phi $ & $\zeta_{BR}$ & $\zeta_{RB} $ \\
  \hline
  None & 57.8598 & 1.675$\pi$& 0.1944 $\pi$ & 0.8997 & 0.1628 \\
  Gaussian & 98.6311 & 1.3747$\pi$& 0.0033$\pi$ &0.8170 & 0.0668\\
  Stable & 10.6854 & 0.2007 $\pi$ & 0.0083 $\pi$& 0.8289 & 0.7124 \\
  T. Stable &42.0838 & 1.2693$\pi$ & 0.0143$\pi$ & 0.9932 & 0.1017\\
  \hline  
\end{tabular}
}
 \caption{Bayesian optimisation results for $U(\psi,\phi,\xi_{BR},\xi_{RB})$ where $\psi,\phi \in [0,2\pi]$ and $\xi_{BR},\xi_{RB} \in (0,1]$ for 500 iterations and T = 2000. $\sigma = 0.05$, $\alpha = 1.5$ and $\lambda$ = 0.75.}\label{tab:optimalvalues}
\end{table}

We note that the Gaussian case obtained a higher scoring optimal value than the deterministic case as a result of either a relativity small number of optimisation iterations for a 4-dimensional parameter space and the algorithm favouring a local maximum. The optimisation algorithm was unable to find a decently scoring set of parameters in the stable L\'evy noise case. We observe nevertheless a broad consistency between the three other cases in that
$\psi>\phi$ and $\zeta_{BR}>\zeta_{RB}$.  In the interests of space we forego here showing plots of the corresponding
time-series for $O_{B,R}$ and $\Delta(t)$ but simply comment that $O_B$ shows generally higher values than $O_R$, and
$\Delta(t)$ is a positive constant for deterministic, ($\Delta\approx 0.6$), and Gaussian $(\Delta\approx 0.01)$ cases.
For the tempered stable case, the expected value of $\Delta$ is approximately $0.06$, however the extreme cases are time-varying both positively
and negatively, not untypical for $1<\alpha<2$. The requirement that $\Delta$ is close to $\phi$ is part of the objective function $U_2$ ensures that Blue comes as close as possible to its objective by not seeking to be as ambitious in its phase ahead of Red as Red seeks in its relation to Blue. The actual values of $\Delta$ match $\phi$ for the deterministic and Gaussian cases. On the other hand, to achieve this Blue agents must seek to couple to Red agents more strongly than Red to Blue; the nature of the internal networks intuitively plays an important role in this, just as Blue's hierarchy requires stronger internal coupling to achieve its level of internal synchronisation. Stable noise is generally destructive of such optima outside the narrow range where one population complies with the intent of the other
(for example, $\phi=-\psi$), as previous results have emphasised. These results are simply a taste of what is possible with such a model. Future work may consider optimisation across more parameters and ultimately across different networks.

\section{Conclusions}

We have explored the two networked populations `Blue-vs-Red' frustrated Kuramoto model subject to tempered stable noise.
Using approximations in the region where the two populations are internally synchronised but seek to lock with respect to each other, we have solved the 
stochastic dynamics using the tempered fractional Fokker-Planck approach. The zero-mode projection under the graph Laplacian of this reveals 
that, whereas for $1<\alpha<2$, noise generally causes an otherwise deterministically locked system to degrade
leading to drift between Blue and Red
clusters, for small
$\alpha<1$  improved locking between Blue and Red takes place. In the situation where deterministically 
the system shows locking, the noise may be such that
stochastically induced drift between Blue and Red is entirely cancelled.
When the deterministic system exhibits drift between
Blue and Red clusters, 
such tempered stable noise can be used to moderate the stochastic drift. 

Through our simulations we see a similar pattern: noise of higher strength or stronger tails can disrupt the
locking between Blue and Red in cases where deterministically the system may be achieving high levels of global synchronisation - with both sets of agents
well synchronised internally and having a fixed frustration in relation to each other. Tempering, naturally, has the effect of restoring locking by moderating the heavy tails. The surprising result we find is that with tempering in the regime where $\alpha<1$ the system may be become less noisy, and thus showing smoother periodic, and
approaching steady-state, behaviour. Again, this is
consistent with Blue and Red either smoothly drifting
in relation to each other or even locking. There is only the
caveat that the analytical result does not apply in the
noisy regime, but provides an indicator or precursor
to it.

Thus both analytically and numerically we obtain a 
surprising result, that the
boundaries of regions of different dynamics (steady-state, periodic, noisy behaviour) of the stochastic
 system are not linear in variation of the parameter $\alpha$ governing the tail of the noise once tempering is introduced.
 In contrast, for purely stable noise reducing $\alpha$ only destroys locking
 between Blue and Red with monotonically increasing severity.
 
 Intuitively, we propose that the
 underlying reason for this behaviour lies
 in an interplay of the ratchet mechanism with tempering and the fundamental change in
 the stable noise probability density
 for $\alpha<1$. As illustrated in 
 \cite{RobKall2017}, for $\alpha<1$
 the density is no longer centred around
 zero but with a heavy tail that is
 increasingly suppressed as $\lambda$ increases. This centering
 becomes the dominant effect
 with tempering, and seemingly acts
 to counter the deterministic drift
 in the ratchet potential.
 
 Importantly, much of the structure of the landscape of the stochastic system, as determined through simulation, can be predicted through the analytically accessible thresholds for instability and periodic behaviour from the deterministic system at least with weak noise strength. Thus analytic understanding of the landscape in the absence of noise remains an important guide to the overall structure of the landscape of dynamics in the regime of weak noise, close to the boundary with noisy behaviour. These insights are then complemented by examining
the Laplacian zero mode projection of the linearised system, where the properties of ratchet potentials under tempered stable noise plays an important, and analytically tractable, role in predicting the weak noise strength dynamics.
This nonlinear mechanism, with tilted ratchet potentials,
has not been articulated before in the context of 
synchronisation on multi-networks.

 Interpreting these results in the context of the competitive system of Blue and Red agents is now interesting. Intuitively it is clear that if two populations both seek to maintain a certain phase advantage over the other it is not possible for both to simultaneously achieve their aims. For different network structures and couplings one side or the other may be able to achieve close to their result within a certain bound. Indeed, our results show that in such interactions `less is more', namely that one side seeking less ambitious phase advantage in relation to the other
 may be able to achieve more optimal results. The landscape method shown here enables
 determination of these boundaries before instability is triggered, and how close to optimality may be achieved. The added dimension of noise, in general, makes this optimal behaviour even more difficult. However, for certain forms of non-Gaussian noise with tempering $\lambda$ for large jumps 
 and far-from-Gaussian power law index $\alpha$ better approach to optimal behaviour may be achieved. Indeed, such behaviour becomes attractive as a mechanism for guiding the behaviour of otherwise complex competitive stochastic systems or
 informing an understanding of the capacity for
 resilience or robustness of the system. As alluded already, future work may broaden the scope of the optimisation to cover the networks themselves. For example, for a given topology, or statistical parametrisation thereof, of Red what is the optimal network within and across for Blue? Returning to our overall aim to apply such models to competitive distributed decision making or social processes, investigations like this with the two-network frustrated Kuramoto model open the door for testing many rather qualitative hypotheses in the organisational, social and management science literature.
 
%
%
%

\section*{Appendix A: Definitions for the linearised Blue-vs-Red model}
We define the quantities $d$ and $L$ used in the main body, drawing basic concepts from graph theory \cite{Boll98}. 
The {\it degree} of Blue agent at node $i \in {\cal B}$ is the number of links from node $i$ to other Blue agents,
\begin{eqnarray*}
d^{(B)}_i \equiv \sum_{j\in {\cal B}} {\cal B}_{ij}, \;\; i \in {\cal B}.
\end{eqnarray*}
The corresponding Blue {\it degree-matrix} ${\cal D}^{(B)}$ is a diagonal matrix with the degrees $d^{(B)}_i$ inhabiting the diagonal entries
\begin{eqnarray*}
{\cal D}^{(B)}_{ij} \equiv d^{(B)}_i \delta_{ij}, \;\;\{i,j\} \in {\cal B}. 
\end{eqnarray*}
The matrices $L$ in Eq.(\ref{linsys}) constitute a {\it graph Laplacian}, where the Laplacian for the Blue population is given by the expression
\begin{eqnarray*}
L^{(B)}_{ij}\equiv {\cal D}^{(B)}_{ij} - {\cal B}_{ij}, \;\;\{i,j\} \in {\cal B}.   
\end{eqnarray*}
Equivalent definitions for $L^{(R)}$ apply to the Red network.

Addressing the corresponding cross-network quantities, we define the degree with which a Blue agent at node $i$ connects to Red agents as,
\begin{eqnarray*}
d^{(BR)}_i \equiv \sum_{j\in {\cal B} \cup {\cal R}} {\cal M}_{ij} = \sum_{j\in {\cal R}} {\cal A}^{(BR)}_{ij},\;\; i \in {\cal B}. 
\end{eqnarray*}
We note that in this instantiation of the model ${\cal A}^{(BR)}$ is the transpose of ${\cal A}^{(RB)}$, and vice-versa (it need not be so, however), so there is symmetry between the total number of degrees between the Blue and Red networks:
\begin{eqnarray}
d^{(BR)}_T = d^{(RB)}_T= \sum_{i \in {\cal B}}d^{(BR)}_i =\sum_{i \in {\cal R}}d^{(RB)}_i .
\label{totaldegree}
\end{eqnarray}
The diagonal degree matrix ${\cal D}^{(BR)}$ which encodes all the Blue to Red links is given by,
\begin{eqnarray*}
{\cal D}^{(BR)}_{ij} \equiv d^{(BR)}_i \delta_{ij},\;\; i \in {\cal B}, \;\; j \in  {\cal B} \cup {\cal R}   ,
\end{eqnarray*}
where the final $M$ diagonal entries of ${\cal D}^{(BR)}$ are zero. This finally leads to the cross-network Laplacian from Blue to Red
\begin{eqnarray*}
L^{(BR)}_{ij}\equiv {\cal D}^{(BR)}_{ij} - {\cal M }_{ij}, \;\; i \in {\cal B}, \;\; j \in  {\cal B} \cup {\cal R}  .   
\end{eqnarray*}
Similar considerations lead to an equivalent expression for the Red to Blue cross-network Laplacian $L^{(RB)}$. 

The Blue and Red network Laplacians  obey the following eigenvalue equations,
\begin{align*}
\sum_{j \in {\cal B}}L^{(B)}_{ij}e^{(B,r)}_j =& \lambda^{(B)}_r e^{(B,r)}_i \;\; i \in {\cal B} , \\ \sum_{j \in {\cal R}}L^{(R)}_{ij}e^{(R,r)}_j =& \lambda^{(R)}_r e^{(R,r)}_i \;\; i \in {\cal R}.
\end{align*}
The graph eigenvalues are well-studied objects \cite{Boll98}. For instance, the \textit{zeroth}-eigenvalue is always zero valued, and the remaining eigenvalues are all real, positive, semi-definite and can be ordered as follows,
\begin{eqnarray*}
0 = \lambda^{(B)}_0  \le \lambda^{(B)}_1 \le \lambda^{(B)}_2 \le \dots \le \lambda^{(B)}_N 
\end{eqnarray*}
where we have used the Blue network as an example. The normalised zero eigenvectors are
\begin{eqnarray*}
e^{(B,0)}_i = \frac{1}{\sqrt{N}} \;\; i \in {\cal B}, \;\; e^{(R,0)}_i = \frac{1}{\sqrt{M}} \;\; i \in {\cal R}.
\end{eqnarray*}
They provide an alternate expression for the Blue and Red centroids:
\begin{eqnarray*}
B =\frac{1}{\sqrt{N}} \sum_{i \in {\cal B}}\beta_i e^{(B,0)}_i, \;\; P = \frac{1}{\sqrt{M}} \sum_{j \in {\cal R}}\rho_j e^{(R,0)}_j.
\end{eqnarray*}

Projections onto the eigenvectors are:
\begin{eqnarray*}
\omega^{(s)} = \sum_{i \in {\cal B}}\omega_i e^{(B,s)}_i, \;\; \bar{\omega} = \frac{1}{N}\sum_{i \in {\cal B}}\omega_i, \;\;d^{(BR)}_s =  \sum_{i \in {\cal B}}d^{(BR)}_i e^{(B,s)}_i,
\end{eqnarray*}
and equivalent expressions hold for the relevant Red quantities. 

Fluctuations $b_i$ and $p_j$ are expanded in the following non-zero \textit{normal-modes}:
\begin{eqnarray}
b_i = \sum_{r \in {\cal B}^E / \{0\}} x_r e^{(B,r)}_i \;\; i \in {\cal B}, \;\; p_j = \sum_{r \in {\cal R}^E / \{0\}} y_r e^{(R,r)}_j \;\; j \in {\cal R}.
\end{eqnarray}

\section*{Appendix B: Networks and frequencies used in numerical calculations} \label{APP_B}
\begin{figure}[h]
\centering
\includegraphics{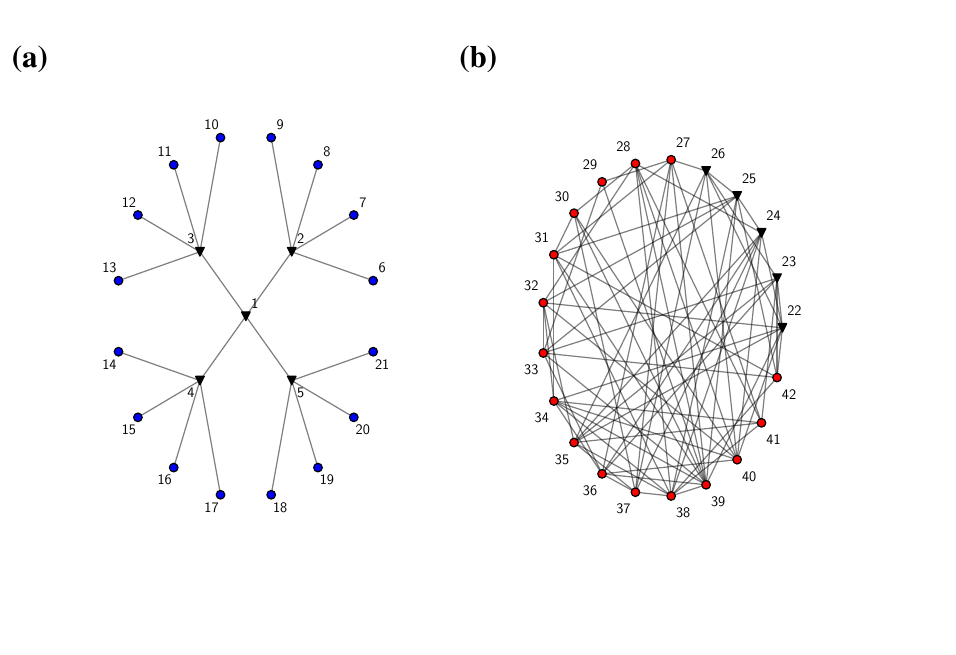}
\caption{ (a): The Blue network and (b): The Red network used in the numerical simulation of Eq.\eqref{BR-eq} with noise.}
\label{fig:BvsR-networks}
\end{figure}

As mentioned in the main body, a tree network was used for Blue and a random graph for Red.
Blue-to-Red interactions are arranged such that each leaf-node of Blue ($i=6,\dots,21$) interacts with the correspondingly labelled Red node ($i=27,\dots,42$), shown as open circles in Fig.\ref{fig:BvsR-networks}. Thus
$d_i^{(BR)}=d_i^{(RB)}= 1 \ \rm{or} \ 0$, for agents engaged, respectively not engaged, with a competitor, but $d^{(BR)}_T=16$.

\begin{figure}[h]
\begin{center}
\begin{subfigure}[b]{1\linewidth}
\centering
\begin{tikzpicture}
    \draw node[name=figure] {\includegraphics[width=\columnwidth]{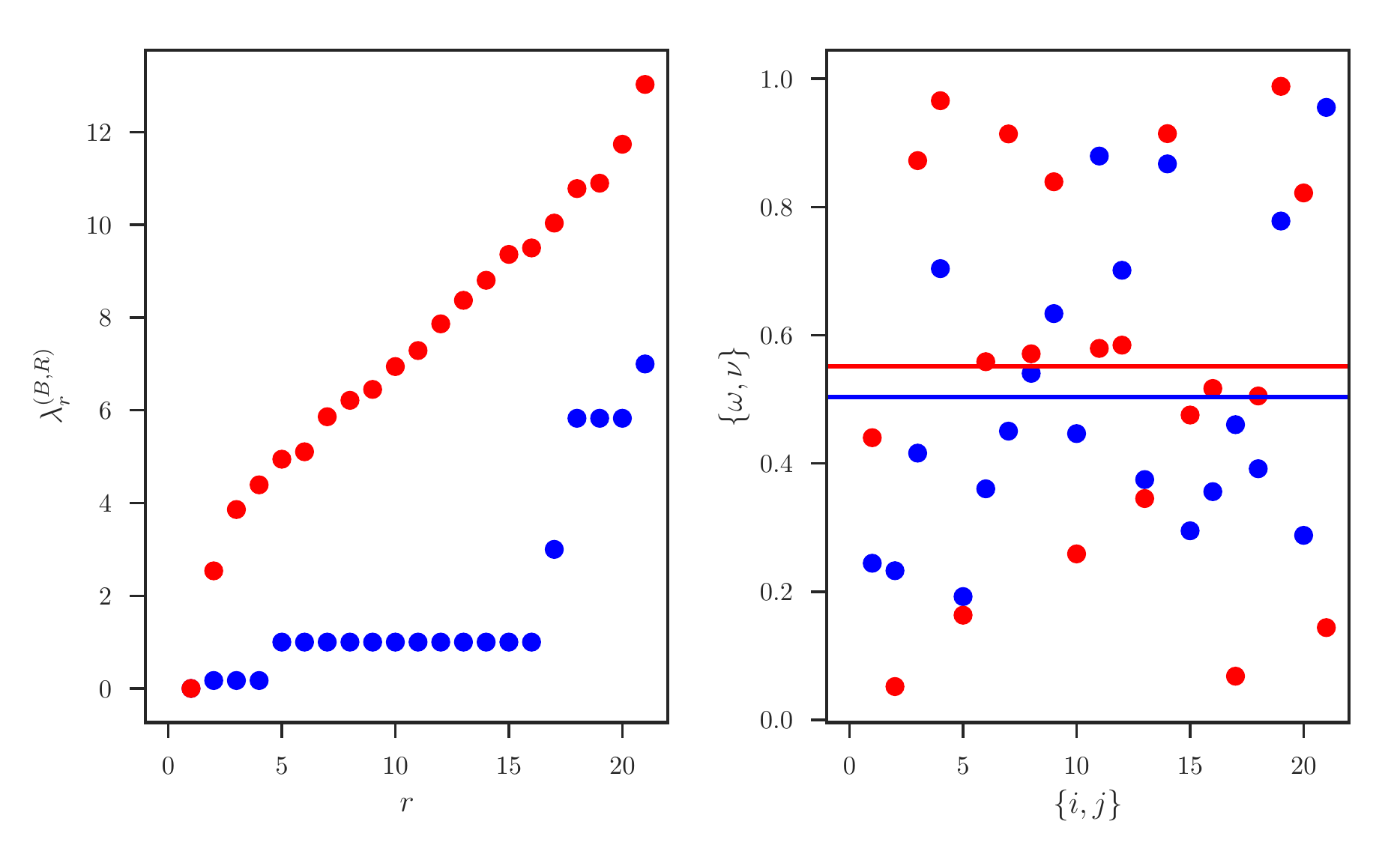}};
    \draw  (figure.north west)  node[anchor=north west,black,yshift=10]{\textbf{\small{(a)}}};
    \draw  (figure.north)  node[anchor=north,black,yshift=10]{\textbf{\small{(b)}}};
\end{tikzpicture}
\end{subfigure}
\caption{(a): The spectrum of the graph Laplacian for Blue and Red networks coloured respectively Blue and Red. 
(b): The frequencies for Blue and Red agents according to the node labelling again coloured blue and red respectively, with solid lines indicating their corresponding means.} 
\label{spectra}
\end{center}
\end{figure}
Given the significance of the graph Laplacians we show their spectra and the frequencies of each agent in Fig.\ref{spectra}. Note here that there are 
many more low lying eigenvalues for $L^{(B)}$ compared to $L^{(R)}$ - a 
consequence of the comparatively poor connectivity of the tree graph
(Fig.\ref{spectra} (a)).


\section*{Acknowledgements}
ACK is supported through a Chief Defence Scientist
Fellowship and is grateful for ANU hospitality.
This research was a collaboration between the Commonwealth of Australia
(represented by the Defence Science and Technology Group) and the Australian
National University through a Defence Science Partnerships agreement.

\end{document}